\documentclass{appolb}
\usepackage{graphicx}
\usepackage{amsmath,hyperref,amssymb}
\def\be{\begin{equation}}
\def\ee{\end{equation}}
\def\bea{\begin{eqnarray}}
\def\eea{\end{eqnarray}}
\def\sp{\;\;\;,\;\;\;}
\def\l{\lambda}
\def\lab{\label}
\def\f{\Phi}

\def\e{\epsilon}
\def\o{\omega}
\def\le{\left}
\def\ri{\right}

\def\half{\frac12}
\def\q{\theta}
\def\cO{{\cal O}}
\def\m{\mu}
\def\n{\nu}

\def\d{\delta}
\def\6{\partial}
\def\de{\partial}

\def\ls{\ell_s}
\def\a{\alpha}
\def\b{\beta}
\def\lab{\label}

\def\l{\lambda}

\def\tr{\textrm{tr}\,}
\def\m{\mu}
\def\n{\nu}
\def\bet{\begin{itemize}}
\def\eet{\end{itemize}}
\def\ben{\begin{enumerate}}
\def\een{\end{enumerate}}
\def\la{\langle}
\def\ra{\rangle} 
\def\eps{\epsilon}
\def\gcs{\Gamma_{\textrm{CS}}}
\def\ncs{N_{\textrm{CS}}}
\def\nn{\nonumber}

\begin{document}

\title{Improved Holographic QCD and the Quark-gluon Plasma%
\thanks{Presented at 56. Cracow School of Theoretical Physics, May 24 --- June 1, 2016, Zakopane, Poland}%
}
\author{Umut G\"ursoy
\address{Institute for Theoretical Physics and Center for Extreme Matter and Emergent Phenomena, Utrecht University, Princetonplein 5, 3584 CC Utrecht, The Netherlands}
\\
}

\maketitle

\begin{abstract}
We review construction of the improved holographic models for QCD-like confining gauge theories and their applications to the physics of the Quark-gluon plasma. We also review recent progress in this area of research. The lecture notes start from the vacuum structure of these theories, then develop calculation of thermodynamic and hydrodynamic observables and end with more advanced topics such as the holographic QCD in the presence of external magnetic fields. This is a summary of the lectures presented at the 56th Cracow School of Theoretical Physics in Spring 2016 at Zakopane, Poland. 
\end{abstract}

\newpage
\tableofcontents 
\newpage

\section{Introduction: AdS/CFT and heavy ion collisions}

Quantum Chromodynamics is well established as the theory of strong interactions that governs the substituents of atomic nuclei, namely the quarks and gluons. Among the salient features of   
QCD, the most important ones are the {\em asymptotic freedom}, {\em confinement} and {\em chiral symmetry breaking}. Even though the theory is well-defined at the level of the Lagrangian, the first aforementioned property, the negativity of the beta-function of QCD, makes it extremely hard to do calculations in QCD in the IR with traditional methods of quantum field theory. Instead, a more fruitful avenue to calculate observables such as the correlation functions of gauge invariant operators, the hadron spectra, and thermodynamic  functions at finite temperature is the lattice QCD. Indeed, placing the theory on a Euclidean lattice with finite spacing can be viewed as the true definition of the theory. Then the observables listed above are obtained with great accuracy from the continuum limit of Euclidean correlation functions. I will not be concerned with the lattice calculations in these notes, apart from presenting a collection of lattice results for comparison purposes. Hence I refer the interested reader to the extensive literature on the subject.  

Having said that, the lattice QCD also has a few disadvantages. The most prominent among these is the fact that calculation of {\em real-time observables} and study of time-dependent phenomena such as the retarded Green's functions, transport coefficients and thermalization are plagued  with systematic and statistical uncertainties. This is because, the lattice QCD being inherently a Euclidean formulation, any quantity that is extracted from a real-time correlator such as the conductivity, shear and bulk viscosity etc require analytic continuation of the Euclidean correlators to real-time which in turn requires the knowledge of full spectral density. For these reasons, an alternative method for calculations of such quantities in the non-perturbative regime is very much in demand. This is especially important in view of applications to dynamics of the quark-gluon plasma produced in the heavy ion collision experiments at RHIC, Brookhaven and LHC, CERN. 

The AdS/CFT correspondence or more generally holography\cite{adscft} provides such an alternative formulation. The correspondence maps the QFT in the limit of large coupling constant, for example the IR regime of QCD-like gauge theories, to a semi-classical theory of gravity in at least one higher dimension and yields an alternative effective and non-perturbative description for such theories. The detailed map to gravity is best understood in the original example \cite{Maldacena} of ${\cal N}=4$ super Yang-Mills theory in 4D, where the gravitational dual is established as the IIB string theory on $AdS_5\times S^5$ background. The next well understood case in 4D are the theories that can be obtained from ${\cal N}=4$ sYM by relevant or marginal deformations. In such theories, generally, there exists the following correspondence between the parameters on the two sides: 
\be\lab{dic1}
g_s \sim g_{YM}^2, \qquad R\ell_s^2 \sim (g_{YM}^2N_c)^{-\half}\, ,
\ee
where $g_s$ is the string coupling constant and $R\ell_s^2$ is the Ricci curvature of the gravitational background in string units, $g_{YM}$ is the coupling constant of the gauge theory, and $N_c$ the rank of the gauge group (the number of colors). The computationally tractable limit of the AdS/CFT therefore corresponds to the 't Hooft limit\cite{tHooft1}: 
\be\lab{thooft}
N_c \to \infty, \quad g_{YM}\to 0, \quad \lambda_t \equiv g_{YM}^2 N_c \gg 1\, ,
\ee 
where the combination $\l_t$ is called the 't Hooft coupling. This limit kills three birds with one stone: it gets rid of the complications arising from string interactions by making $g_s$ small; it reduces the string theory that effectively contains arbitrarily high derivative terms in the effective action to two-derivative Einstein's gravity by making the curvature small; it focuses on the strong (effective) coupling limit of the gauge theory that is the non-perturbative regime we are interested in.  I will only consider this limit in the rest of these notes and explain the construction of effective holographic theories for QCD in section \ref{holoQCD}. But before we come to that we should ask: {\em what do we want to learn from holographic QCD?}

\begin{figure}[h!]
 \begin{center}
\includegraphics[scale=0.9]{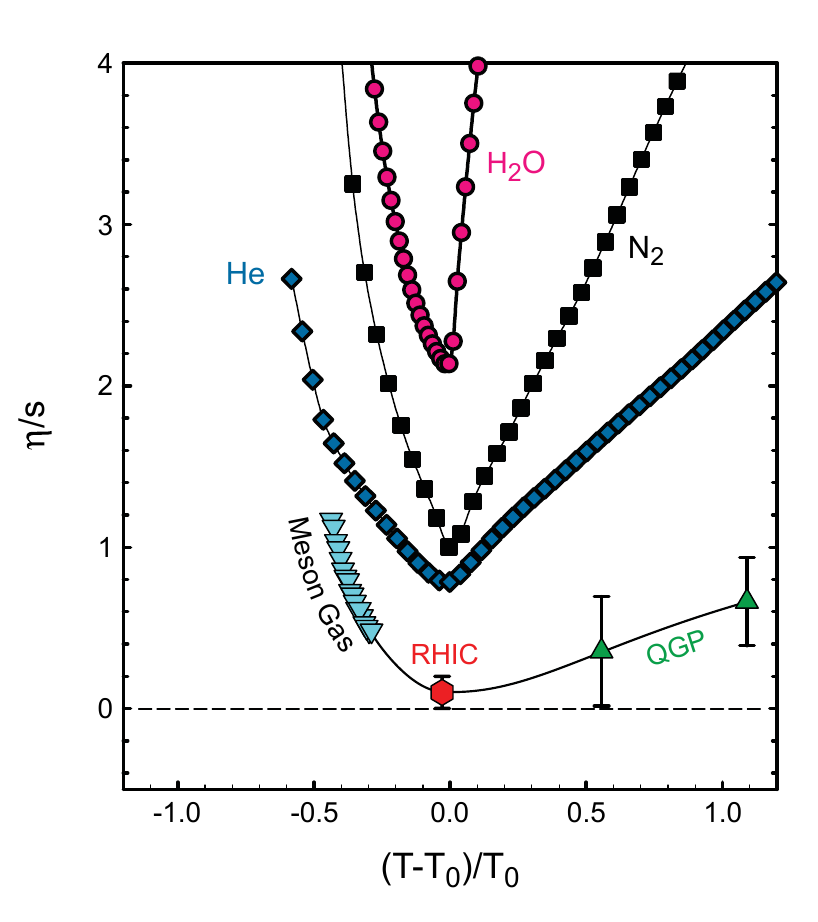}
 \end{center}
 \caption[]{Comparison of the shear viscosity to entropy ratio of QGP with various ordinary matter.}  
\label{fig1}
\end{figure}

One of the main objectives of such an effective theory is to understand the real-time dynamics in the {\em quark-gluon plasma} produced at the heavy ion collisions at RHIC, Brookhaven and LHC, CERN. Heavy ion collisions are gateways to extreme phenomena in nature. The QGP is the most extreme fluid we find in the universe: it has an extremely small viscosity ($\eta/s \sim 0.08-2$), that is very close to an ideal fluid with vanishing viscosities, see figure \ref{fig1}; it is produced at extremely high temperatures (about 450-600MeV); and the largest magnetic fields known (about   $10^{18}-10^{19}$ Gauss) in the universe are generated in the off-central heavy ion collisions. Another extremity is the fact that this fluid is very strongly coupled. This for example can be inferred from the fact that the plasma has a very small shear viscosity: a perturbative QCD calculation instead gives $\eta/S \propto \lambda^{-2}/\ln(1/\lambda)$ for small $\lambda$ in the large N limit. However, comparison of HIC data to hydrodynamics simulations lead to a value $\eta/s \sim 0.08-2$ which greatly disagrees with the perturbative result that diverges in the weak coupling limit. In these notes we start with the assumption that large N QCD at strong coupling yields a better approximation to calculate the observables of the system and we explain how to determine these observables using holographic methods. One can list the kind of observables we are interested in, in an order of increasing difficulty as follows: 

\begin{itemize}
\item Firstly we want to extract the spectrum of hadrons in the $T=0$ vacuum state. As explained in section \ref{holoQCD} below, holographic QCD can capture at most spin-2 operators, hence we will calculate the spectra of glueballs and mesons\footnote{Baryon spectra in improved holographic QCD has not been calculated and it is an open problem.} in  and match to the available lattice QCD results section \ref{ihqcdvac}.  

\item Next level in difficulty is to calculate the thermodynamics of the system. We shall discover that generically there exists a first order confinement-deconfinement transition at some finite temperature. In the holographic dual the confined state corresponds to the so-called ``thermal gas'' and the deconfined state to the black-brane geometries. We shall then calculate the thermodynamic functions in section \ref{thermo}, such as the free energy, entropy and energy density as a function of T in the deconfined state, again comparing with available lattice QCD data.   

\item The next level is to consider the small 4-momenta expansion in hydrodynamics. The zeroth order in this expansion is completely determined by the thermodynamic quantities. At the first order there appears two non-trivial transport coefficients, the shear and the bulk viscosities, which we will again calculate in the holographic model  in section \ref{hydro} and compare with available data. The Chern-Simons decay rate is another transport coefficient that appears in the CP-odd sector of the theory that we calculate as well, in section \ref{finiteB}. 

\item Another set of important observables in the physics of QGP consists of the hard probes. These are highly energetic quarks produced during the early phase after the collision and since they do not thermalize due to their high energy --- that is to say, they can travel through the plasma losing a portion of their energy, yet can make it to the detector --- the quantities that characterize their energy loss, such as the {\em Jet quenching parameter} and {\em diffusion coefficients}, contain crucial information on the QGP.  Determination of these quantities in holographic QCD will be discussed in section \ref{hardprobes}. 

\item Finally we shall discuss calculation of the new observables that arise in the presence of an external magnetic field B in section \ref{finiteB} that is the generic situation in off-central heavy ion collisions. 

\end{itemize}  

This list also serves as a plan of this review. We open the review in the next section with an introduction to holographic QCD theories in general and end it with a discussion of the topics that are left out present a look ahead.  We provide quantitative evidence for success of a particular holographic QCD model in these notes, called the {\em improved holographic QCD} \cite{ihqcd1,ihqcd2,ihqcd3} and will mostly work with this model.  To keep this review short we do not derive all but some of the results and refer to literature for derivations. There exists some reviews on the improved holographic QCD: see \cite{Gursoy-review} for a comparison of thermodynamics of ihQCD with other existing holographic QCD models and \cite{ihqcd-review} for an extensive review of the subject. 

\begin{figure}[h!]
 \begin{center}
\includegraphics[scale=0.9]{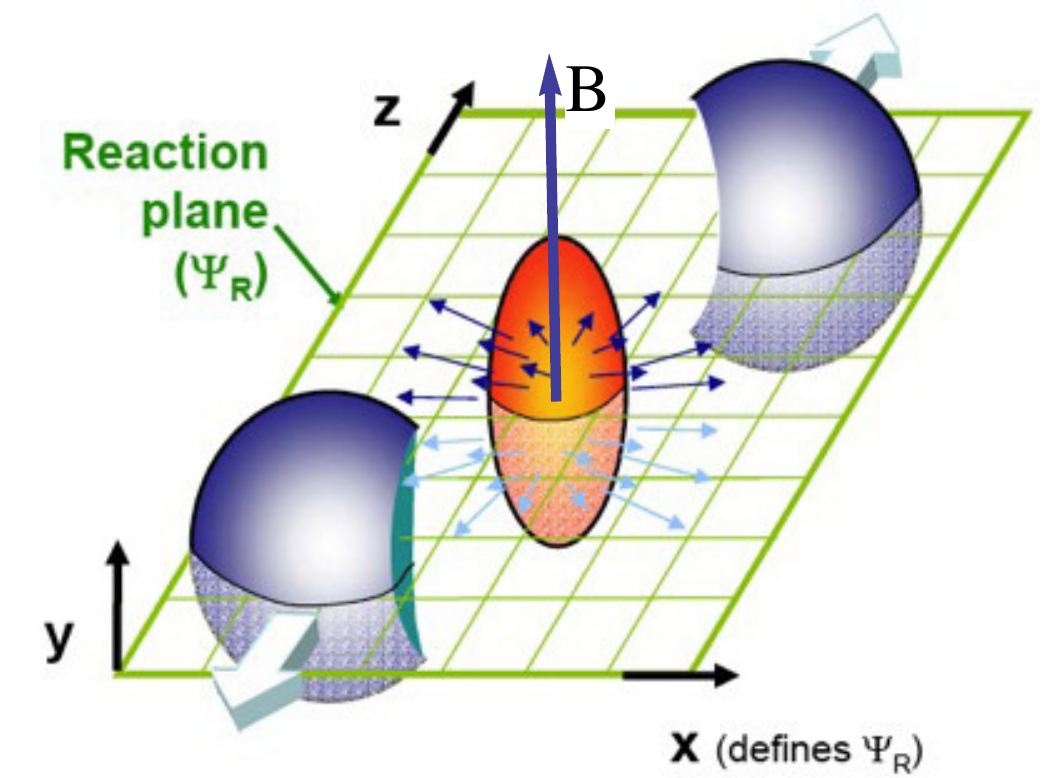}
 \end{center}
 \caption[]{Production of the QGP in the heavy ion collisions.}  
\label{fig2}
\end{figure}


\section{Holographic QCD theories} 
\lab{holoQCD}

{\it Top-down approach}: QCD as well as other confining gauge theories are different than the ${\cal N}=4$ sYM theory and its deformations  and the holographic duality map in this case is much less understood. The top-down approach to holographic formulation of QCD-like theories starts from a certain D-brane set-up in string theory, such as $N_c$ D4 branes wrapped on an $S^1$ in IIA string theory in 10D \cite{Witten1} and taking the so-called {\em decoupling limit} \cite{adscft-review} that replaces the D-brane set-up with a gravitational background of geometry and the various form fields.  
This approach has later been generalized to include the flavor dynamics in QCD by adding D8 flavor branes \cite{SakaiSugimoto1}, \cite{SakaiSugimoto2}. Even though such a top-down approach to holography has the enormous advantage of providing a precise dictionary between the QFT on the D-branes and the dual gravitational quantities, it usually results in a theory that is {\em different} than only QCD or pure Yang-Mills theory, but also contains an additional sector in its Hilbert space spanned by infinitely many operators that arise from the KK-modes in the 5 extra dimensions. See \cite{Gursoy-review} for example for a short discussion. Actually the ${\cal N}=4$ sYM is not different in this perspective, as it also contains operators arising from the extra $S^5$, but these operators turn out to be in precise correspondence with operators with higher conformal dimension \cite{adscft-review} and moreover there exists a well-controlled limit of low energy, where a subsector of the Hilbert space that contains the ${\cal N}=4$ super-multiplets of the energy-momentum tensor and flavor currrents can be identified and be put in correspondence with the low-lying gravitational fields. One is not as lucky with the non-conformal, confining gauge theories, essentially because in such theories there always exists an additional energy scale analogous to the dynamically generated IR energy scale $\Lambda_{QCD}$ in QCD that breaks conformality. The Hilbert space of such theories contain operators of arbitrarily large spin and scale dimensions, all proportional to  
$\Lambda_{QCD}$, and there exist no parametric separation in this Hilbert space of the aforementioned low-lying operators from the rest. Because, unlike in a conformal theory the energy scale is not a moduli, and one cannot tune to IR to achieve such decoupling of the low-lying operators. This problem corresponds in the dual language to the fact that both the pure Yang-Mills sector and the KK-states mentioned above are governed by the energy scale $\Lambda_{QCD}$. One needs to take limit of small radius of the cycles in the transverse space in order to decouple these KK-operators from pure Yang-Mills\footnote{See \cite{GursoyNunez} for a suggestions to bypass this problem. The problem always shows up in different guises however, \cite{Gursoy1}.}, and this limit results in large curvatures, necessitating inclusion of higher derivative terms in the dual string action. All in all, it is fair to say that, one needs full higher derivative string theory to study QCD-like theories in the top-down approach.  

{\it Bottom-up approach}: Faced with the difficulties of the top-down approach, different, more direct ways to capture the IR dynamics of QCD in holography have been sought for since mid 00s. It is hard to point to a single reference for this approach, some of the oldest and most notable papers being \cite{PolchinskiStrassler}, \cite{KupersteinSonnenschein}, \cite{KlebanovMaldacena}, \cite{Bigazzi}, \cite{Erlich}, \cite{Pomarol}, \cite{Karch}, \cite{Csaki}, \cite{ihqcd1, ihqcd2}, \cite{GubserMimic1}. The basic idea is to give up the ambitious goal of finding a precise holographic dual to QCD, but to construct an IR effective theory to capture the IR dynamics of relevant and/or marginal operators in the theory. 

The early bottom-up models \cite{Erlich}, \cite{Pomarol}, sometimes called the ``hard-wall'' models consisted of an $AdS_5$ space terminating at a hard-wall at some location in the deep interior, to introduce the scale $\Lambda_{QCD}$ and effectuate breaking of conformal symmetry. The main advantage of this model is its simplicity, calculations being almost identical to AdS. However, it leads to unrealistic results when applied to QCD such as vanishing trace anomaly, vanishing bulk viscosity, completely unrealistic behavior of thermodynamic functions in T, etc. It also leads to various uncertainties in the hadron spectra due to the various possible boundary conditions one can impose at the hard cut-off. It also has the unrealistic feature of having a quadratic spectrum $m_n^2 \propto n^2$ for large excitation number $n$. The ``soft-wall'' model was invented in \cite{Karch} to overcome these difficulties. In these models the background consists of the $AdS_5$ metric and a dilaton field whose profile is chosen by hand to obtain realistic features. The main purpose of \cite{Karch} was to describe well the ``mesonic'' physics that follows from the space-filling ``flavor'' branes embedded in this geometry. The model indeed fulfils this purpose, however it leads to unrealistic features in the ``glue'' sector and in thermodynamics. See the short review \cite{Gursoy-review} where a comparison of the ``hard-wall'', ``soft-wall'' and improved holographic models is provided. Almost all of these undesired problems are solved in the ``improved holographic QCD'' models. These can be thought of making the soft-wall theory dynamical: in these models, instead of starting with a background designed by hand one finds the desired background by minimizing Einstein's gravity coupled to a scalar field. Below we explain the general construction of such theories.   

{\it Improved holographic QCD:} There exist various indications in the QCD literature using arguments based on the sum-rules \cite{sumrules} and the operator product algebra that a sector of relevant and marginal {\em low-lying} operators can be treated separately from the rest of the Hilbert space of operators. Now the task is somewhat simpler: 
to construct an effective theory that correctly captures the physics that involves these low-lying operators in QCD using the basic ingredients from holography. The theory we aim at is $SU(N_c)$ gauge theory in the large $N_c$ limit. We should then ask the question 
{\em what should be the minimal ingredients of the holographic dual of such a theory?} 
\begin{itemize}

\item First of all we need one additional ``holographic'' dimension $r$ corresponding to the RG energy scale in the dual gauge theory. Therefore the theory we look for is in general a solution to a 5D non-critical string theory. We know very little about non-critical string theories. However, as we only aim at the IR physics where the coupling constant is large, we expect to be able to approximate this theory by a two-derivative gravitational action. The higher derivative corrections are then expected to be important only in the UV. 

\item There are three relevant/marginal operators in the large $N_c$ limit: the stress tensor $T_{\m\n}$, the scalar glueball operator $\tr F^2$ and the axionic glueball operator $\tr F \wedge F$. The other operators that one can construct out of the gluon fields $A_\mu^a$ have higher scale dimension in the IR. Moreover, as we discuss in section \ref{finiteB} the physics of the last operator, $\tr F \wedge F$ is suppressed by $1/N_c$ in the 't Hooft limit, hence can be treated as a perturbation on the background of the first two operators. Using the general AdS/CFT dictionary, $T_{\m\n}$ should be dual to the 5D metric $g_{\m\n}$ and the operator $\tr F^2$ should correspond to the dilaton field $\Phi$ in the 5D bulk. The operator $\tr F^2$ couples to the Lagrangian as $1/g_{YM}^2 \tr F^2 $ and in general in string theory (also in non-critical sting theory) the coupling constant $g_{YM}^2 = g_s$ is determined by the asymptotic value of the ``dilaton'' field that is a massless scalar field. The massless bulk fields correspond to marginal operators in the dual field theory, which is indeed the case for the operator $\tr F^2$ in the UV.  
Therefore the minimal theory we look for is an Einstein-dilaton theory with a dilaton potential $V(\Phi)$. 
   
\item In order to apply the rules of AdS/CFT we need the solutions to approach the $AdS_5$ space-time asymptotically near the conformal boundary. However we do not want AdS isometries all the way to the deep interior of the space-time. In particular we want the scaling isometries be broken. In QCD-like confining gauge theories, the corresponding scaling   
symmetry is  broken by the running coupling constant. Since the coupling constant corresponds to the dilaton field, and since the RG energy scale is related to the holographic coordinate $r$, energy scale dependence of the coupling constant translates into $r$ dependence of $\Phi$. To achieve such a non-trivial dependence, one then needs a non-trivial potential $V(\Phi)$ for the dilaton\footnote{A constant potential would lead to a pure $AdS_5$ space with constant dilaton that would then correspond to a conformal field theory instead.}. This potential should then be in correspondence with the beta-function of the dual field theory (see below for details). The consistency of this restriction to the low lying subsector of operators, and the fact that the physics of this sector is determined by the beta-function follows from the trace Ward identity 
\be\lab{Wardid} 
T^\m_\m = \frac{\beta(\l)}{4\l^2} \tr F^2\, .
\ee

\item Another physical requirement in the kind of theories we want to study is the linear confinement of quarks, that the potential energy between a test quark and an anti-quark is $V_{q\bar{q}} = c\, L + \cdots$ for $L\gg 1$ where $L$ is the distance between the test charges. In the holographic dual the test quarks are realized as end-points of open strings on the boundary. Therefore linear confinement translates into the statement that the  Nambu-Goto  action of this probe string behaves linear in $L$ for large distances. As shown below, this requirement restricts the large $\Phi$, IR behavior of the dilaton potential to be of the form: 
\be\lab{VIR} 
V(\Phi) \propto e^{\frac43 \Phi} \Phi^P, \quad P>0, \quad \textrm{or}\quad  V(\Phi) \propto e^{Q \Phi},\,\, Q>4/3 \qquad \Phi\gg 1\, .
\ee

\item The construction above applies to the gauge theories in the large-N limit with a finite number of flavors. The flavor sector in this theory corresponds to the space-filling D4 branes \cite{Bigazzi, Casero}. Contribution of the flavor branes to the total gravitational action is proportional to the number of flavors $N_f^2$. In the limit $N_c\to\infty$, $N_f$ finite this contribution is proportional to $N_f/N_c\to 0$ (see equation (\ref{action})), and these branes can be treated perturbatively. For many interesting applications to QCD however, we need a more realistic value $N_f/N_c =1$ or $2/3$. To capture this behavior in the large N limit then one needs to consider the Veneziano limit
\be\lab{Veneziano} 
N_c, N_f \to \infty,\qquad N_f/N_c = \textrm{finite}\, .
\ee   
In this limit the flavor branes cannot be treated as a perturbation. Instead one should consistently solve the coupled gravitational system of $g_{\m\n}$, $\Phi$ and the low lying fields on the flavor branes. The latter are given by a complex ``open tachyon'' field $T$, gauge-fields on the flavor branes $A_{L,\mu}^a$ and on the anti-flavor branes $A_{R,\mu}^a$. The tachyon $T$ corresponds to the quark-anti-quark condensate operator $\bar{q} q$ and the gauge fields correspond to the currents of flavor symmetry $SU_L(N_f) \times  SU_R(N_f)$. There are then extra physical requirements on the flavor section of the holographic dual from chiral symmetry breaking and the flavor anomalies. This will be discussed in section \ref{flavor}. 
\end{itemize} 
 
 \section{Improved holographic QCD - construction of the theory}
 \lab{ihqcdvac}

As motivated above, we take the following Einstein-dilaton action as our starting point\footnote{The unconventional normalization of the dilaton kinetic term is motivated by the underlying non-critical string theory in 5D \cite{ihqcd1}. This can be brought back to the conventional form with a $1/2$ by the rescaling $\phi \to\sqrt{3/8} \phi$ in the following formulae.}: 
\be\lab{action} 
S = M_p^3 N_c^2 \int \sqrt{-g} d^5 \le( R - \frac43 (\6\Phi)^2 + V(\Phi)\ri) + \textrm{GH} + S_{ct}\, ,
\ee
where $M_p$ is the Planck energy scale of the 5D theory (that will be fixed below) and we made the $N_c$ dependence of the action explicit. Here ``GH'' term is the Gibbons-Hawking term that is included to make the variational problem of the metric well-defined  on geometries with  boundary, and the last term is the standard counter-term action, necessary to obtain 
a finite value for the on-shell action on geometries with infinite volume such as the asymptotically AdS space-times we are interested in. The GH term is given by
\begin{equation}
   {\cal S}_{GH} =2M_p^{3} \int_{\partial M}d^4x \sqrt{h}~K
    \label{app1}\end{equation}
    with
    \be
K_{\m\n}\equiv  -\nabla_\mu n_\nu = {1\over 2}n^{\rho}\partial_{\rho}h_{\m\n}\sp K=h^{ab}K_{ab}
\label{app2}\ee
where $h_{ab}$ is the induced metric on the boundary and $n_{\m}$ is the (outward directed) unit
normal to the boundary. We will not need the precise form of the counterterm action in (\ref{action}) in the following, but it is well known \cite{Skenderis,Papadimitriou2}. 

Both the dilaton and the metric functions will be assumed to depend on the holographic coordinate $u$ which runs from the boundary at $u=-\infty$ and the origin at $u=u_0$. In the {\em vacuum state}, at vanishing temperature the boundary theory enjoys the Lorentz symmetry $SO(3,1)$ which should be reflected in the isometries of the corresponding gravity solution. Hence the ansatz for the metric can be taken with no loss of generality as, 
\be\lab{metTG} 
 ds^2 =du^2 + e^{2A(u)}\eta_{\mu\nu}dx^\mu dx^\nu\, .
 \ee
The Einstein's equations then reduce to 
\be\lab{EE1} 
A'' = -\frac49 (\Phi')^2, \qquad 3 A'' + 12 A^{'2} = V(\Phi)\, .
\ee
The equation of motion of the dilaton can be derived from these two equations. 
\subsection{UV asymptotics}
\lab{UVexp}

We demand that the metric asymptotes to AdS near the boundary: 
\be\lab{asads} 
A(u) \to -u/\ell +\cdots, \qquad u\to -\infty\, .
\ee
We note that the first equation in (\ref{EE1}) requires that the derivative $A'$ is monotonically decreasing. This fact can be traced back to the null-energy condition in the 5D space-time and directly related to the c-theorem in the dual QFT \cite{FreedmanGubser}. But there is more to conclude \cite{ihqcd2}: $A' = -1/\ell<0$ as $u\to -\infty$ from (\ref{asads}) by  requirement of asymptotically AdS space-time. Then, the condition that $A'$ is monotonically decreasing with increasing $u$ leads to the fact that $A(u)\to -\infty$ at some point $u=u_0$ and this point corresponds to curvature singularity \cite{ihqcd2}. Such possible singularities were classified in \cite{ihqcd2}. One can make sense of such singularities in the context of holography \cite{GubserGBN} and this is explained below in detail. 

The second equation in (\ref{EE1}) requires $V\to 12/\ell^2$ on the boundary. This is the value of the cosmological constant corresponding to $AdS_5$ space-time and it constitutes the leading term of the dilaton potential in the UV limit. Now we want to determine the subleading terms in this limit by making connection to the operator $\tr F^2$ dual to $\Phi$ in the corresponding field theory. There are basically two options: 
\ben
\item Approximate the scaling dimension of $\tr F^2$ (that is exactly marginal in the UV) by some number close to but smaller than 4, $\Delta=4-\epsilon$. Then the corresponding field has a mass given by the usual AdS/CFT formula 
\be
m^2\ell^2 =  \Delta(4-\Delta)\, , 
\ee    
in our conventions. In this case the UV limit of the potential reads 
\be\lab{UV1}
V = \frac{12}{\ell^2} + \frac{m^2}{2} (\Phi- \Phi_0)^2 + \cdots  
\ee
and the UV fixed point corresponds to the value $\Phi=\Phi_0$. This choice is advocated in \cite{GubserMimic1,GubserMimic2} and has the advantage of being a more familiar in the AdS/CFT context. In principle, we understand very well the holographic renormalization in such a case \cite{Skenderis}. However it does not correspond to real QCD where the operator is marginal rather than relevant in the UV. It also has various other disadvantages as the corresponding vacuum  solutions can be unstable \cite{GursoyWilke}. 

\item Take $\Delta = 4$ exactly. In this case the dilaton field is massless and the UV asymptotics of the dilaton potential will be qualitatively different than the case above. This case mimics better the running of the coupling constant and the dimension $\Delta$ in QCD, and it is this theory we will be calling the {\em improved holographic QCD}. Below we explain how to fix the UV asymptotics of the potential using the known beta-function of pure SU(N) theory. Holographic renormalization in this non-standard case is also worked out in detail in \cite{Papadimitriou2}. 

\een

The perturbative beta-function of the SU(N) gauge theory in the large N limit, with {\em quenched} fundamental flavors, is given by 
\be\lab{beta} 
\beta(\l) = \frac{d\l}{d\ln E} = -b_0 \l^2 - b_1 \l^3 +\cdots 
\ee
in the limit $\l\ll 1$, i.e. in the UV. Here the first two beta-function coefficients 
\be\lab{b0b1}
b_0 = \frac{22}{3(4\pi)^2},\qquad b_1 = \frac{51}{121}b_0^2\, ,
\ee 
are scheme-independent and positive definite implying asymptotic freedom of the theory. The higher order coefficients are scheme-dependent as can be shown by a redefinition of $\l$. Now we want to connect this UV story to holography near the boundary. Clearly, the holographic theory is not to be trusted in the far UV, when $\l\ll 1$ and therefore when the higher derivative corrections to gravity---which we want to neglect here---are important. Indeed we shall not trust the theory in the far UV limit, however we may still use the identification with running of the perturbative QCD theory to provide {\em initial conditions} for the holographic RG flow. The initial conditions set at small $\lambda$ determine the behavior of the theory at intermediate and strong $\l$, that is, in the IR, the regime expected to be trustable in holography. 

The question now is: how do we make the connection between the field theory quantities such as the 't Hooft coupling $\l$ and the RG energy scale E  and the corresponding quantities in the dual gravitational theory? As mentioned above, the dilaton, more precisely $\exp{\Phi}$ couples to the operator $\tr F^2$ on a probe D3 brane in the gravitational background \cite{adscft-review}, hence its non-normalizable mode should be associated with the  't Hooft coupling and its normalizable mode should be associated with the VeV $\la \tr F^2 \ra$. On the other hand the energy scale $E$ should be related to the conformal factor scale $\exp{A}$ in the metric (\ref{metTG}) \cite{PeetPolchinski}. The motivation for this identification comes from the fact that the energy of a state at location $u$ in the interior of the geometry, measured by an asymptotic observer involves the factor $\exp{A}$ because of the gravitational red-shift determined by $g_{tt}$ \cite{adscft-review}. Therefore we are motivated to make the identifications\footnote{There is the possibility of including a constant multiplicative factor in the first identification \cite{ihqcd5}, which we set to 1.}  
\be\lab{idsUV}
\l = \exp{\Phi(u)}, \qquad \ln E = A(u)\, .
\ee
Here the second choice fixes a particular holographic renormalization scheme. See \cite{ihqcd1} for a discussion of all scheme dependences in these identifications. With these identifications one finds, 
\be\lab{betaX}
\beta(\l) = \l \frac{d\Phi}{dA} \equiv 3\l X(\Phi)\, ,
\ee
where we defined the {\em scalar variable} \cite{ihqcd4} 
\be\lab{Xdef} 
X(\Phi) \equiv \frac{d\Phi}{3dA} = \frac13 \frac{\Phi'(u)}{A'(u)}\, .
\ee
It is related to the ``fake superpotential'' $W(\Phi)$ in the gravitational theory by $X = -3/4 \, d\ln(W)/d\Phi$ \cite{Freedman-review,ihqcd4}. One can easily derive, see appendix \ref{appA}, the equation of motion for the scalar variable $X$ defined above, starting from Einstein's equations (\ref{EE2}): 
\be\lab{Xeq}
\frac{dX}{d\f} = -\frac43~(1-X^2)\le(1+\frac{3}{8}\frac{1}{X}\frac{d\log V}{d\f}\ri)\, .
\ee
We will assume that the solution of this equation, $X$ is negative definite throughout the full range of $\Phi$:
\be\lab{Xneg}
X(\phi) < 0\, .
\ee
This corresponds to the assumption that there is no IR fixed point in the theories we want to consider.   Then,  we learn from the definition (\ref{Xdef}) that $\f'>0$, since $A'<0$ as we explained above. Consistently, we will assume that the coupling constant in the dual field theory grows indefinitely towards the IR. Hence the dilaton diverges at the origin: 
\be\lab{phiIR}
\phi(u) \to \infty, \qquad u\to u_0\, .
\ee
We will see below what happens when these requirements are loosened.

Through equations (\ref{beta}), (\ref{betaX}) and (\ref{Xeq}) then one obtains the desired UV expansion of the dilaton potential as, 
\be\lab{VUV} 
V(\f) = \frac{12}{\ell^2}\le( 1 + v_0 e^\f + v_1 e^{2\f} + \cdots\ri) \qquad v_0 = \frac89 b_0,\,\, v_1 = \frac{1}{81}\le(23 b_0^2 + 36 b_1\ri)\, .
\ee 
This determines the UV asymptotics of the ihQCD potential. We note the one-to-one correspondence between the non-perturbative beta function and the scalar variable $X$ in (\ref{betaX}). This correspondence also carries over to a correspondence between the beta-function and the dilaton potential through (\ref{Xeq}) but there is a catch: one still has to fix an integration constant in solving (\ref{Xeq}) that will be important in determining the correct correspondence of $V$ with the non-perturbative beta-function. We will need IR information to fix this below.  

Given (\ref{VUV}) one obtains the near boundary asymptotics of the background by solving (\ref{EE1}). It is more illustrative to present this expansion in another coordinate system, 
\be\lab{metTGcon} 
 ds^2 =e^{2A(r)}(dr^2 + \eta_{\mu\nu}dx^\mu dx^\nu)\, 
 \ee
related to (\ref{metTG}) by $u = \int^r dr e^{A(r)}$. In this frame the boundary is at $r=0$ and the expansion of the background reads
\bea
\lab{metUV}
e^{A(r)} &=& \frac{\ell}{r}\le[ 1 + \frac49 \frac{1}{\ln[r\Lambda]} -\frac{4b_1}{9b_0^2} \frac{\ln[-\ln [r\Lambda]]}{\ln^2[r\Lambda]}+\cdots \ri]\, ,\\
b_0 e^{\f(r)} &=& -\frac{1}{\ln[r \Lambda]} + \frac{b_1}{b_0^2}\frac{\ln[-\ln [r\Lambda]]}{\ln^2[r\Lambda]}+\cdots
\eea
Here $\Lambda$ is an integration constant, associated to the running of the coupling in the dual field theory and will be identified with the IR scale $\Lambda_{QCD}$.  Let us also write down the Einstein's equations evaluated on the ansatz (\ref{metUV}), that can be obtained from (\ref{EE1}) by the aforementioned change of variables: 
\be\lab{EE2tg} 
\ddot{A} - (\dot{A})^2  = -\frac49 (\dot{\f})^2\, \qquad 3 \ddot{A} + 9 (\dot{A})^2 = e^{2A}V(\Phi)\, .
\ee
Here dot denotes derivative with respect to $r$.

\subsection{IR asymptotics}
\lab{IRexp} 

\begin{figure}[h!]
 \begin{center}
\includegraphics[scale=0.9]{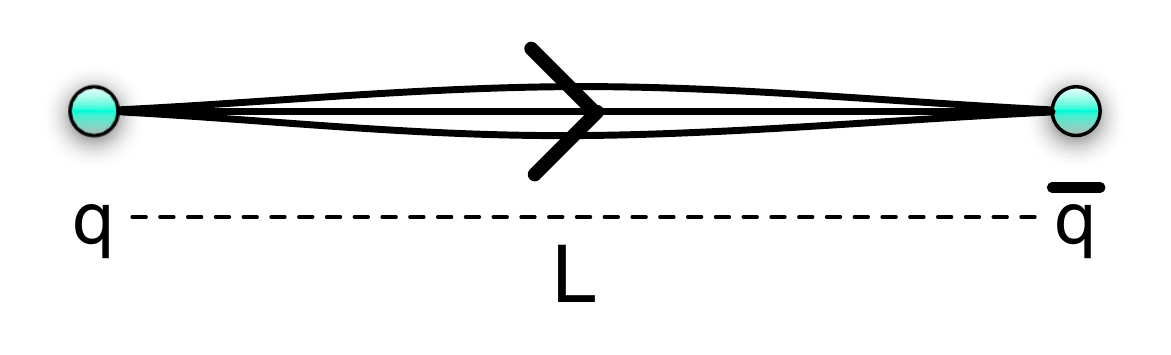}
 \end{center}
 \caption[]{Linear confinement in QCD-like confining theories.}  
\label{fig3}
\end{figure}

IR asymptotics of the dilaton potential is determined by the requirement of quark confinement. In QCD-like confining theories the potential between a test quark and a test anti-quark goes linearly like 
\be\lab{Vqq}
V_{q\bar{q}}(L) = \sigma_0 L + \cdots, \qquad L\gg 1/\Lambda\, ,
\ee
for a large separation $L$ between them. Here $\sigma_0$ is the QCD string tension. Linear quark confinement can be qualitatively understood in terms of a gluon flux tube connecting the quark and the anti-quark, see figure \ref{fig3}. A simple calculation based on ``Gauss' law" in this case shows that the potential energy is proportional to the distance $L$. This is to be contrasted with the electric flux in QED that emanates from a test charge towards all directions. In this case Gauss' law determines the electric potential proportional to inverse of the distance. 
This quark-anti-quark potential is dual on the gravity side to the action of a string with endpoints at the locations $x=0$ and $x=L$ \cite{Maldacena2, Rey}:
\be\lab{Vst}
t V_{q\bar{q}}(L) = S_{NG} - S_{ct} = \frac{1}{2\pi\ell_s^2} \int_0^t d\tau d\sigma \sqrt{-det\, g_{\a\b}}\, - S_{ct}\, ,
\ee
where we denote the space-time coordinates by $X^\mu$ and we have chosen the gauge $X^0= \tau$.  $\ell_s$ is the string length scale. One also typically chooses $\sigma = X^1 = x$. The world-sheet metric is $h_{\a\b} = \6_\a X^\mu \6_\b X^\nu G^s_{\m\n}$, where $G^s$ is the background metric in the {\em string frame}, related to the metric (\ref{metTGcon}) in the Einstein frame as:
\be\lab{metst} 
 ds^2_{st} =e^{2A_s(r)}(dr^2 + \eta_{\mu\nu}dx^\mu dx^\nu), \qquad A_s(r) = A(r) + \frac23 \f(r)\, .
 \ee
There are two important points one has to take into account. Firstly, we included a counter-term action in (\ref{Vst}) because the on-shell string action diverges on asymptotically AdS space-times. There is a standard way to determine this counter-term action \cite{Kinar} and the detailed calculation for our background is presented in \cite{ihqcd2}. Secondly, in the presence of a non-trivial dilaton profile, one has to remember that there is an additional term 
\be\lab{Sdil}
S_s = \int d\sigma\d\tau \sqrt{-g} R^{(2)} \Phi(X^\mu)\, ,
\ee
where $R^{(2)}$ is the world-sheet Ricci scalar. Typically this term is topological, counting handles on the closed string, but it is more complicated in the presence of a non-trivial dilaton. This term is calculated in Appendix C of \cite{ihqcd2} and shown that it does not modify the qualitative results discussed below.   

The generic mechanism that gives rise to the behavior (\ref{Vqq}) from (\ref{Vst}) is as follows \cite{Witten1}: when the geometry ends at a specific point $r=r_0$ deep in the interior (this can correspond to a singularity \cite{Kinar}) then the tip of the string hanging from the boundary to the interior will get stuck at this locus because this is how it will minimize its energy. As one takes the end points further apart in the limit $L\to\infty$ then there will be a contribution from this tip proportional to $L$. This is how the hard-wall background of \cite{Erlich, Pomarol} manages to confine quarks: the tip of the string gets stuck at the location of the hard-wall, since the geometry ends there.

\begin{figure}[h!]
 \begin{center}
\includegraphics[scale=0.7]{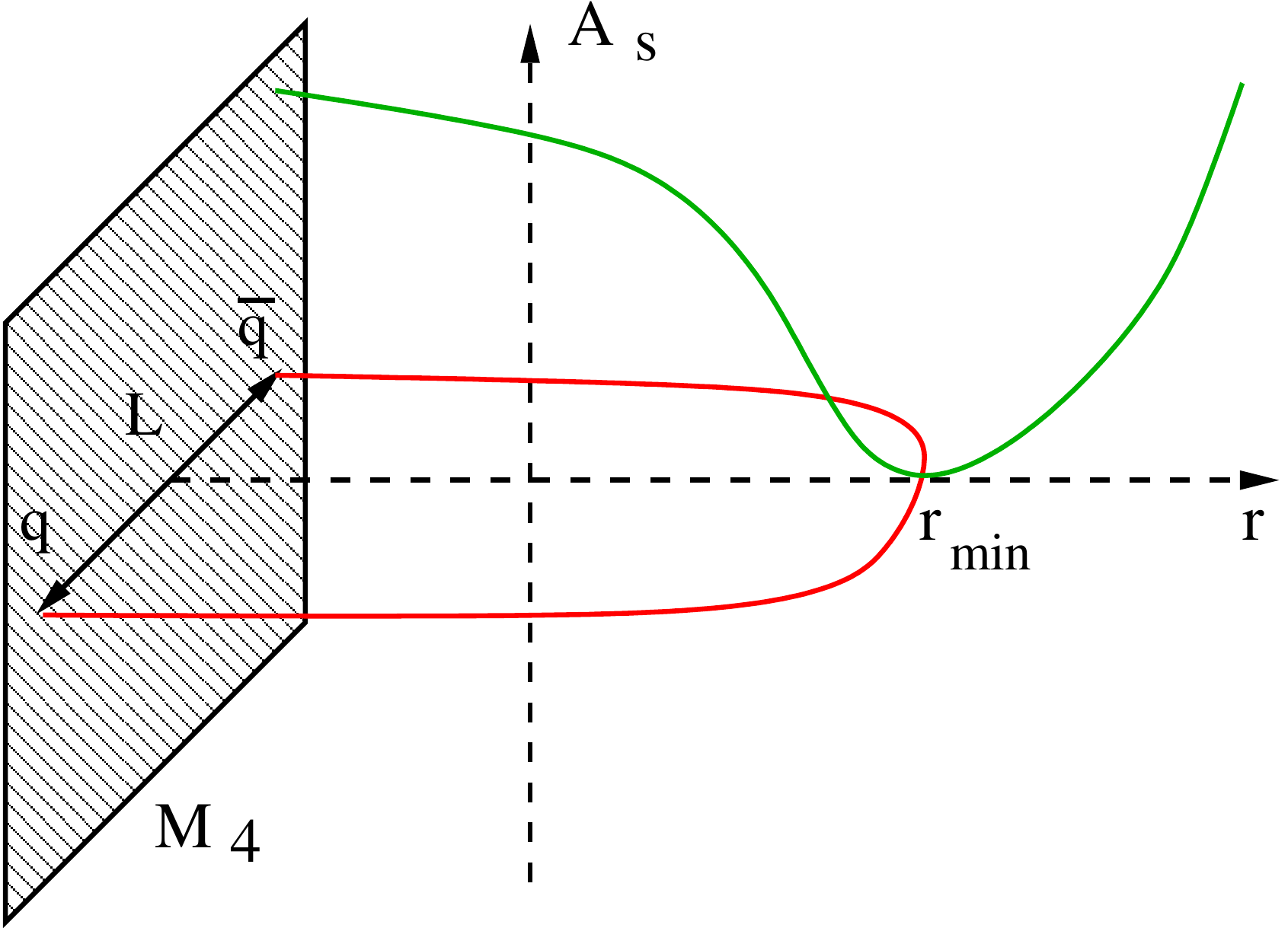}
 \end{center}
 \caption[]{Linear confinement through a minimum of the string-frame scale factor in holographic QCD theories.}  
\label{fig4}
\end{figure}

 This mechanism is generalized  in \cite{ihqcd2} where it is shown that the tip of the string still gets stuck and the action becomes proportional to $L$ in the large $L$ limit, also when the string-frame scale factor $e^{2A_s}$ has a minimum at some location $r=r_{min}$. This is pictorially described in figure \ref{fig4}.  
A simple calculation \cite{ihqcd2} shows that the QCD string tension $\sigma_0$ in (\ref{Vqq}) is related to the tension of the string in 5D as 
\be\lab{tension} 
\sigma_0 = \frac{e^{2A_s(r_{min})}}{2\pi\ell_s^2}\, .
\ee
This mechanism is the most general one that leads to linear quark confinement and a finite QCD string tension, since the original mechanism described above can be obtained from the limit $r_{min}\to r_0$. 

Now, the question is how does this requirement translate into a condition on the dilaton potential? From the UV asymptotics in section \ref{UVexp} it is clear that the string frame scale factor $A_s = A +2\f/3$ in (\ref{metst}) goes to infinity on the boundary. It is also clear from this section that $A_s$ starts decreasing from the boundary towards the interior. In order to acquire a minimum at $r_{min}$ it should start increasing again. Assuming for simplicity that there is a single minimum of the function $A_s$ then this requires $A_s$ diverge as one approaches the IR end point of the geometry  $r\to r_0$ (or $u\to u_0$) where $r_0 > r_{min}$.  For this to happen, as can be seen clearly from (\ref{metst}), we have to require 
\be\lab{cond}
\frac{dA}{d\f}> -\frac23 \rightarrow X< -\frac12\, .
\ee
as $r\to r_0$. A more careful analysis \cite{ihqcd2} shows that
\be\lab{conf} 
\lim_{\Phi\to\infty} \le(X + \frac12\ri)\Phi = K,\qquad 0\geq K \geq -\infty\, .
\ee
This means that for linear confinement to take place the scalar variable $X$ should approach  $-1/2$ from below with the rate $K/\f$.  

Solving (\ref{Xeq}) in the limit $\Phi\to \infty$ we find that this can only happen if $X$ flows to one of the fixed points of the differential equation (\ref{Xeq}) as $\f\to \infty$:  
\bea
\lab{case1}
\textrm{I.} \qquad X & \to & -\frac38 \frac{V'(\f)}{V(\f)}\bigg|_{\f=\infty}\, , \\ 
\lab{case2}
\textrm{II.} \qquad X & \to & -1 \, , \\ 
\lab{case3}
\textrm{III.} \qquad X & \to & +1\, .
\eea 
 The first case happens only when the potential is dominated by an exponential term $V\to \exp{4\f/3}$ in the large $\f$ region. Both the second and the third case are generic: starting from an initial value\footnote{We want the initial value $X_0$ negative since $X$ is negative in the UV, as shown above.} $X_0<0$ at $\f = \f_0$ and solving the equation (\ref{Xeq}) numerically in the region $\f>\f_0$ one finds that typically it either flows to -1 or +1. The important exception is when the potential is asymptotically exponential as mentioned above, and one fine tunes the initial conditions $X_0$ such that case I holds. The third case requires $X$ pass from 0 that, according to (\ref{betaX}) implies that the corresponding theory  flows to a fixed point. This is not we want from QCD-like confining theories, hence we disregard this case. The second case also turns out to be problematic. There is a curvature singularity at $\f=\infty$ and this singularity is not of acceptable type according to \cite{GubserGBN}. As we show in detail in section \ref{thermo}, this singularity is of {\em good type} if and only if it corresponds to the special solution in case I. Hence, this requirement uniquely fixed the integration constant of equation (\ref{Xeq}). 

This is a special case because this requirement restricts the IR asymptotics of the dilaton potential to 
\be\lab{VIR1} 
V(\f) \to V_{\infty}\, e^{\frac43 \f}\f^{-\frac83 K}, \qquad \f\to \infty\, ,
\ee
where $V_\infty$ is some constant. 
The IR background geometry now follows from a particular choice of the constant $K$ \cite{ihqcd2}. The particular case $K = -\infty$ corresponds to the case where the asymptotic value\footnote{One can easily see from (\ref{Xeq}) that $X= -1$ is an attractive fixed point and $X$ cannot go below this value \cite{ihqcd1}. A more strict condition on this exponent  comes from analyzing the spectra \cite{ihqcd2}.} . In this case the asymptotics of the potential should be chosen as, 
\be\lab{VIR2} 
V(\f) \to V_{\infty}\, e^{-\frac83 X(\infty)\f}, \qquad \f\to \infty\, ,
\ee
that translates into the second option in equation (\ref{VIR}). 

Solving Einstein's equations in this limit, with the requirement (\ref{VIR1}) one finds one finds \cite{ihqcd2} that there are two classes of confining IR geometries in the coordinate frame (\ref{metTGcon})  depending on whether $K$ is smaller or larger than $-3/8$. One finds for the Einstein frame scale factor: 
\bea\lab{smallK}
A &\to& - C r^\a,\qquad -\frac38 < K \leq 0, \qquad K\equiv -\frac38 \frac{\a-1}{\a} \\ 
\lab{largeK}
A &\to& -C(r_0-r)^{-\tilde{\a}}, \qquad -\infty < K < -\frac38, \qquad K \equiv -\frac38 \frac{\tilde{\a}+1}{\tilde{\a}}\\
A &\to&  \delta \log(r_0 - r), \qquad X(\infty) = \frac23\sqrt{1 + 1/\delta}< -\frac12\, ,
\eea
where $C$ is an integration constant determined in terms of $\Lambda$. 
In particular (\ref{metTGcon}) has a curvature singularity at a finite locus $r=r_0<\infty$ when $K<-3/8$ and at infinity $r_0=\infty$ when $0\geq K>-3/8$. The asymptotics of the dilaton reads 
\be\lab{dilIR}
\f(r) \to -\frac32 A(r) + \frac34 \ln|A'(r)| + \cdots
\ee
where $A(r)$ is given above.

\subsection{Curvature singularity} 

As we have seen there is a singularity in the deep interior at the origin $u=u_0$ or $r=\infty$ in the solutions we consider in this paper. The dilaton diverges and the conformal factor $\exp A$ vanishes at this point. One can see that this corresponds to an actual curvature singularity in the Einstein frame by computing the Ricci scalar. One finds that $R$ behaves in the Einstein and string frames as 
\be\lab{Rsing}
R \sim e^{-2A} A^{'2}, \qquad R_s \sim e^{-2A_s} A_s^{'2}\, ,
\ee 
where $A_s$ is defined in (\ref{metst}). In the solutions that we study, i.e. the ones with the IR asymptotics in (\ref{smallK}) one finds 
\be\lab{Rsing1}
R \sim e^{2Cr^\a}r^{2(\a-1)}, \qquad R_s \sim \frac{1}{r^{\a+1}}\, .
\ee 
Since $\a>1$ we find that there is a curvature singularity in the Einstein frame, however there is no singularity in the string frame. As we consider these backgrounds as embedded in string theory whose low energy effective action is naturally given in the string frame, we conclude that the IR limit of the holographic theory is trustable. Instead the string frame Ricci scalar near the boundary diverges, leading to the conclusion that we can only trust the theory up to a UV cut-off that $r_{UV}$ that is determined by demanding $R_s(r_{UV}) \sim {\cal O(1)}$. There is another independent curvature invariant in the theory that is $g^{\m\n} \6_\m \f \6_\n \f$ and one can easily show \cite{ihqcd2} that it scales exactly as the Ricci scalar above. 

One should also worry about the diverging dilaton in embedding to string theory.  The asymptotic value of exponential of the dilaton corresponds to the string coupling constant $g_s$ which we want to keep small to ignore string loop corrections. This is taken care of by the large $N$ limit: in fact when we wrote down (\ref{action}) we factored out the dependence on $N_c$ by rescaling $\exp(\f)$ by $N_c$. This rescaling should be performed in the action in the string frame, and it is not possible to see it in (\ref{action}). The scaling exactly produces the $N_c^2$ factor in front of the action once one changes to the Einstein frame \cite{ihqcd1}. Therefore the actual dilaton in string theory hence the corresponding string coupling scales like $g_s \sim N_c^{-1} \exp(\f)$  that vanishes everywhere, if one takes the large $N_c$ limit first. 

\subsection{Constants of motion and parameters} 
\lab{consts}
Let us briefly discuss the integration constants in the system of differential equations that lead to the ihQCD backgrounds. We have a third order system\footnote{A way to cast these equations in the form of 3 first order equations is described in Appendix \ref{appA}.} in (\ref{EE2}). These constants can be regarded as the value of the fields $\f$, $A$ and $X$ (defined in (\ref{Xdef})) at a reference point $r_f$. As we have seen above, and as we  further discuss below, the integration constant of the $X$ equation of motion (\ref{Xeq})  should be fixed by the requirement of an acceptable singularity in the deep interior. The remaining constants of motion $A(r_f)$ and $\f(r_f)$ should be given physical meaning. It is obvious from the ansatz  (\ref{metTG}) that the former corresponds to a rescaling of the volume of the boundary space-time, hence we can say it corresponds to the volume. Since we shall consider infinite volume we will only be interested in volume independent quantities, such as densities, e.g. entropy density, energy density etc. Therefore the integration constant $A(r_f)$ will decouple in physical results. On the other hand the integration constant $\f(r_f)$ is physical and it corresponds to the confinement scale $\Lambda_{QCD}$ in the dual field theory. This is related to the constant $\Lambda$ that appears in the UV expansion in (\ref{metUV}). One can fix this constant of motion either by fitting the actual value of $\Lambda_{QCD}$ or  equivalently by matching the first excited glueball mass, as we discuss below. 

In addition, we have included an overall constant $M_p$ in the action (\ref{action}). This will be fixed once we discuss solutions at finite temperature. The on-shell action corresponds to the free energy of the dual field theory that is a pure glue gauge theory, whose free energy in the large $T$ scales as $F = const \times T^4$. Hence $M_p$ will be fixed by this constant. Finally, there is  the string length scale $\ell_s$ that appears in (\ref{tension}) which also suggests a way to fix this value by matching to the tension of glue flux between the quarks that can be computed in the lattice. 

\subsection{A choice for the potential} 

The IR (large $\exp \f$ ) and UV (small $\exp \f$ ) asymptotics of the dilaton potential is completely fixed by the physical requirements we discussed above. As we also discussed, there is a one-to-one correspondence between the non-perturbative beta function of the field theory and the the dilaton potential\footnote{More accurately the correspondence is with the function $X$ or the fake superpotential $W$. The correspondence with the dilaton potential becomes unique after fixing the integration constant in equation (\ref{Xeq}) by the acceptable singularity condition discussed above.} up to field redefinitions and a choice of renormalization scheme. Therefore, in principle one could be able to fix the entire dilaton potential if one knew the full non-perturbative beta function of the theory. Here we take a more practical approach and pick one particular choice that possesses all the features described above: 
\be\lab{pot} 
V(\f) = \frac{12}{\ell^2}\le\{1 + V_0 e^{\f} + V_1 e^{\frac43\f}\le[\log\le(1+ V_2 e^{\frac43\f} + V_3 e^{2\f}\ri)\ri]^{1/2}\ri\}\, .
\ee
The 4 parameters of the potential will be fixed by comparison to the glueball spectrum below and thermodynamics in the next section. 

\subsection{The glueball spectra} 
\lab{spectrum}

The particle spectrum in AdS/CFT is given by the finite energy excitations around the gravitational background that are normalizable both near the boundary and at the origin. Here we first discuss general features of the particle glueball spectra in ihQCD, see \cite{Gluerev} for a review of the glueball spectrum calculations in AdS/CFT. 

The action for the fluctuations can be obtained by expanding (\ref{action}) to quadratic order in fluctuations. Alternatively one can fluctuate the background equations of motion to linear order. It is important to work with diffeo-invariant combinations of fluctuations. For example the transverse traceless metric fluctuation   $\delta g_{\m\n}$,
\be\lab{fluc1} 
 g^{\m\n} \delta g_{\m\n} = 0, \quad k^\m \delta g_{\m\n} = 0 \, ,
\ee
is invariant under a diffeomorphism of the r direction but the fluctuation of the $\f$ field mixes with the fluctuaiton of the trace of the metric as we will see below. Assuming that $\xi(r,x)$ is a diffeo-invariant fluctuation, the quadratic term in the expansion of (\ref{action}) in $\xi$ can generically be written as 

\be\lab{genac} S[\xi] \sim \int
dr d^4x ~e^{2B(r)}\left[ \left(\de_r\xi\right)^2 +
\left(\de_i\xi\right)^2 + M^2(r) \xi^2 \right]\, ,\ee where $B(r)$
and $M^2(r)$ are functions depending on the background and on the
type of fluctuation in question. We look for 4D mass eigenstates 
\be
\xi(r,x)  = \xi(r)\xi^{(4)}(x), \qquad \Box \xi^{(4)}(x) = m^2\xi^{(4)}(x).
\ee
of the fluctuation equation
\begin{equation}\lab{scaff}
\ddot{\xi}+2\dot{B}\dot{\xi}+ \Box_4 \xi - M^2(r)\xi = 0\, .
\end{equation}
This equation can be put in Schrodinger form 
\begin{equation}\lab{schro}
-\frac{d^2}{dr^2}\psi + V_s(r) \psi = m^2\psi,
\end{equation}
with
\begin{equation}
\lab{gluepot}
V_s(r) = \frac{d^2B}{dr^2}+\le(\frac{dB}{dr}\ri)^2 + M^2(r)\, ,
\end{equation}
by redefining 
\be\label{wave}
\xi(r) = e^{-B(r)}\psi(r)\,.
\ee
We now demand that the energy of the fluctuation $\xi$ is finite. For example the kinetic term from 
(\ref{genac}) gives 
\be
\left(\int dr
e^{2B(r)} |\xi(r)|^2 \right) \int d^4x
\left(\de_\m\xi^{(4)}(x)\right)^2 = \left(\int dr |\psi(r)|^2
\right)\int d^4x  \left(\de_\m\xi^{(4)}(x)\right)^2\,.
 \ee
Demanding that this is finite then leads to the standard square-integrability condition 
\be\lab{normali}
\int dr
|\psi(r)|^2 < \infty.
\ee
in the Schrodinger problem. 

Next, we note that the  equation (\ref{schro}) can be written as:
\be
\left(P^\dagger P + M^2(r)\right)\psi = m^2\psi, \qquad P = (-\de_r + \dot{B}(r))\, .
\ee
This means that the spectrum will be non-negative provided that $M^2\geq 0$. We note that $M^2 = 0$ for fluctuations of the metric and bulk gauge fields.  

We now ask the question whether the 4D spectrum is {\em gapped} or not. If there is a massless mode, $m^2=0$ solution to  (\ref{schro}) then clearly it can only exist when $M^2=0$. In this case the solution to (\ref{schro}) reads 
\be\label{zeromodes}
\psi_0^{(1)}(r) = e^{B(r)}, \qquad \psi_0^{(2)} = e^{B(r)} \int_0^r dr' e^{-2B(r')}\, .
\ee
We want to know if these solutions satisfy (\ref{normali}). Near the asymptotically AdS boundary we universally have $B \sim 3/2 A$ and $A \sim -log(r) + \cdots$. Therefore the first solution above cannot be normalizable near the the boundary. We should then look for the second one. This is normalizable near the boundary but it is not near the origin \cite{ihqcd2} for an arbitrary choice of the dilaton potential, as long as there is a singularity there. Therefore we cannot find normalizable solutions with $m^2=0$. The only way the mass gap may vanish is that there is a continuous spectrum starting from $m^2 = 0^+$. This however requires the potential $V_s$  in (\ref{gluepot}) vanishes as $r\to \infty$. From (\ref{smallK}) we learn that, as $r\to \infty$: \be A(r) \sim - \left({r\over
R}\right)^{\a}, \ee therefore \be\lab{potIR} V(r) = \dot{B}^{2}(r) +
\ddot{B}(r) \sim R^{-2}\left({r\over R}\right)^{2(\a-1)}. \ee 
We therefore find that  the mass gap condition is $\a \geq 1$ that is, very remarkably\footnote{Note that the two calculations are completely independent. Quark confinement comes from analyzing the NG action of the string and mass gap comes from linear fluctuations around the classical background.}, precisely {\em
the same condition we found independently demanding quark confinement}. If
we require $\a>1$ strictly, we moreover obtain a purely discrete
spectrum, since then  $V(r) \to +\infty$ for large $r$. If $\a=1$ the
spectrum becomes continuous for $m^2 \geq V(r\to\infty)$. 

Moreover, a WKB analysis of the potential (\ref{gluepot}) \cite{ihqcd2} gives an asymptotic spectrum for large excitation number $n\gg 1$ as 
 \be\lab{slope} m ~\sim
~\Lambda ~n^{\a-1\over \a}. 
\ee 
In particular we have ``linear
confinement'' ($m^2 \sim n$) if $\a =2$ which is what we choose from now on. 

One can determine the glueball spectrum  by solving (\ref{scaff}) numerically. For this one shoots from the boundary starting from the solution in the asymptotically AdS background and demanding the solution does not diverge and become normalizable near the origin by tuning the parameter $m^2$. One finds a discrete set of $m^2$ that is identified with the glueball spectrum. The result for the few low-lying modes are compared with the lattice results of \cite{Meyer} in the table below.  The spin-2 glueball $2^{++}$ spectrum is obtained from the transverse-traceless fluctuation (\ref{fluc1}) which satisfies (\ref{scaff}) with $B = 3A/2$ and the spin-0 glueball  $0^{++} $ spectrum is obtained from the diffeo-invariant combination \cite{ihqcd2}
\be\lab{fluc2}
\xi = \xi_0 - \frac{1}{3X}\delta \f\, ,
\ee
where $\xi_0$ is the trace part of the metric fluctuation, $\delta\f$ is the fluctuation of the dilaton and $X$ is the function defined in (\ref{Xdef}). This combination satisfies (\ref{scaff}) with $B = 3A/2 + \log |X|$.  
\begin{center}
\begin{tabular}{|c|c|c|c|}
\hline
   $J^{PC}$  &   Lattice (MeV) &   Our model (MeV) &  Mismatch \\
\hline
 $ 0^{++}$  & {\it 1475 (4\%)} &{\it 1475} & 0  \\
  $2^{++}$ &  2150 (5\%) & 2055 & 4\%  \\
  $0^{++*}$ & {\it  2755 (4\%) } & {\it 2753}  & 0 \\
  $2^{++*}$ & 2880 (5\%) & 2991 & 4\%  \\
  $0^{++**}$ & 3370 (4\%) & 3561  & 5\%   \\
 $0^{++***}$ & 3990 (5\%) & 4253 & 6\%    \\
\hline
\end{tabular}
\end{center}
Here the masses in italic are fine-tuned according to the lattice results by fixing the integration constant $A(r_f)$  in section \ref{consts} and a combination of the parameters $V_1$ and $V_3$ in (\ref{pot}). The rest are predictions.

\section{Thermodynamics and the confinement/deconfinement transition}
\label{thermo} 
At finite temperature the state of the theory is obtained by minimizing the Gibbs free energy $F = E - TS$. By the AdS/CFT correspondence \cite{Witten1} this free energy equals the gravitational action 
\be\lab{free1}
F/T = S[on-shell]\, , 
\ee
evaluated on the background solution with Euclidean time compactified: 
\be\lab{Euc} 
\tau = i t, \qquad \tau \sim \tau + 1/T\, .
\ee 
There exist two solutions with the same AdS asymptotics near the boundary. The first one is just the ``thermal gas'' solution (\ref{metTGcon}) heated up to temperature T: 
\be\lab{sol1} 
 ds^2 =e^{2A_0(r)}(dr^2 + d\tau^2 + \delta_{ij}dx^i dx^j),\qquad \f = \f_0(r)\, , 
 \ee
with the identification $\tau \sim \tau + 1/T$. The confinement analysis we presented above for the vacuum solution (\ref{metTGcon}) obviously goes through for (\ref{sol1}). Therefore we learn that this solution just corresponds to a finite temperature gas of the fluctuations in the confined theory, in other words (\ref{sol1}) corresponds to a glueball gas at temperature $T$. 

\subsection{Black-brane solution} 

There also exists the possibility of a black-brane solution with a non-trivial blackening factor $f$: 
\be\lab{sol2} 
 ds^2 =e^{2A(r)}(\frac{dr^2}{f(r)} + f(r) d\tau^2 + \delta_{ij}dx^i dx^j)\, ,
\ee
where the function $f(r)$ vanishes at some point in the interior that corresponds to the {\em horizon}: 
\be\lab{horizon}
f(r_h) = 0\, .
\ee 
What state does the black-brane solution correspond to? It is clear from the discussion on confinement in section \ref{IRexp} that this solution corresponds to a state with color charges {\em deconfined}.  This is because as you pull the end points of the test string apart from each other the tip of the string will move towards the interior of the geometry and beyond some point $L = L_{max}$ the tip of the string will reach the horizon and dissolve. Therefore there will be no linear confinement of the test quark charges. This will happen provided that $r_h< r_{min}$. As we show below the location of the horizon $r_h$ is related to the temperature and the small $r_h$ regime is attained for larger temperatures. Thus we conclude that the black-brane solution for small enough $r_h$ (large enough T) corresponds to the {\em deconfined (plasma) phase}. In the large-N limit  we consider here, this is a plasma of gluons.\footnote{In section \ref{flavor} we discuss the Veneziano limit where the number of flavors also taken to infinity. In this theory the black-brane phase will correspond to a plasma of gluons {\em and} quarks.} 

The Einstein's equations evaluated on this ansatz are, 
\be\lab{EE2} 
\ddot{A} - (\dot{A})^2  = -\frac49 (\dot{\f})^2, \qquad 3 \ddot{A} + 9 (\dot{A})^2  + 3 \dot{A} \frac{\dot{f}}{f}= \frac{e^{2A}}{f}V(\Phi),\qquad \ddot{f}+ 3\dot{A}\dot{f}=0\, .
\ee
We note that these equations reduce to (\ref{EE2}) when one sets $f=1$, as in (\ref{sol1}). 

Let us now count the number of constants of motion in this problem, that will parametrize our black-brane solutions. This can be done most efficiently by reformulating Einstein's equations in terms of scalar variables, just as in (\ref{Xdef}). Here we have to define two such scalar variables: 
\be\lab{XYdef} 
X(\Phi) = \frac13 \frac{\dot{\Phi}(r)}{\dot{A}(r)},\qquad Y = \frac14 \frac{\dot{f}(r)}{f(r)\dot{A}(r)}\, .
\ee
As shown in appendix \ref{appA}, the Einstein's equations then get reduced to only two first order system of equations that are coupled \footnote{As shown in appendix \ref{appA} once X and Y are determined the background functions $\f$, $A$ and $f$ can be obtained by a single integration.}:
\bea\lab{X1eq}
\frac{dX}{d\f} & = & -\frac43~(1-X^2 + Y)\le(1+\frac{3}{8}\frac{1}{X}\frac{d\log V}{d\f}\ri)\, ,\\
\lab{Y1eq}
\frac{dY}{d\f} & = & -\frac43~(1-X^2 + Y) \frac{Y}{X}\, .
\eea
As clear from the definition in (\ref{XYdef}) $Y$ should diverge at the horizon, because $f$ vanishes whereas $\dot{A}$ (because otherwise one has a curvature singularity at the horizon) and $\dot{f}$ (because this is proportional $T$) should be finite. $\f$ should also be finite at the horizon which means that $Y$ should diverge like $(r_h-r)^{-1} \sim (\f_h-\f)^{-1}$ where $\f_h$ is the value of dilaton at the horizon. In addition $dX/d\f$ should also be finite at the horizon, otherwise there will be a curvature singularity there. Now, using that $Y$ diverges as $(\f_h-\f)^{-1}$ at $\f_h$, from (\ref{X1eq}) we find that the value of $X$ is completely fixed in terms of the dilaton potential as
\be\lab{Xh}
X_h\equiv X(\f_h) = - \frac38 \frac{d\log V}{d\f}\bigg|_{\f_h}\, 
\ee
which then, using (\ref{Y1eq}) also determines the behavior of $Y$ near the horizon as,
\be\lab{Yh}
Y \to \frac{Y_h}{\f_h-\f}, \qquad Y_h = -\frac34 X_h\, .
\ee 
This means that, regularity at the horizon fixes one of the integration constants in the system (\ref{X1eq},\ref{Y1eq}), leaving a single integration constant that is $\f_h$. This constant is related to the temperature of the system as we show below. Solving the rest of the first order differential equations for $\f$, $A$ and $f$ in appendix \ref{appA}  we have 3 more integration constants. For asymptotically AdS space we need to require $f\to 1$ at the boundary, which fixes the integration constant of the $f$ equation. The one for the $\f$ equation can be identified with $\lambda_{QCD}$ in the dual theory, just as in the discussion in section \ref{consts}. The one for the $A$ equation is again related to the volume of the dual theory, which is scaled away in dimensionless quantities. Hence we obtain only two non-trivial integration constants, $\Lambda_{QCD}$ and $T$. The former for the black-brane solution should be identified with the analogous integration constant in the vacuum solution since these are different states in the same theory. Hence the entire thermodynamics will be determined in terms of the dimensionless parameter $T/\Lambda_{QCD}$. In particular the free energy of the system will be a non-trivial function of, and only of $T/\Lambda_{QCD}$. 

We are now at a stage to fix the {\em good singularity} condition mentioned below equation (\ref{case1}). We claimed there that this condition uniquely fixes the IR asymptotics of the confined solution to be (\ref{case1}). As demonstrated in \cite{GubserGBN} a strong version of the good singularity condition requires that the TG solution be obtained from a BB solution in the limit the horizon marginally traps the singularity. But we learned from (\ref{Xh}) that the value of $X$ at the horizon is completely fixed in terms of the dilaton potential. Hence sending $r_h\to \infty$ ($\f_h \to \infty$) we learn that (\ref{case1}) should be satisfied by the TG solution to have a good singularity. 

\subsection{Temperature, entropy, gluon condensate and conformal anomaly} 

The temperature associated with the black-brane solution is obtained by the standard argument of Hawking 
requiring absence of a conical singularity at the horizon in the Euclidean solution. This fixes the period of the Euclidean time cycle as: 
\be\lab{temp} 
T = -\frac{1}{4\pi} \dot{f}(r_h)\, .
\ee
Large temperatures correspond to $r_h\to 0$ where the blackening factor approaches to that of AdS-Schwarzchild: 
\be\lab{fAdS}
f_{AdS} = 1 - \frac{r^4}{r_h^4}\, ,
\ee
from which we determine the relation between $T$ and $r_h$ at large values of $T$ as: 
\be\lab{largeT} 
T  = \frac{1}{\pi r_h}, \qquad T\to \infty\, .
\ee
The entropy is given by the area of the horizon divided by 4 times the Newton constant: 
\be\lab{entropy}
S = \frac{area}{4G_N}  = 4\pi M_p^3 N_c^2 e^{3A(r_h)} V_3\, ,
\ee 
where $V_3$ is the spatial volume spanned by coordinates x, y, z and we used the fact that $G_N$ is determined from (\ref{action}) as $16 \pi G_N  = 1/ (M_p^3 N_c^2)$.  At large T the BB solution approaches to AdS with $\exp A \to \ell/ r$ resulting in 
\be\lab{largeS} 
S/T^3 \to  4\pi^4 (M_p\ell)^3 N_c^2 V_3, \qquad T\to\infty\, ,  
\ee
 where we used (\ref{largeT}). 
 
Now let us discuss the gluon condensate and the conformal anomaly. For this we first need to present the UV expansion of the metric functions in the black-brane solution: 
\bea\lab{Aexp} 
e^{A(r)}  &=& e^{A_0(r)} \le( 1 + G \frac{r^4}{\ell^3} + \cdots \ri), \qquad r\to 0\\
\lab{phiexp} 
\f(r)  &=& \f_0(r) + e^{A_0(r)}  + \frac{45G}{8} \frac{r^4}{\ell^3}\log\Lambda r + \cdots, \qquad r\to 0\\
\lab{fexp} 
f(r)  &=& 1 - \frac{C}{4} \frac{r^4}{\ell^3} + \cdots, \qquad r\to 0
\eea
Here $G$ and $C$ are integration constants of the black-brane that depend on $r_h$. Interpretation of $G$ is clear. Since this is coming from the difference of the normalizable terms in $\f$ and $\f$ is dual to the operator $\tr F^2$ it is identified with the difference of the VeVs of this operator between the plasma state and the confined state. A careful calculation (see section 4 of \cite{ihqcd4}) yields: 
\be\lab{vev} 
\la \tr F^2 \ra_{BB} - \la \tr F^2 \ra_{TG} = - \frac{240}{b_0} M_p^3 N_c^2\, G\, ,
\ee  
where $b_0$ enters through (\ref{metUV}). Similarly, one can interpret $G$ as the excess of conformal anomaly between the plasma and confined phases \cite{ihqcd4}:   
\be\lab{anomaly} 
\la T^\mu_\mu  \ra_{BB} -  \la T^\mu_\mu \ra_{TG} = 60M_p^3 N_c^2\, G\, .
\ee  
Let us mention in passing that the expressions (\ref{vev}) and (\ref{anomaly}) perfectly obeys the expected Ward identity 
\be\lab{Ward} 
T^\mu_\mu = \frac{\beta(\lambda)}{4\lambda^2} \tr F^2\, \,
\ee
near UV \cite{ihqcd4}. Thus we learn that $G$ is the gluon condensate in the plasma phase normalized by its vacuum value. 

So what is $C$? To work this out first note that the last equation in (\ref{EE2}) can be analytically solved to obtain 
\be\lab{fsol} 
f(r) = 1 - \frac{\int_0^r e^{-3A(r)}dr}{\int_0^{r_h} e^{-3A(r)}dr}\, ,
\ee
where we fixed the integration constants in the solution requiring $f\to 1$ at the boundary and that it vanishes at the horizon. Expanding this expression near the boundary, where $\exp A \to 1/r$, and comparing to (\ref{fexp}) we find 
$$ C =  \frac{1}{\int_0^{r_h} e^{-3A(r)}dr} \, .$$
On the other hand, using the formulae for the temperature and entropy in (\ref{temp}) and (\ref{entropy})
we obtain   
$$\frac{1}{\int_0^{r_h} e^{-3A(r)}dr} =  TS / (M_p^3 N_c^2V_3)\, .$$
Thus we obtain the interpretation of constant $C$ in (\ref{fexp}) as the enthalpy density 
\be\lab{C}
C = T s /M_p^3\, , 
\ee
where we define little s as the density per gluon $s  = S/N_c^2 V_3$.

\subsection{Deconfinement transition}

\begin{figure}[h!]
 \begin{center}
\includegraphics[scale=1.4]{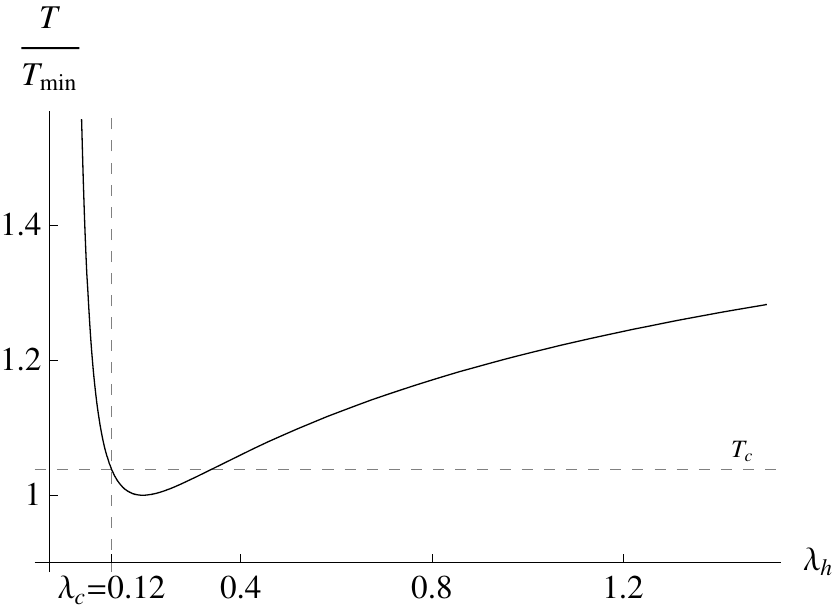}
 \end{center}
 \caption[]{Temperature as a function of $\exp \f_h$ in the ihQCD model. }  
\label{figTlh}
\end{figure}

Now the obvious question is which of the solutions above, (\ref{sol1}) or (\ref{sol2}) minimize the free-energy. We answer this question by calculating the difference of on-shell actions
\be\lab{difac} 
\Delta S = S[\textrm{sol2}] - S[\textrm{sol1}] = \Delta F/T\, .
\ee  
When $\Delta S<0$ the plasma state wins, when $\Delta S>0$ the confined state wins. Calculation of this difference is non-trivial. Here I will only highlight the important points in the calculation sparing the reader from the details which can be found in Appendix C of \cite{ihqcd4}. First of all, one has to note that there are two contributions to the difference, the one coming from the Einstein-Hilbert action and the other from the Gibbons-Hawking term in (\ref{action}). Thus we write each term in the difference as $S  = S_{EH} + S_{GH}$. Both contribute to the difference non-trivially. Second, both of these contributions can be expressed in terms of the boundary asymptotics of the background functions. This is obvious for the GH term (\ref{app1}), but also true for the EH term. The latter is because, upon use of the background equations of motion one can express the integrand in the EH term as a total derivative: 
\be\lab{EH}
S_{EH}  = 2 M_p^2 V_3 \beta \int_\eps^{r_h} \frac{d}{dr} \le(\dot{A}(r) f(r) e^{3A(r)}\ri)\, .
\ee   
where $\beta = 1/T$ from the Euclidean time integral and $V_3$ the volume of boundary space from the spatial integration and $\eps$ is a UV cut-off that we will take to zero in the end. The contribution from the horizon vanishes as $f=0$ there and the other background functions are finite. Thus, 
\be\lab{EH2}
S_{EH}  = -2 M_p^2 V_3 \beta \dot{A}(\eps) f(\eps) e^{3A(\eps)}  \, .
\ee   
On the other hand the GH term can be calculated by substituting in (\ref{app1}) the metric ansatz: 
\be\lab{GH}
S_{GH}  =  M_p^2 V_3 \beta e^{3A(\eps)} f(\eps) \le(8 \dot{A}(\eps) + \frac{\dot{f}(\eps)}{f(\eps)} \ri)\, .
\ee   
The third thing to note is that both (\ref{GH}) and (\ref{EH2}) are divergent in the limit $\eps\to 0$. This is expected, it only corresponds to the usual UV divergence in the QFT coming from the bubble diagrams contributing to the free energy. This divergence is perfectly cancelled in the difference (\ref{difac}) because both states should contain the same UV divergence. Therefore the subtraction in (\ref{difac}) can be thought of as a regularization scheme. To ensure that the UV divergences cancel, one needs to demand that the background functions become the same near the boundary. Comparison of the metrics (\ref{sol1}) and (\ref{sol2}) then yields the conditions: 
\be\lab{match}
\beta_0 e^{A_0(\eps_0)} = \beta e^{A(\eps)} \sqrt{f(\eps)}, \qquad V_3^0 e^{3A_0(\eps_0)} = V_3 e^{3A(\eps)}, \qquad \f_0(\eps_0) = \f(\eps)\, ,
\ee
that come from matching the time-cycles, the space-cycles and the dilatons at the cut-offs and we allowed for different values for length of these cycles and position of the cut-offs in the black-brane and the thermal gas solutions. The latter is necessary in order to leave freedom to keep the integration constants $\lambda$ in (\ref{metUV}) and the analogous UV expansion of the black-brane function the same \cite{ihqcd4}.  One can now calculate the difference (\ref{difac}) using (\ref{EH2}) and (\ref{GH}) both for the BB and the TG solutions, requiring (\ref{match}), substituting the near boundary expansions (\ref{Aexp}), (\ref{phiexp}) and (\ref{fexp}) and taking the limit $\eps\to 0$ to obtain the finite result: 
\be\lab{deltaF} 
\Delta F  = \frac{1}{\beta} \Delta S = M_p^3 N_c^2 V_3 \le( 15 G(T) - \frac{1}{4} Ts\ri)\, .
\ee
Furthermore, one can calculate the energy difference in the two states using the ADM mass formula (see \cite{ihqcd4} for details) as
\be\lab{energy} 
\Delta E  = M_p^3 N_c^2 V_3 \le( 15 G(T) + \frac{3}{4} Ts\ri)\, .
\ee
Combining (\ref{deltaF}) and (\ref{energy}) we learn that the system nicely satisfies the Smarr relation $F = E - TS$ as it should. We finally note that the functions $G$ and $s$ in the expression for the free energy depends on the integration constant $r_h$ (or $\f_h$) as they are obtained from the near boundary expansions of the background functions. To obtain the expression in $T$ one still has to relate $r_h$ (or $\f_h$) to T. This can be done by calculating (\ref{temp}) by substitution of the numerical solutions. One obtains figure \ref{figTlh} where we show $T$ as a function of $\exp \f_h$ for convenience. 
Two comments are in order. First of all we see that the black-brane solutions only exist above a minimum temperature $T = T_{min}$ that depends on the particular model. Below this temperature there exists only the thermal gas solution and it dominates the ensemble. Second, we see that for any $T> T_{min}$ there are two black brane branches one with a large value of $\phi_h$ (or $r_h$) and one with a small value of $\phi_h$. The BB with smaller value of $r_h$ has a bigger event horizon since, as we showed in the previous section, $A(r_h)$ is a monotonically decreasing function and the event horizon is proportional to $\exp 3A(r_h)$. Therefore we call the solution with smaller $r_h$ the {\em large black-brane} and the solution with larger $r_h$ the {\em small black-brane}. As we show below, the latter solution is always subdominant in the ensemble, whereas the former one, the large BB corresponds to the true plasma phase in the theory.

Now, we can come back to the question we asked above: which phase minimizes $F$ at a given $T$. In equation (\ref{deltaF}), the gluon condensate $G$ is a positive definite quantity. On the other hand the entropy term is negative definite. It is therefore conceivable that there exists a critical temperature $T_c$ where $\Delta F$ vanishes.  At very high temperatures the entropy term normalized by $T^4$  should go to a positive constant given by (\ref{largeS}). On the other hand the difference in the gluon condensate $G(T)/T^4$ should vanish as it should approach the same value in the plasma and the confined phases in the UV.  This means that at large T the plasma phase wins. The question then is, whether or not $\Delta F$ becomes positive at small T. The answer is in the affirmative and can be obtained by calculating (\ref{deltaF}) numerically as in \cite{ihqcd5}. 

\begin{figure}[h!]
 \begin{center}
\includegraphics[scale=1.4]{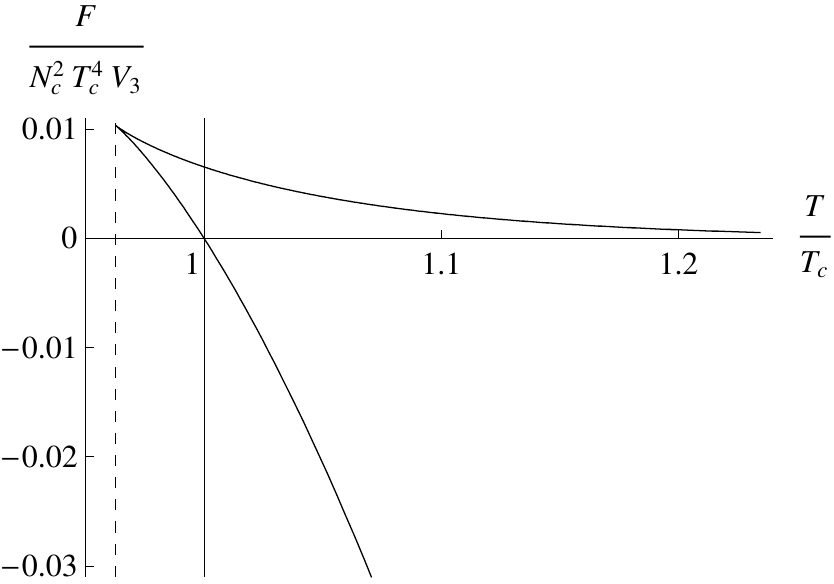}
 \end{center}
 \caption[]{Difference of free energies between the plasma and the confined phase $\Delta F$ as a function of T.}  
\label{figfree}
\end{figure}

One finds the picture  shown in figure \ref{figfree} for the free energy. 
The axis in this figure $F = 0$ corresponds to the free energy of the thermal gas solution. This is because, as discussed above, the thermal gas solution is obtained by sending $r_h\to\infty$ (on the small BB branch) in figure \ref{figTlh}. In this limit the horizon area shrinks to zero yielding vanishing entropy. Similarly the ADM mass of the BB also vanishes yielding vanishing E. Then from the  Smarr formula we have $F(TG) = 0$. We also see the presence of the aforementioned two BB branches in this figure. They exist above $T=T_{min}$ and the one with positive $F$ is the small BB. As we see this branch is always sub-dominant in the ensemble. The branch below is the large BB branch and we also see that it crosses the x-axis at a particular temperature $T=T_c$ that is higher than $T_{min}$. Therefore we obtain the holographic description of the {\em deconfinement transition} in our holographic model.  We also see that this is a {\em first order} phase transition as expected in large N QCD. 

You should be asking how did we fixed the parameters of the model, particularly the parameters in (\ref{pot}) to obtain this figure. As mentioned around equation (\ref{VUV}) $V_0$ and $V_2V_1$ is fixed by matching the first two scheme-independent beta-function coefficients in the pure $SU(N)$ theory. As also mentioned at the end of section  \ref{spectrum} we fix a combination of $V_1$ and $V_3$ to match the second glueball mass in \cite{Meyer}. We can now fix the other combination of $V_1$ and $V_3$ by matching the entropy density with the lattice result of \cite{Boyd} at a fixed temperature $T=2T_c$. The best fit turns out to be $V_1 = 14$, $V_3 = 170$. The only non-trivial integration constant (apart from T) in the background solutions is $\Lambda$ that is fixed by matching the first glueball mass as explained in section \ref{spectrum}. The only quantity yet to be fixed is the Planck mass $M_p$. We can fix this from equation (\ref{largeS}) by matching the entropy of pure $SU(N)$ theory in the large N large T limit as \cite{ihqcd4}   
\be\lab{Mp}
(M_p \ell)^3 = \frac{1}{45\pi^2}\, .
\ee
Having fixed all the parameters in the model, the rest is prediction to be tested against lattice data.  In particular one obtains 
\be\lab{Tc}
T_c = 247\, \textrm{MeV}\, 
\ee 
for the transition temperature, that compares very well with the lattice data \cite{Lucini}. For the latent heat  $L_h = \Delta E(T_c) = T_c \Delta S(T_c)$ at the transition we find 
\be\lab{Lh}
L_h = 0.31 N_c^2 T_c^4 \, , 
\ee 
that also matches very well the lattice data at large $N_c$ \cite{Lucini}. 

\begin{figure}[h!]
\includegraphics[scale=0.3]{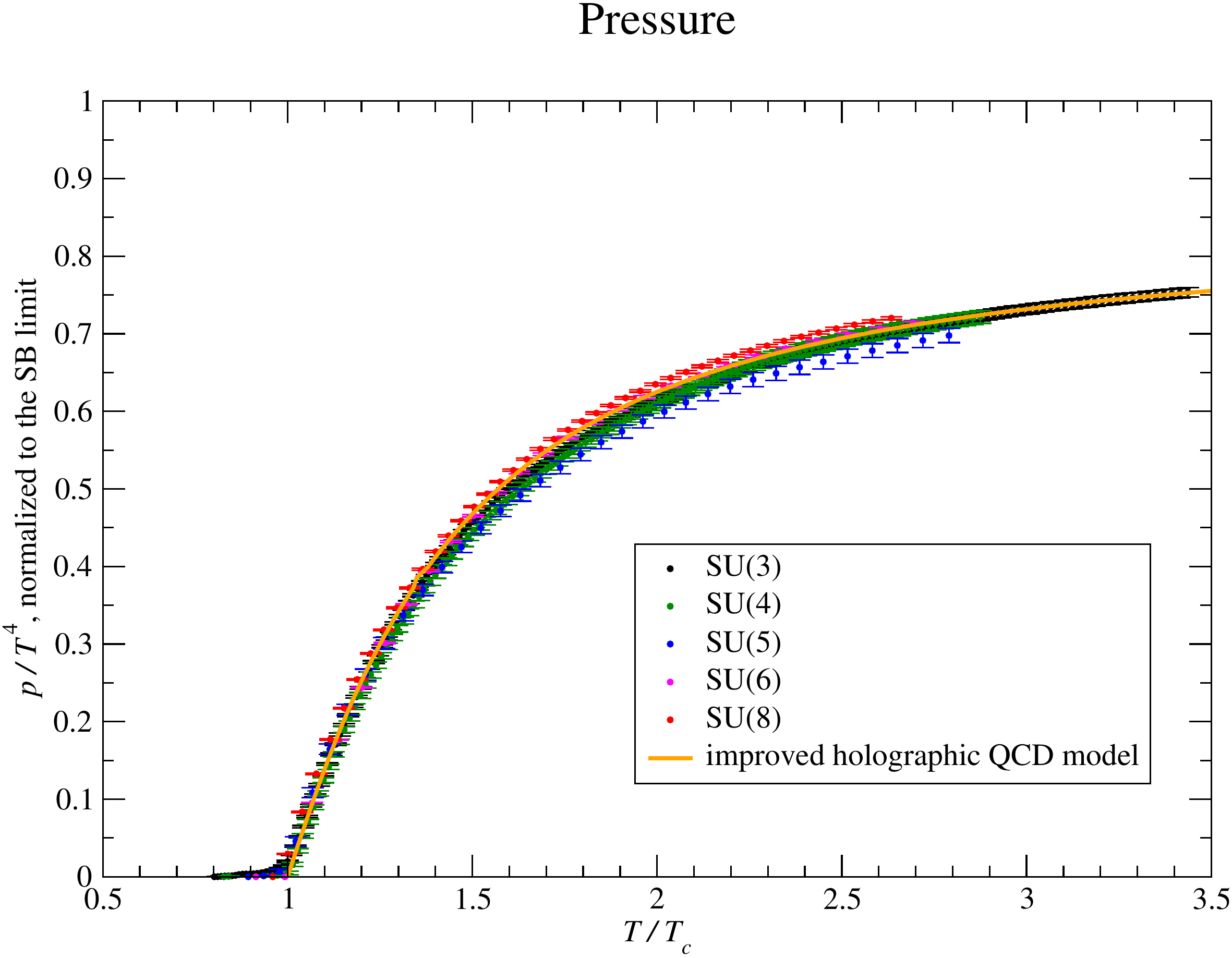}
\includegraphics[scale=0.3]{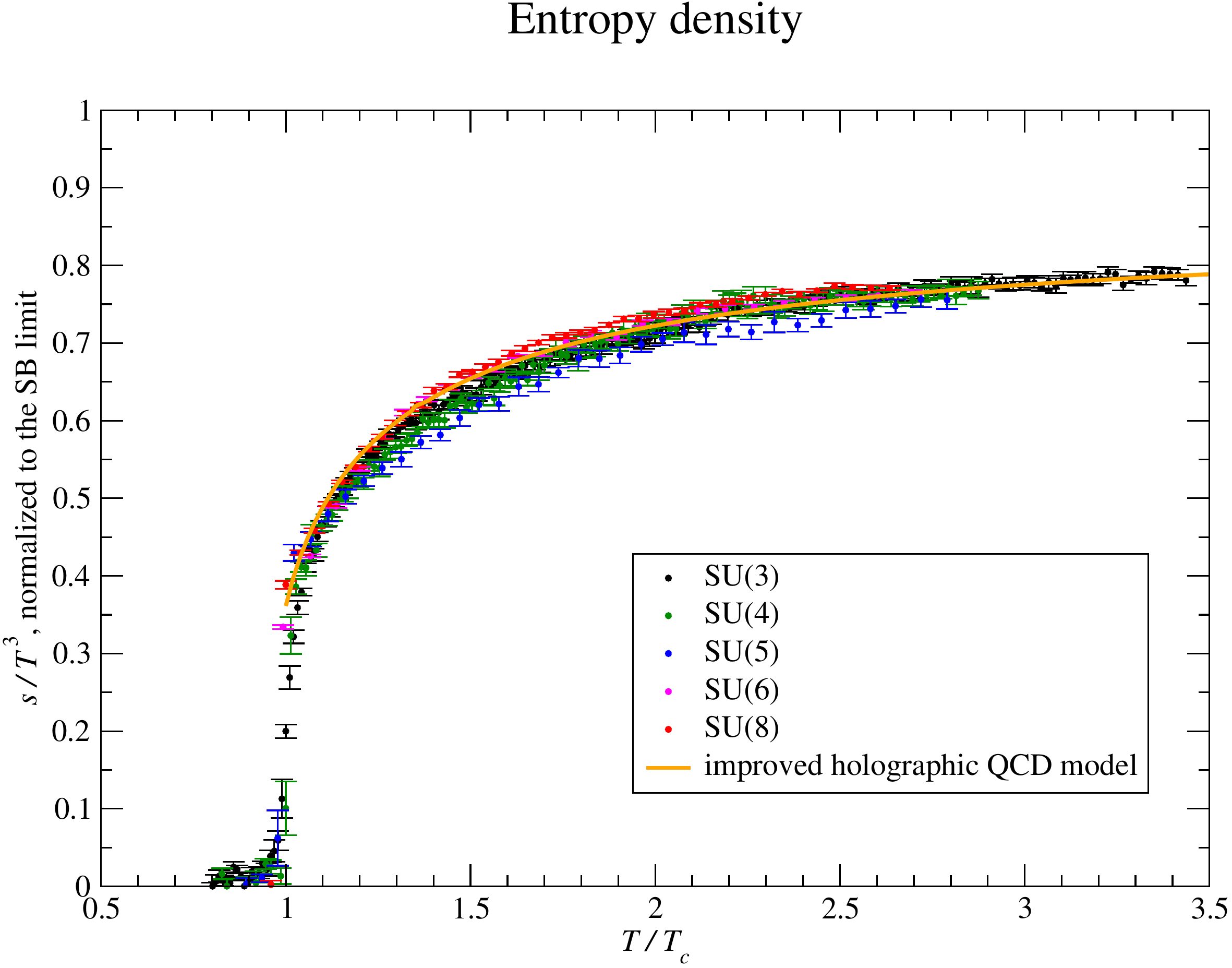} \\
\includegraphics[scale=0.3]{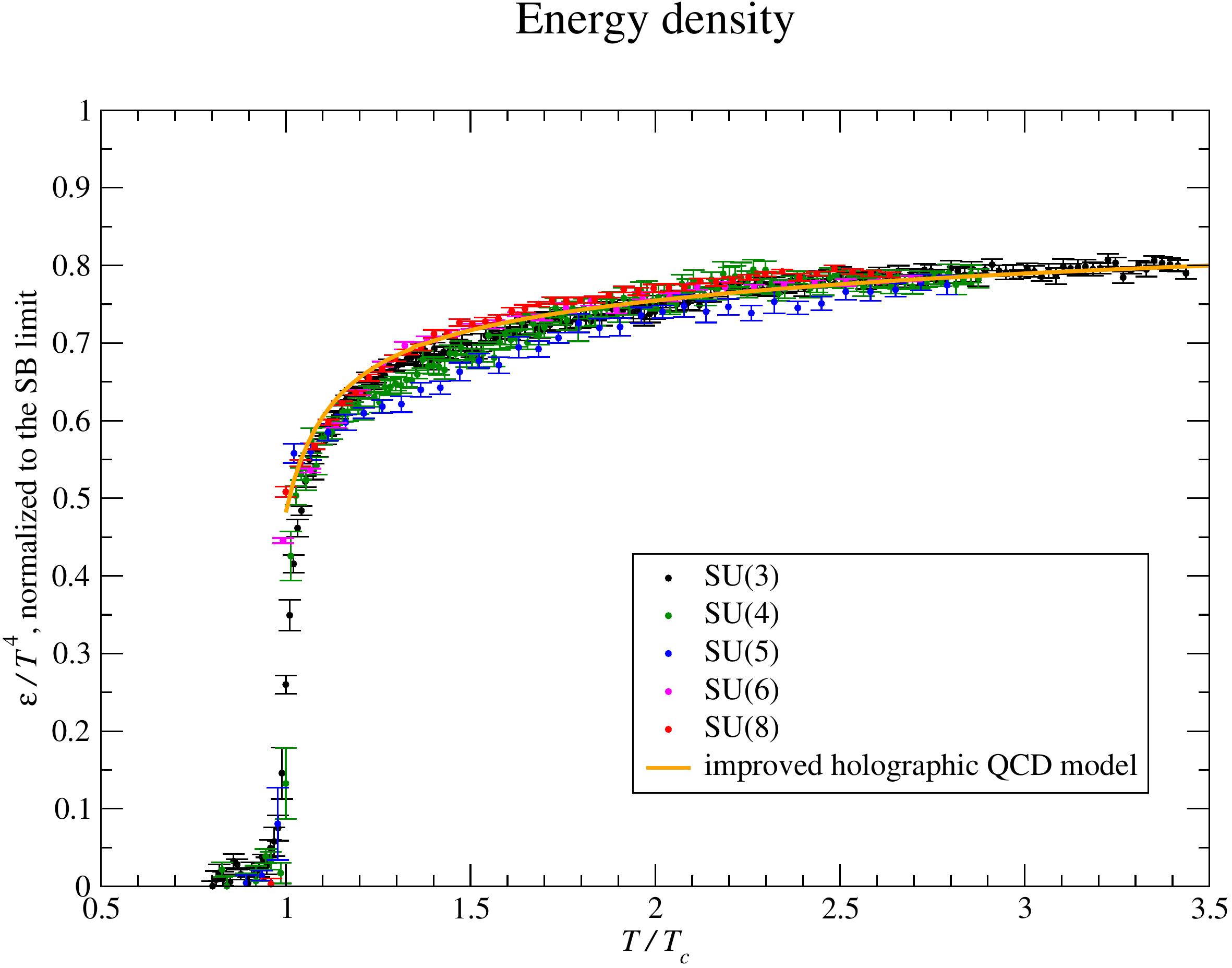}
\includegraphics[scale=0.3]{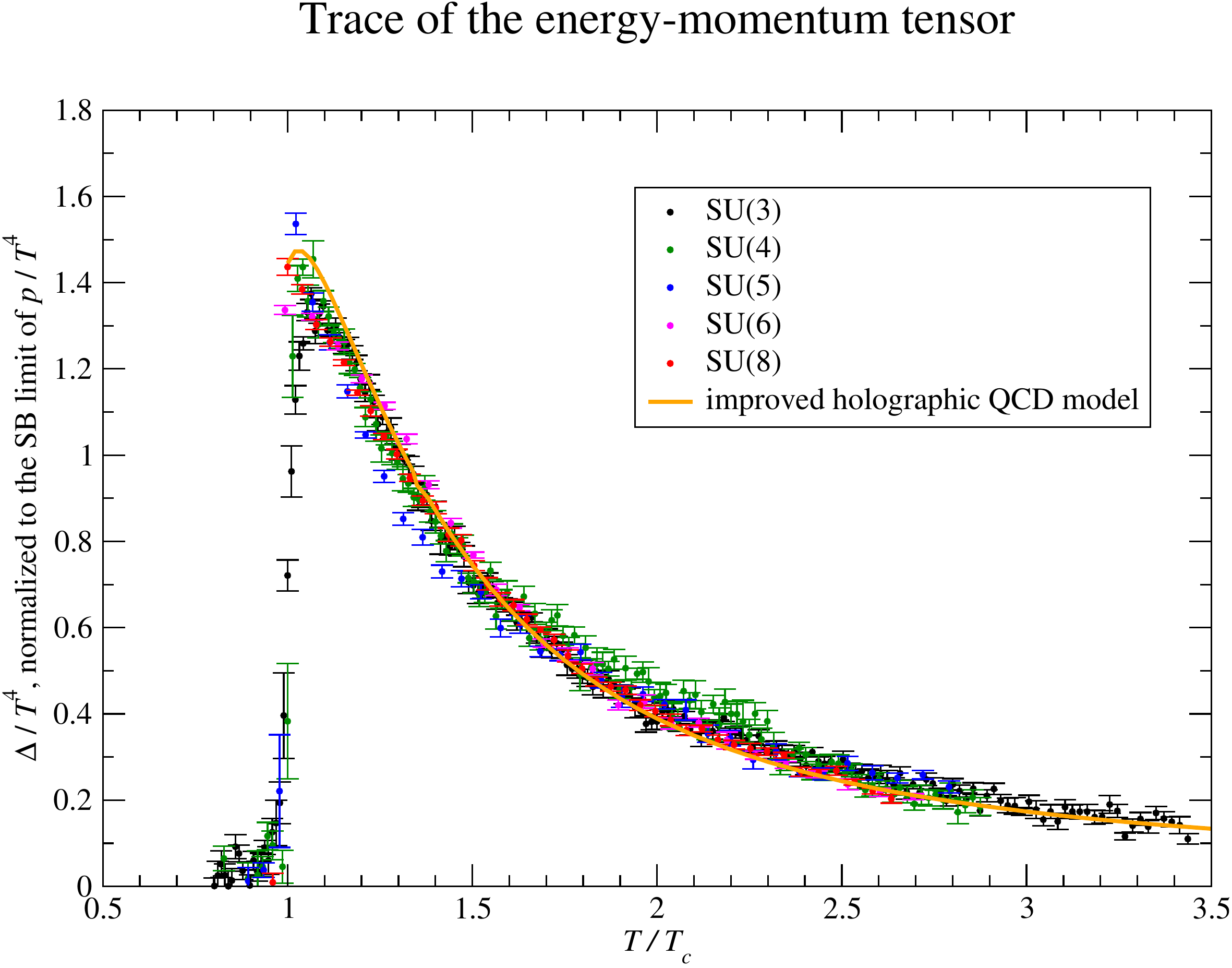} 
 \caption[]{Comparison of lattice data of \cite{Panero} for various values of $N_c$ with ihQCD (orange curves). All thermodynamic functions are densitiesm and further normalized by a factor of $N_c^2$ and an appropriate power of $T$ to make dimensionless quantities. Pressure $P = - F$ for extensive systems. }  
\label{figthermo}
\end{figure}

It is very instructive to compare the thermodynamic functions obtained from ihQCD with the existing lattice studies. In particular \cite{Panero} studied the thermodynamic functions of pure $SU(N)$ theory at various values of $N_c$ and compared his data with our findings. This comparison is shown in figures \ref{figthermo}. 
We observe two important features in these plots. First, when appropriately normalized, the thermodynamic quantities collapse on a single curve modulo small errors. This means that these properly normalized thermodynamic functions exhibit very weak dependence on the number of colors $N_c$. Thus, our results that are necessarily valid at $N_c\to \infty$ are not supposed to be bad at all! Second, we observe that the thermodynamic functions coming from the ihQCD model matches this curve  perfectly!  

\section{Flavor sector}
\lab{flavor}

So far we discussed the construction of the holographic theory only in the glue sector. This description is valid in the limit $N_c\to\infty$ when the number of flavors $N_f$ is kept finite. This is because in the large $N_c$ limit one can consistently ignore the fermion loop corrections in the Feynman diagrams. In real QCD however one typically considers $N_f=3$ for light flavors  corresponding to up, down and strange quarks and $N_c=3$ with ratio 1. Hence, one expects a better approximation to real QCD with light flavors in the large-N limit, by taking also the number of flavors to infinity, keeping the ratio finite: 
\be
\label{vlim}
N_f, N_c \to \infty\,,\quad x={N_f \over N_c}=\mathrm{fixed} \, , \quad \lambda=\frac{g_\mathrm{YM}^2 N_c}{8\pi^2} =\mathrm{fixed} \, .
\ee
This is called the Veneziano limit. We keep the ratio $x$ as a free parameter in what follows, the actual value for real QCD with light flavors corresponding to $x=1$ (for up, down and strange) or $x=2/3$ (for up and down quarks). The theory with flavors is naturally richer: in the massless quark limit (that we consider here)  there is the global  $U(N_f)_L\times U(N_f)_R$ flavor symmetry that rotates the left and right handed quarks separately. The vector $U(1)_V  = U(1)_{L+R}$  part of this symmetry corresponds to the baryon number under which $u$, $s$ and $\bar{d}$ quarks carry  charge $+2/3$ and  $\bar{u}$, $\bar{s}$ and $d$ quarks carry  charge $-1/3$. The other diagonal $U(1)_A = U(1)_{L-R}$ is anomalous and non-conserved. Furthermore, the remaining flavor symmetry $SU(N_f)_L\times SU(N_f)_R$ is {\em spontaneously broken}  to $SU(N_f)_{L+R}$, because of the non-trivial expectation value of the quark condensate $\la \bar q q \ra$  in the vacuum state.  

As discussed in the Introduction, the improved holographic QCD theory is capable of reproducing all of these salient features. The flavor sector in the holographic theory is introduced through the flavor branes \cite{Paredes1, Paredes2, Casero} embedded in the geometry. These are space-filling  $N_f$  D4-branes and $N_f$ $\bar{D}4$-branes in the 5D bulk. In the Veneziano-limit the energy-momentum tensor of these flavor branes become comparable to the Planck mass $M_p^3N_c^2$ in (\ref{action}), hence one has to take into account their backreaction on the background. This means that one has to solve the Einstein's equations that arise from the full action: 
\be\lab{fullaction} 
S = S_g + S_f\, ,
\ee 
where the glue part $S_g$, is given in (\ref{action}) and the effective DBI action on the flavor branes read \cite{Casero, Jarvinen:2011qe}:
\begin{eqnarray}
\lab{actf}
S_f=-\frac{1}{2}M_{p}^3 N_c\mathbb{T}\mathrm{r}\!\!\int d^5x\left(V_f(\lambda, T^\dag T)\sqrt{-\det \mathbf{A}_L} +V_f((\lambda, T T^\dag) \sqrt{-\det \mathbf{A}_R}\right),
\end{eqnarray}
where $\mathbb{T}\mathrm{r}$ denotes the ``super-trace'' on the non-Abelian branes \cite{Paredes1, Paredes2, Casero}, the fields $\mathbf{A}$ are given by 
\begin{eqnarray}
\mathbf{A}_{L\mu\nu}&=&g_{\mu\nu}+w(\lambda,T)F^{L}_{\mu\nu}+\frac{\kappa(\lambda,T)}{2}\left[(D_\mu T)^{\dag}(D_\nu T)+(D_\nu T)^{\dag}(D_\mu T)\right],\nonumber \\
\mathbf{A}_{R\mu\nu}&=&g_{\mu\nu}+w(\lambda,T)F^{R}_{\mu\nu}+\frac{\kappa(\lambda,T)}{2}\left[(D_\mu T)(D_\nu T)^{\dag}+(D_\nu T)(D_\mu T)^{\dag}\right],
\end{eqnarray}
and the covariant derivative is given by 
\begin{equation}
D_\mu T=\partial_\mu T+iTA^L_\mu -iA^R_\mu T.
\end{equation}
Here, $A_L$ and $A_R$ denote the gauge fields living on the flavor D-branes corresponding to the global flavor symmetry $U(N_f)_L\times U(N_f)_R$ with $F^L$ and $F^R$ the corresponding field strengths. $T$ is a complex scalar, called the {\em open string tachyon}, that transforms as a bifundamental under this flavor symmetry and corresponds to the quark mass operator $\bar q q$. 
Following \cite{Paredes1, Paredes2, Casero}  (inspired by Sen's action for the open string tachyon \cite{Sen})  we choose the tachyon potential as 
\begin{equation}\lab{Vf}
V_f(\lambda,TT^{\dag})=V_{f0}(\lambda)e^{-a(\lambda)TT^{\dag}}\, .
\end{equation}
This form of the tachyon action was motivated in \cite{Paredes1, Paredes2, Casero}  by reproducing the expected spontaneous symmetry breaking and the axial anomaly of QCD. Then $V_{f0}$, $w$ and $\kappa$ are new potentials (in addition to $V$ in (\ref{action})) that, in the bottom-up approximation should be fixed by phenomenological requirements as in the previous sections. The theory is further developed in the subsequent works in \cite{Jarvinen:2011qe, Arean:2012mq, Arean:2013tja, Alho:2013hsa, Iatrakis:2014txa, Alho:2015zua, Jarvinen:2015ofa, Arean:2016hcs}. 
One typically also makes a simplifying assumption and takes $\kappa(\lambda,T)$ and $w(\lambda,T)$ independent of $T$. The potentials $V_{f0}(\lambda)$, $a(\lambda)$, $\kappa(\lambda)$ and $w(\lambda)$ are constrained by requirements from the low energy QCD phenomenology, such as chiral symmetry breaking and meson spectra \cite{Arean:2013tja}. A judicious choice for these potentials are presented in Appendix \ref{appB}. 

For equal quark masses (that we take zero in this section) for all $N_f$ flavors, one can further make the simplification by choosing a diagonal tachyon field
\begin{equation}
T=\tau(r) \mathbb{I}_{N_f},
\end{equation}
that corresponds to $N_f$ light quarks with the same mass in boundary field theory. As mentioned above, $\tau(r)$ is holographically dual to the quark mass operator and its non-trivial profile is responsible for the chiral symmetry breaking on the boundary theory. The boundary asymptotics of this function, for the choice of potentials given in appendix \ref{appB} is
\be
\tau(r) \simeq m_q r(-\log \Lambda r)^{- \rho}+\langle {\bar q} q \rangle r^3 (-\log \Lambda r)^{\rho}
\label{tauuv}
\ee
the power $\rho$ is to be matched to the anomalous dimension of ${\bar q} q$ and the QCD $\beta$-function (see~\cite{Jarvinen:2011qe,Arean:2013tja} for details). In this work we only consider massless quarks $m_q=0$ so the non normalizable mode of the tachyon solution vanishes, thus providing a boundary condition for the $\tau$ equation of motion. 

Calculation of flavor current correlators in the holographic theory follow from fluctuating the bulk gauge fields $A^a_L$ and $A^a_R$ in (\ref{actf}) where the small $a$ index corresponds to non-Abelian flavor. We will not be interested in these correlators in this review. However we will be interested in studying the effects of a non-vanishing quark chemical potential $\mu$ on the QGP. This chemical potential can be introduced through the boundary value of the $U(1)_V$, $a=0$, part of the bulk gauge fields as
\be\lab{chempot} 
A^V_{\nu} = \frac{A^0_{L,\nu} +A^0_{R,\nu}}{2} \to (\mu, 0, 0, 0, 0), \qquad r\to 0 \, ,
\ee 
where $\mu$ corresponds to the $\nu= 0$ component. Therefore we can finally simplify the flavor action by setting all  $A_L$ and $A_R$ to zero except (\ref{chempot}):
\be
\label{actfs}
S_f =-x\, M^3 N_c^2 \int d^5x\, V_f(\l,\tau) \sqrt{- \mathrm{det}\left(g_{\m\n} + w(\l)\, F^V_{\m\n} + \kappa(\l)\, \partial_{\m} \tau \,\partial_{\n} \tau\right) } \, .
\ee
We shall not discuss the physics that follows from this action in detail here. The meson spectrum (obtained by studying fluctuations of the bulk gauge fields), the quark condensate (obtained by studying the profile of $\tau$) etc are all studied in detail in the references listed above. Here, we only want to summarize the qualitative effect of a non-vanishing $\mu$ on the phase diagram. 

\begin{figure}
 \begin{center}
\includegraphics[scale=1]{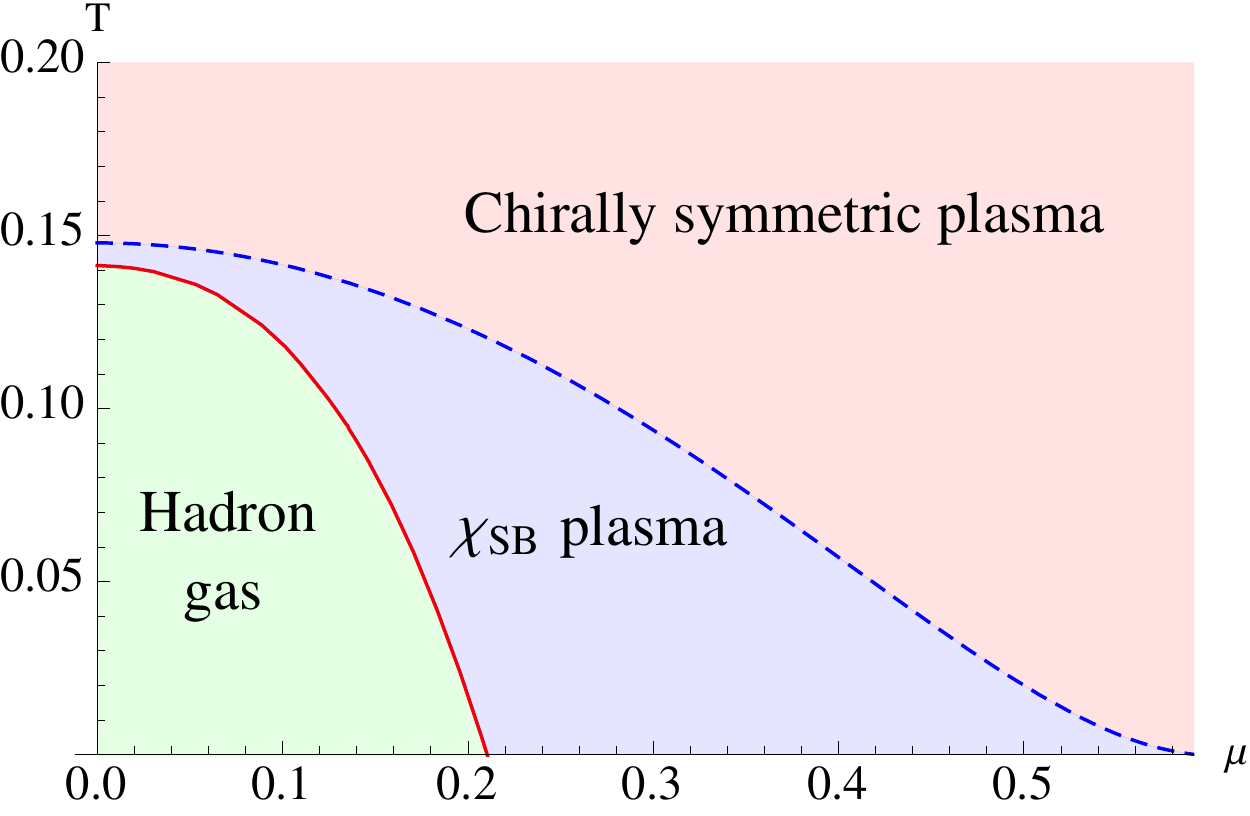}
 \end{center}
 \caption[]{The phase diagram of the ihQCD theory in the Veneziano limit with finite quark chemical potential. Figure taken from \cite{Alho:2013hsa}.}   
\label{pdmu}
\end{figure}

The qualitative picture that arises from (\ref{actfs}) and (\ref{action}) in (\ref{fullaction}) is summarized in figure \ref{pdmu} taken from \cite{Alho:2013hsa}. 
We observe the possibility of three phases in this diagram. First of all the confined phase denoted by ``hadron gas'' in the figure continues to exist for $\mu\neq 0$ in the small temperature regime. This phase holographically corresponds to the thermal gas solution in the previous section, generalized for $\mu \neq 0$.   On top of this phase, we observe two separate phases for larger values of the temperature. The phase denoted by $\chi SB$  corresponds to a deconfined quark-gluon plasma with a non-vanishing value of the quark condensate. Therefore this phase is a quark-gluon plasma where the chiral symmetry is broken $SU(N_f)_L\times SU(N_f)_R\to SU(N_f)_{L+R}$. Holographically, this phase corresponds to the black-brane phase of the previous section accompanied by a non-trivial vector bulk field (\ref{chempot}) and a non-trivial profile for the tachyon field $\tau(r)$. The hadron gas phase is separated from the $\chi SB$ phase by a first order phase separation curve $T_c(\mu)$ (red, solid)  in figure \ref{pdmu}. Finally, when one cranks up T further, the quark-condensate melts trough a second-order phase transition (blue, dashed curve) at $T_\chi(\mu)$ and one obtains a deconfined state where the chiral symmetry is restored. This phase holographically corresponds to a generalization of the black-brane background of the previous section for finite (\ref{chempot}) and $\tau=0$. In section \ref{finiteB} we shall see how this phase diagram is altered for vanishing chemical potential $\mu=0$ but a finite external magnetic field B turned on instead.

\section{Hydrodynamics and transport coefficients}
\label{hydro}

The next level in increasing difficulty in our treatment of the quark-gluon plasma is hydrodynamics. Thermodynamics of the previous section should be embedded in this theory that has a bigger range of applicability, in particular it also encompasses the physics of transport and dissipation. Hydrodynamics is a theory organized in a derivative expansion, that is an expansion in powers of momentum compared to an intrinsic scale in the system such as the mean free path in systems with quasi-particle excitations, $k\ell_{mfp}$ or compared to temperature $k/T$ in systems, such as our strongly interacting plasma, where no particle-like excitations exist. Each term in this derivative expansion is determined by conservation laws, such as the energy-momentum and charge conservation in the plasma. Therefore, in some sense one can think of hydrodynamics as the IR effective theory of these conserved charges.   
\subsection{Generalities}
In this section, we consider hydrodynamics of the neutral glue plasma, hence the only non-trivial conservation equation is the energy-momentum conservation: 
\be\lab{emcon} 
\nabla_\mu T^{\m\n} = 0\, .
\ee
These are 4 equations and we need to express the solution in terms of 4 unknowns. In this case these 4 unknown functions of space-time (with metric $g_{\m\n}$) can be taken as the 4-velocity field of the fluid $u^\mu$ and temperature: 
\be\lab{unknown} 
u^\mu(x),\quad g_{\m\n} u^\m u^\n = -1; \qquad T(x)\, .
\ee
Then we need a {\em constitutive relation} to express $T^{\m\n}$ in terms of these unknowns. In relativistic hydrodynamics, to zeroth order in momentum, the only symmetric two-index objects are $g_{\m\n}$ and $u^\m u^\n$ therefore one can directly write: 
\be\lab{cons1}
T^{\m\n}_0 = u^\mu u^\n (\epsilon + p) + g^{\m\n} p \, ,
\ee
where we parametrized the coefficients in terms of energy $\eps$ and pressure $p$ of the fluid\footnote{It is often useful to express quantities in the rest frame $u^\m = (1,0,0,0)$ where indeed $T^{00} = \eps$ and $T^{ii} = p$.}.  This term at zeroth order in the derivative expansion corresponds to an {\em ideal} relativistic fluid. Energy and pressure as a function of temperature should be defined using microscopic properties of the theory, and we already did this in the previous section. 

The next term in the derivative expansion corresponds to dissipative terms\footnote{One of the most recent advances in the study of QGP involve anomalous transport. These terms are argued to produce no dissipation and they are represented by introducing new terms in the hydrodynamic expansion \cite{SonSurowka}. We will omit anomalous transport in this discussion.} and at this order, for a neutral plasma we have only two such terms corresponding to {\em shear} and {\em bulk} deformations. The derivation can be found in standard textbooks and review papers\footnote{I find the discussion in \cite{Romatschke} particularly nice.} and they read: 
\be T^{\m\n}_1 = P^{\m\a}P^{\n \b} \le[ \eta \le(\6_{\a}u_{\b}+\6_{\b}u_{\a}-\frac23g_{\a\b}\,
\6\cdot u\ri)+ \zeta\, g_{\a\b}\,\6\cdot u\ri]\, , 
\ee
where $P^{\a\b}$ is the projector on the plane transverse to $u^\a$: 
\be\lab{Pab}
P^{\a\b} = g^{\a\b}  + u^{\a} u^{\b}\, ,
\ee
and the coefficients $\eta$  and $\zeta$ are called the  ``shear viscosity" and the ``bulk viscosity" respectively. They characterize the response of the fluid to shear (traceless) and volume (trace) deformations of the energy-momentum tensor. The derivative expansion goes on like this and one encounters more and more transport coefficients at higher orders. 

The transport coefficients, in our case only the shear and bulk viscosity, are supposed to be determined from microscopic properties of the fluid. According to the linear response theory, the first order change in the expectation value of an operator ${\cal O}_B$ due to a deformation of the Lagrangian of the  system by an operator ${\cal O}_A$ is given by the retarded Green's function of the operators ${\cal O}_B$ and ${\cal O}_A$: 
\be\lab{defr}
{\cal L} \to {\cal L}  + \int {\cal O}^A \delta \phi_A  \qquad \Rightarrow \qquad \la {\cal O}^B\ra = G_R^{BA} \delta\phi_A\, ,
\ee
where the retarded Green's function is given by 
\be\lab{retG}
G^{BA}_R(\o,\vec{k}) = -i \int d^4x e^{-i k\cdot x} \theta(t) \la[\cO^A(t,\vec{x}),\cO^B(0,\vec{0}) ]\ra \, .
\ee
The last average is a thermal average.  In our case we are interested in deformations of the energy-momentum tensor due to metric deformation that itself couple to the energy-momentum tensor, hence both ${\cal O}_A$ and ${\cal O}_B$ are $T_{\m\n}$ and the shear and the bulk viscosities are obtained in the limit  
\be\lab{viscos}
\eta \le( \delta^{il}\delta^{km} +\delta^{im}\delta^{kl} - \frac23 \delta^{ik}\delta^{lm}\ri) + \zeta\delta^{ik}\delta^{lm} =  \lim_{\o\to 0}\frac{i}{\o} G_R^{ik,lm}(\o,\vec{0})\, ,
\ee
with momentum $\vec{k}$ set to zero. Thus, the shear viscosity can be read off from the $(12,12)$ and the bulk viscosity can be read off from the $(11+22+33,11+22+33)$ components of the Green's function of the energy-momentum tensor. 
\subsection{Shear viscosity}
\begin{figure}[h]
\includegraphics[scale=0.26]{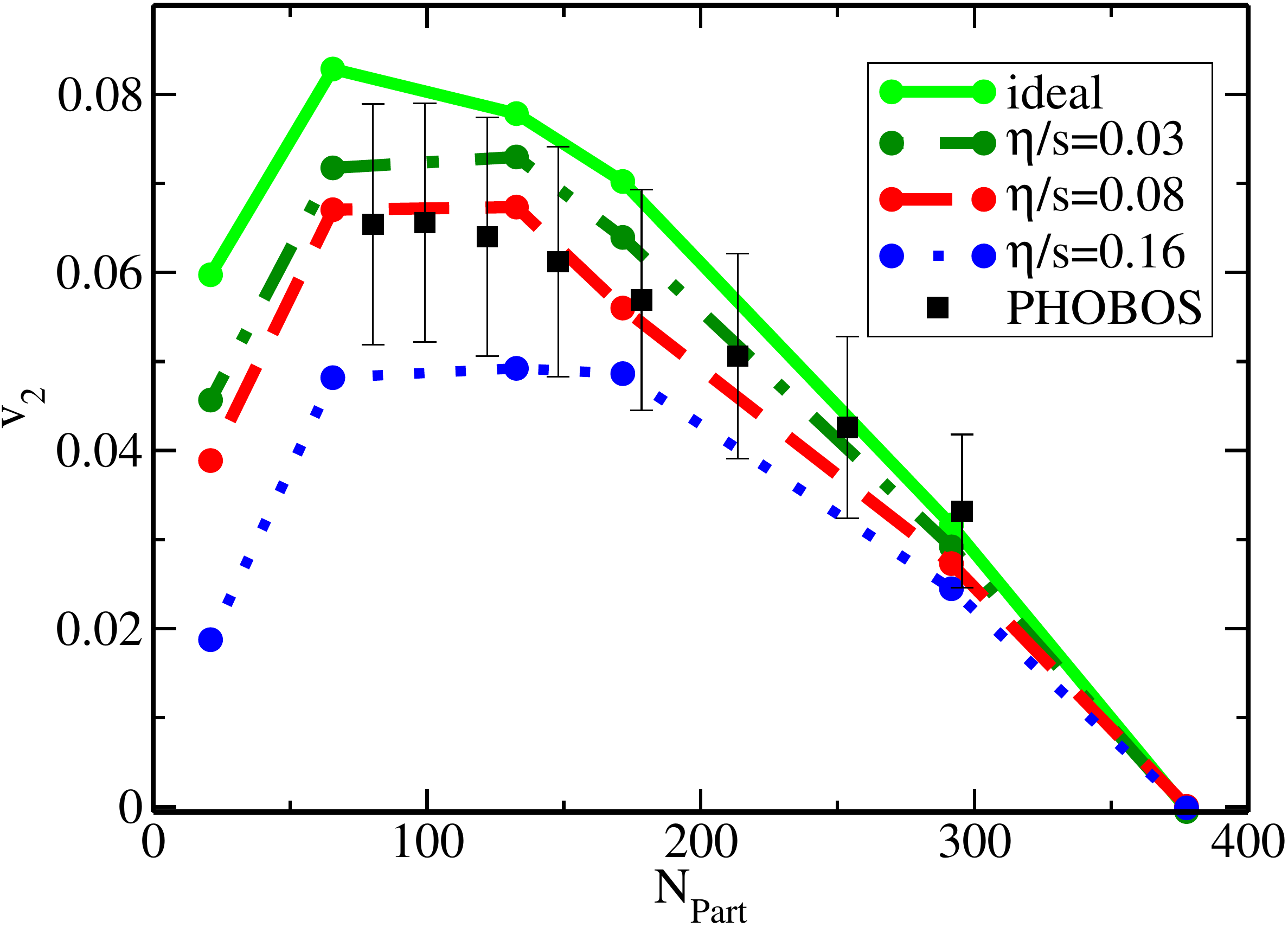}
\includegraphics[scale=0.26]{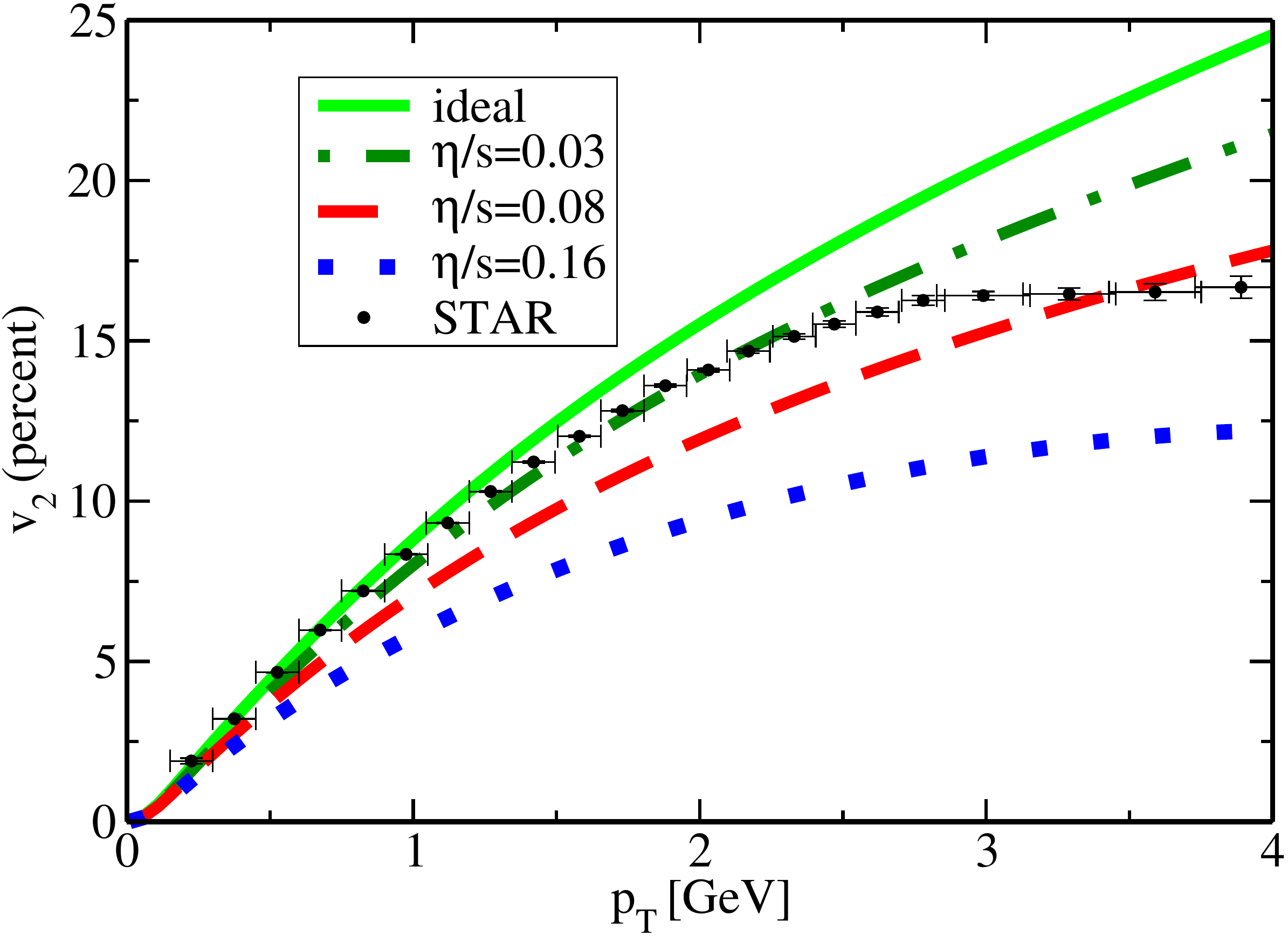}
\caption{Comparison of hydrodynamic simulations for the elliptic flow parameter $v_2$ of the hadron spectrum to actual data at RHIC for the various input values of the shear viscosity. Data agrees well with the AdS/CFT result (\ref{shear}).}
\label{figv2}
\end{figure}

In the strong coupling limit this two point function is calculated by the AdS/CFT prescription. For example, for the shear viscosity, one has to solve the equation of motion for the fluctuation $\delta g_{xy}(r,\o)$ for $\vec{k}=0$, with infalling boundary conditions at the horizon $r_h$ and non-normalizable boundary condition at the boundary: 
\be\lab{bcflucs}
\delta g_{xy} \to (r_h - r)^{-i\frac{\omega}{4\pi T}}, \quad r\to r_h, \qquad \delta g_{xy} \to r^4, \quad r\to 0\, .
\ee
The fluctuation equation for the $(x,y)$ component for the metric (\ref{sol2}) is given by
\be\lab{sfluc}
\ddot{\delta g}_{xy} + \dot{\delta g}_{xy}( 3\dot{A} + \frac{\dot{f}}{f} ) + \frac{\omega^2}{f} \delta g_{xy} = 0\, .  
\ee
The result of this calculation for the shear viscosity is well-known \cite{PSS, KSS}. For any two-derivative gravity theory the answer is fixed by universality at the horizon \cite{Buchel:2003tz}, regardless of the details of the field content or the potentials as: 
\be\lab{shear} 
\frac{\eta}{s}  = \frac{1}{4\pi} \approx 0.08\, ,
\ee
where $s$ is the entropy density. The aforementioned universality arises in the $\omega\to 0$ limit of equation (\ref{sfluc}) as the mass term vanishes in this limit \cite{Buchel:2003tz}. The result (\ref{shear}) corresponds to an extremely small shear viscosity. This result is to be compared with the perturbative QCD result
\be\lab{pshear} 
\frac{\eta}{s}  \propto  -\frac{1}{\l_t^2 \log \l_t}\, ,
\ee
where $\l_t$ is the 't Hooft coupling in large N QCD. This result becomes very large in the small coupling limit. On the other hand the AdS/CFT result (\ref{shear}) agrees much better with the hydrodynamic simulations where one tunes $\eta$ as an input parameter to match the hadron spectrum obtained from these hydro simulations to actual QGP spectrum, see figure \ref{figv2}. The result shown in figure \ref{figv2}  is for the {\em elliptic flow parameter} defined as the second moment of the hadron spectrum in the azimuthal angle $\phi$ on the interaction plane. Thus, one has strong indications that the  QGP produced in these experiments are in fact strongly coupled. 
\subsection{Bulk viscosity}
The fluctuation equation for the volume deformation on the other hand is given by 
\be\lab{sflucvol}
\ddot{\delta g}_{ii} + \dot{\delta g}_{ii}( 3\dot{A} + \frac{\dot{f}}{f} + 2\frac{\dot{X}}{X}) + (\frac{\omega^2}{f}  - \frac{\dot{f}}{f}\frac{\dot{X}}{X} ) \delta g_{ii} = 0\, .  
\ee
This equation does not exhibit any universality at the horizon, because of the presence non-vanishing mass term in the limit $\omega\to 0$ and the result, that is a non-trivial function of $T$,  indeed depends on the  choice of the potential in (\ref{action}). For the choice (\ref{pot}) ihQCD theory gives \cite{GursoyBulk} the plot given in figure \ref{figbulk}.  
\begin{figure}
\includegraphics[scale=1.56]{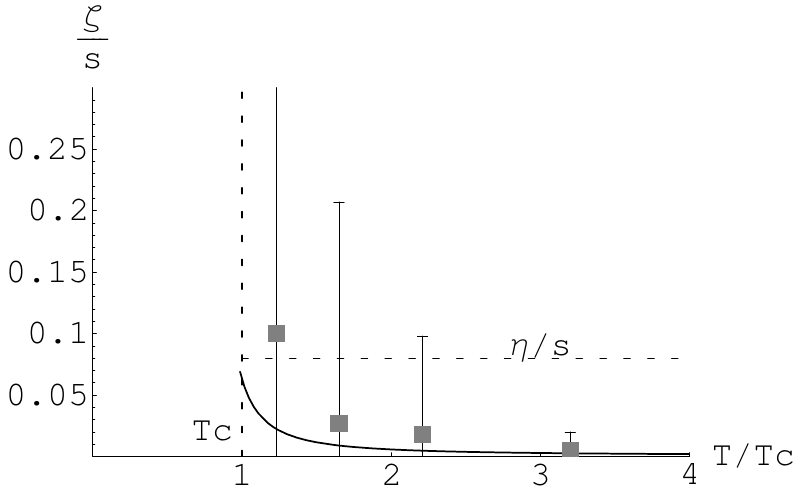}
\caption{Result of the ihQCD calculation for the bulk viscosity compared with the lattice data of \cite{MeyerBulk}.}
\lab{figbulk}
\end{figure}

In this plot we compare our result with the lattice QCD calculation of \cite{MeyerBulk}. The latter calculation involves large systematic and statistical errors. These errors are due to the fact that, to obtain a real-time correlation function such as (\ref{retG}) from the lattice, one needs to analytically continue the Euclidean correlators, that necessitate the knowledge of the entire spectral density of QCD associated with the energy-momentum tensor \cite{MeyerBulk}, an information that we do not have.  The ihQCD result quantitatively agrees with another holographic model for QCD \cite{GubserMimic2}. We observe two features in figure \ref{figbulk}. First, the bulk viscosity increases towards the deconfinement transition at $T=T_c$. Second, the ratio $\zeta/s$ vanishes at very large temperatures, a result qualitatively consistent with  perturbative QCD. 
\begin{figure}[h]
\includegraphics[scale=0.4]{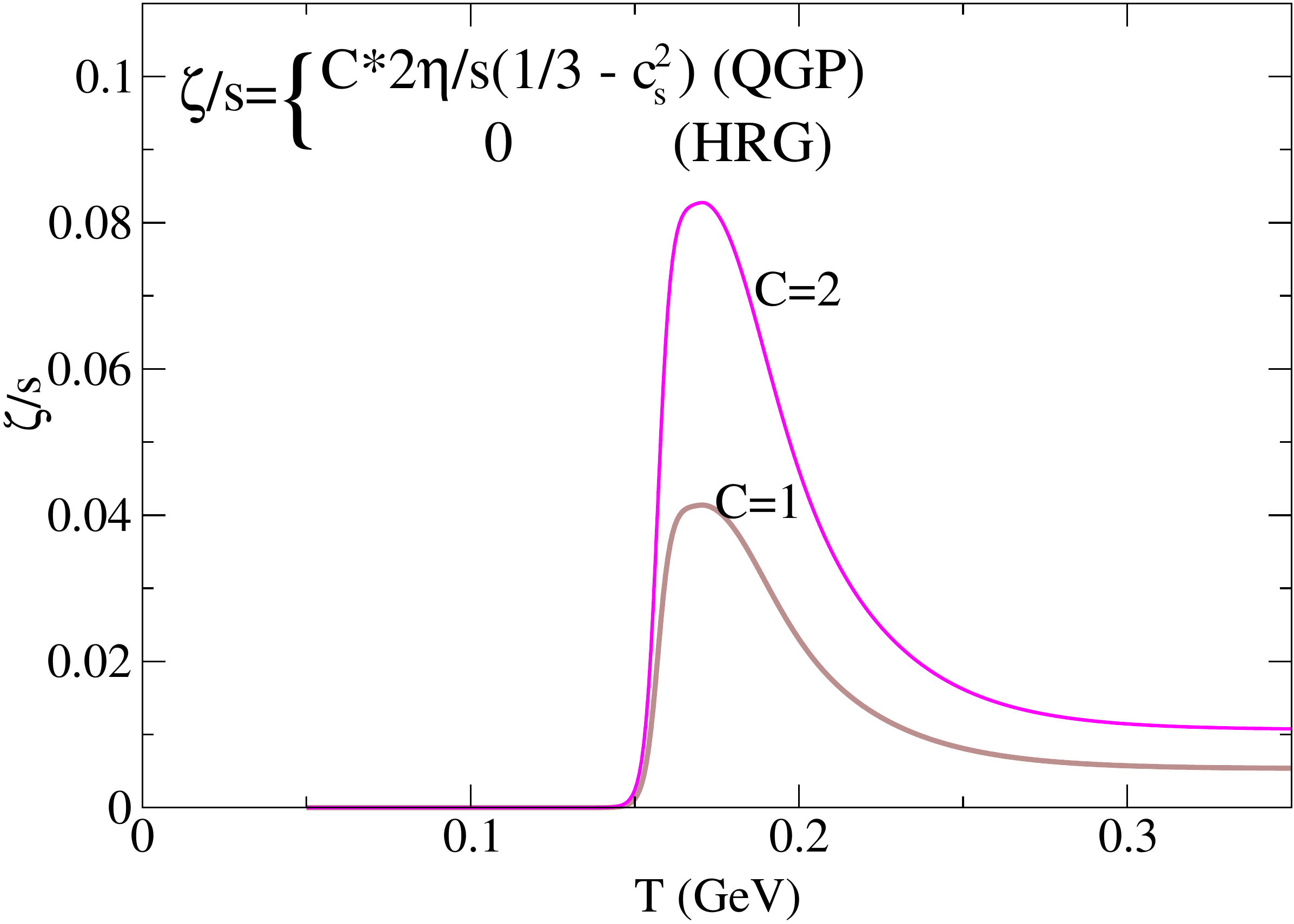}\\
\includegraphics[scale=0.4]{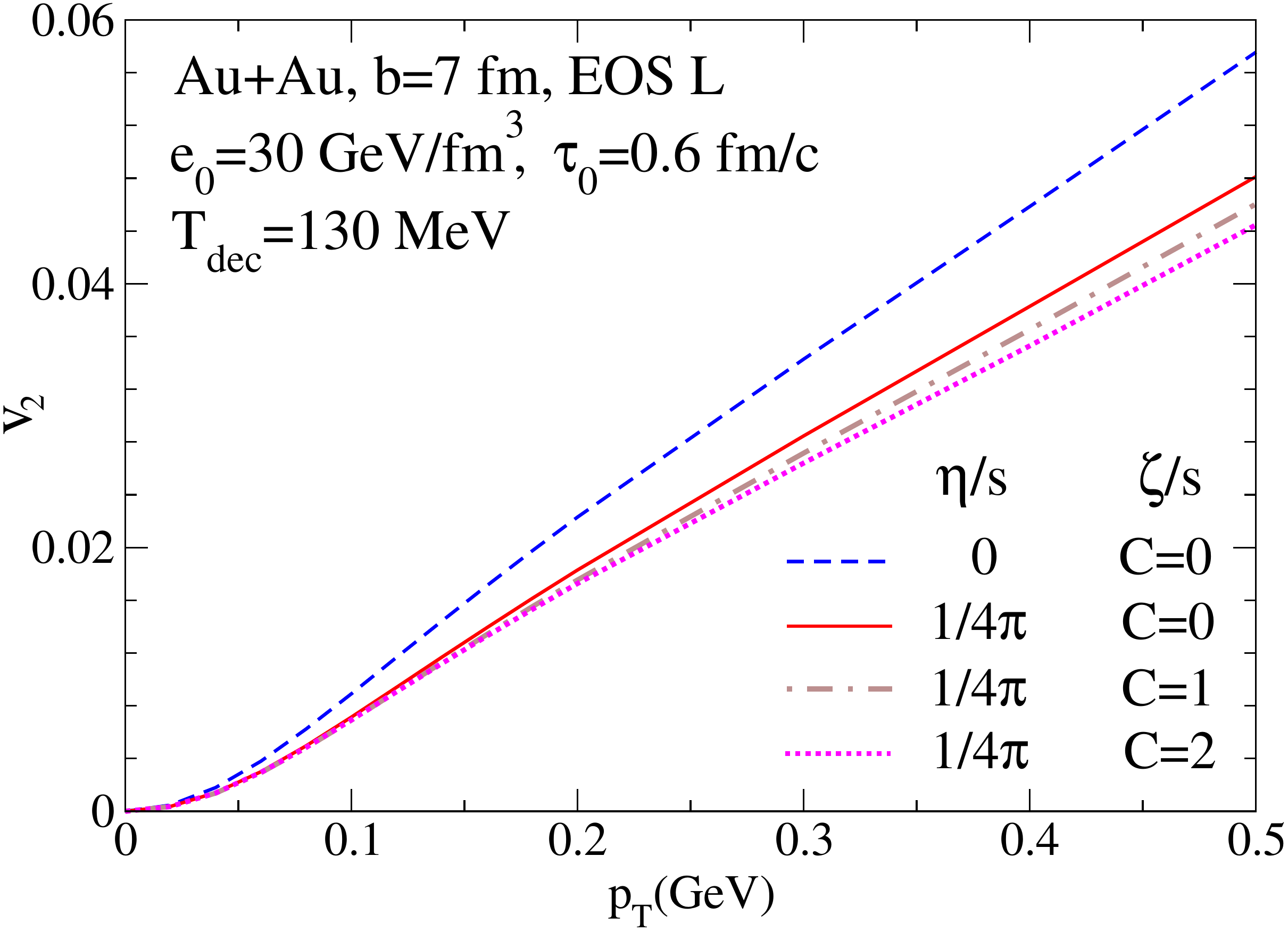}
\caption{Effect of the bulk viscosity on the elliptic flow parameter (the second moment of the spectrum) obtained from RHIC data. Top: parametrization of the  trial bulk viscosity profile. Bottom: comparison with data.}
\label{bulkQGP}
\end{figure}

How much does a non-trivial bulk viscosity affects the hadron spectrum in the heavy ion collision experiments? In figure \ref{bulkQGP} we show a plot taken from  the study \cite{Heinz} comparing the different elliptic flow parameters $v_2$ obtained by the hydrodynamic simulations with varying profiles for $\zeta$ (parametrized by the function on top of the first figure) to data at RHIC, showing that a small bulk viscosity such as figure \ref{figbulk} indeed affects the spectrum, albeit not as much as the shear viscosity.    
\begin{figure}
\includegraphics[scale=0.55]{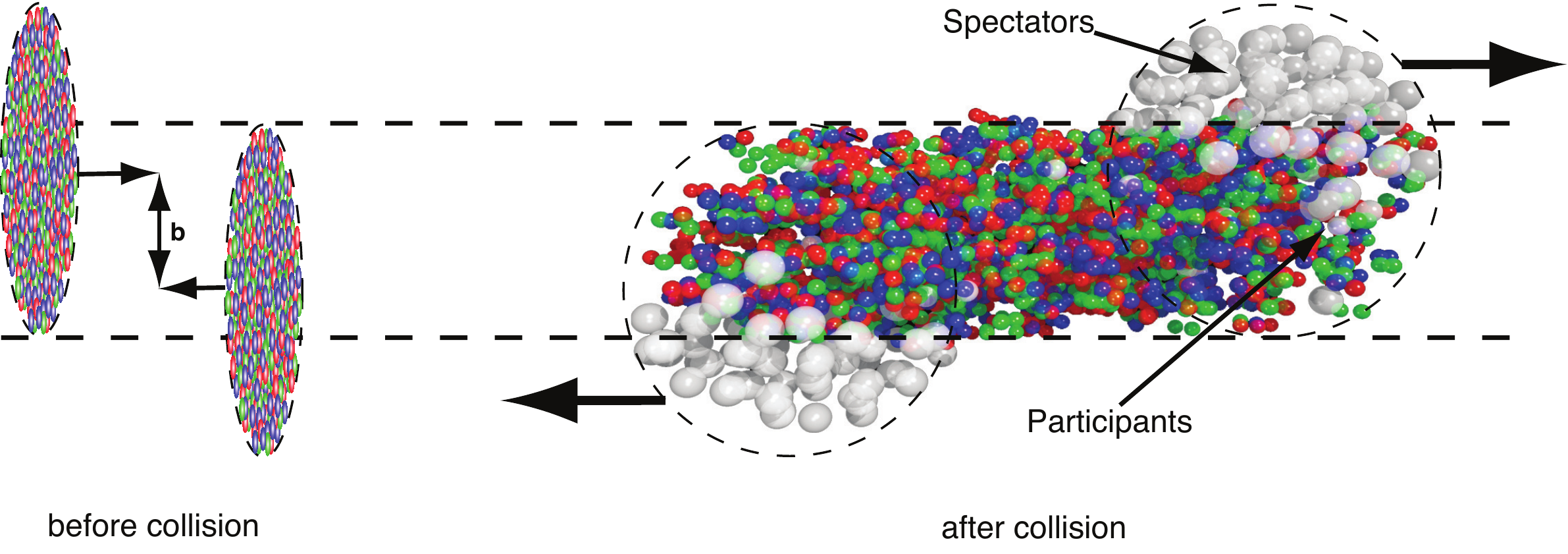}
\caption{Schematic description of hard probes (the ``spectator'' (hard) ions depicted as white balls) only weakly interact with the QGP (''participants'' depicted by colored balls) and provide a measure for energy and momentum dissipation in the plasma.}
\lab{hard}
\end{figure}
\newpage 
\section{Hard probes}
\label{hardprobes} 

Another class of important observables in the heavy ion collisions involve energy and momentum dissipation experienced by the highly energetic ``hard'' quark probes when traveling through the plasma, see figure \ref{hard}. 
\subsection{Generalities}
The hard probes undergo energy loss and momentum broadening when they travel trough the plasma. There are at least two mechanisms this can happen. One is through emission of soft gluons, ``gluon brehmstrahlung''. 
This phenomenon is first studied in the context of AdS/CFT for the conformal ${\cal N} = 4$ super Yang-Mills plasma in \cite{Wiedemann}. It is also studied in the context of the improved holographic QCD in \cite{GursoyLangevin}. Here we will not explain this phenomenon in detail and we will instead focus on another mechanism that leads to energy-momentum loss: the drag force and the statistical Langevin force the hard probes experience when they travel trough the QGP. 
\begin{figure}
\begin{center}
\includegraphics[scale=0.6]{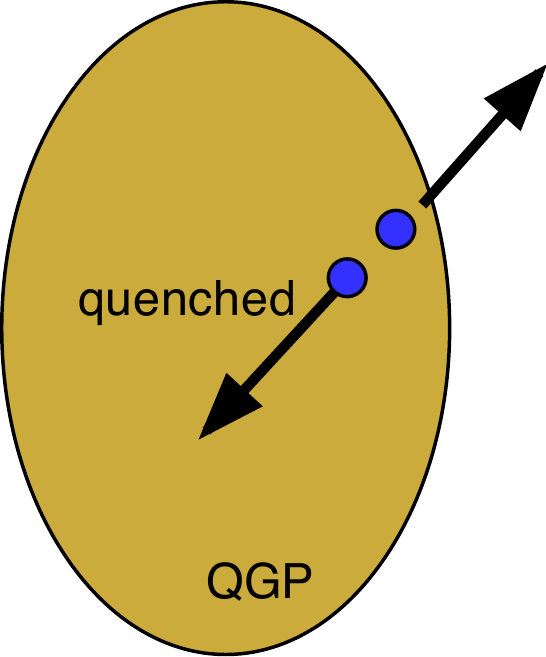}
\end{center}
\caption{Schematic description the jet-quenching phenomenon.}
\lab{jetq1}
\end{figure}
One can write down a phenomenological equation of motion as for the drag force under these two forces as: 
\be\lab{hardforce}
\frac{dp^i}{dt} = -\eta_D^{ij}(\vec{p}^2) p_j + \xi^i(t), \qquad \langle \xi^i(t)\xi^j(t')\rangle = \kappa^{ij} \delta(t-t')\, , 
\ee
where $p^i$ is the spatial momentum of the hard probe, $\eta_D^{ij}$ is a drag coefficient associated with the general drag exerted upon the probe by the QGP, $\xi^i$ is the statistical Langevin force encapsulating the effects of small kicks from fluctations of the quarks and gluons in the plasma, modelled by Brownian motion, and $\kappa^{ij}$ are the diffusion constants representing the white noise associated with the Brownian motion in the plasma.  

One observable that is directly related to the diffusion constants in (\ref{hardforce}) is the so-called ``jet-quenching parameter''. This phenomenon is associated with two back-to-back quarks created close to the boundary of the plasma: as schematically represented in figure \ref{jetq1}, one quark easily gets out through the boundary, but its partner lose energy and momentum, having to travel through the entire plasma. 
This phenomenon is indeed observed in the heavy ion collisions. In figure \ref{jetq2} we show an actual event observed at the LHC. As one can see the lucky quark jet gets out of the plasma finally depositing its energy-momentum at the calorimeters, but its partner is gone missing depositing all of its energy-momentum in the plasma. 

The jet-quenching parameter associated with this phenomenon can be defined by the average transverse momentum lost by the quark-probe per length of flight $D$ as\footnote{See \cite{Wiedemann} for an alternative definition associated with another physical mechanism, ``gluon Brehmstahlung'' in the QGP.}  
\be\lab{qhat}
\hat{q} = \frac{\la p_\perp^2\ra}{D}  = 2\frac{\kappa_\perp}{v} \, ,
\ee
where the second equation follows from a standard calculation \cite{GursoyLangevin} using the equation of  motion (\ref{hardforce}) with $v$ being the average velocity of the hard-probe. 
\begin{figure}
\includegraphics[scale=0.2]{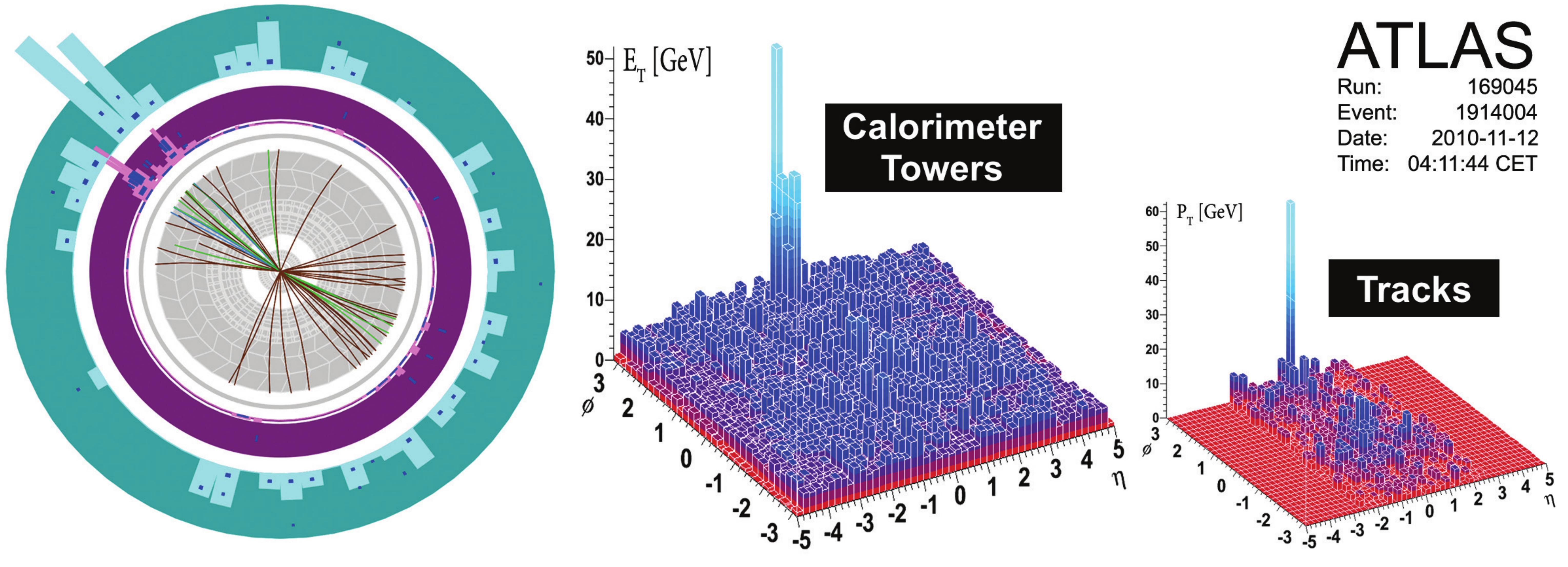}
\caption{A jet-quenching event observed in the heavy ion collisions at LHC.}
\lab{jetq2}
\end{figure}

How do we describe this phenomenon in the holographic dual theory? As we described in the Introduction, an infinitely massive (probe)  quark is associated to the end points of open strings ending on the boundary of the geometry and extending through the interior of the bulk. Then the hard probe moving through the plasma with velocity $v$ corresponds to the ``trailing string'' \cite{Gubsertrail, Kozcaz}, shown in figure \ref{fighard}. 
\begin{figure}
 \begin{center}
\includegraphics[scale=0.6]{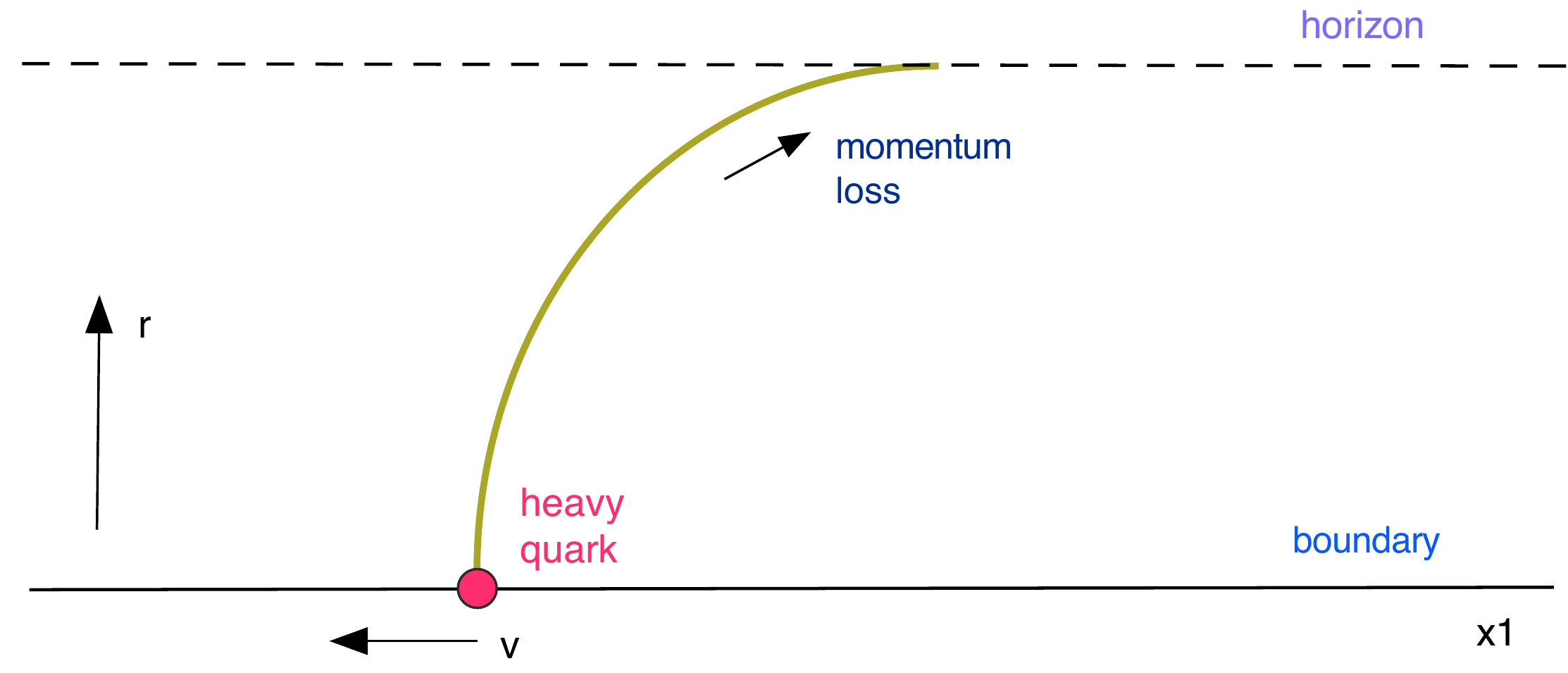}
 \end{center}
 \caption{A trailing string solution in the holographic background that describes the hard-probe traveling through the QGP with velocity $v$. The string loses its momentum to the horizon of the black-brane background depicted by the dashed line.}
 \lab{fighard}
\end{figure}
Given the background geometry, it is a standard exercise to solve the equation of motion of the string that follows from the string action (\ref{Vst}) with the boundary condition $X^1 = v t$ at $r=0$. One can then make an ansatz 
\be\lab{stansz} 
X^1 = v t   + \rho(r), \qquad X^i = 0\quad  (i\neq 1)\, ,
\ee 
and compute the tail $\rho(r)$ from the string equation of motion. We shall not reproduce this calculation in detail here but mention the important points. The original calculation for the AdS background (conformal plasma) can be found in \cite{Gubsertrail, Kozcaz}, a general discussion can be found in \cite{Giecold:2009cg} and the calculation for the ihQCD background (ignoring flavors) can be found in \cite{GursoyBulk}. 

First of all, when one calculates the metric on the world-sheet of the string (\ref{stansz})  i.e. $h_{\a\b} = \6_\a X^\mu \6_\b X^\nu G_{\m\n}$ embedded in the black-brane background (that corresponds to the plasma state) that is denoted by $G_{\m\n}$ here, one generically finds a horizon on the world-sheet, the world-sheet metric being:
\be\lab{wsmet}
ds^2 = b^2\le[ -(f(r) - v^2) d\tau^2 + \frac{dr^2}{f(r)- v^2e^{4A(r_s)-4A_s(r)}}\ri]\, ,
\ee
where $f(r)$ is the blackening factor in metric (\ref{sol2}) and $A_s(r)$ is the string-frame conformal factor in (\ref{metst}) for (\ref{sol2}). That is to say we have a ``black world-sheet''. This is not to be confused with the horizon of the background geometry that is located at $r=r_h$, shown by the dashed line in figure \ref{fighard}. This world-sheet horizon is instead at a location 
\be\lab{wshor} 
r=r_s \qquad \textrm{where}\qquad f(r_s) = v^2\, ,
\ee
where $f(r)$ here is the blackening factor in (\ref{sol2}). We depict the generic geometry of the world-sheet in figure \ref{figws}.  
\begin{figure}
 \begin{center}
\includegraphics[scale=0.4]{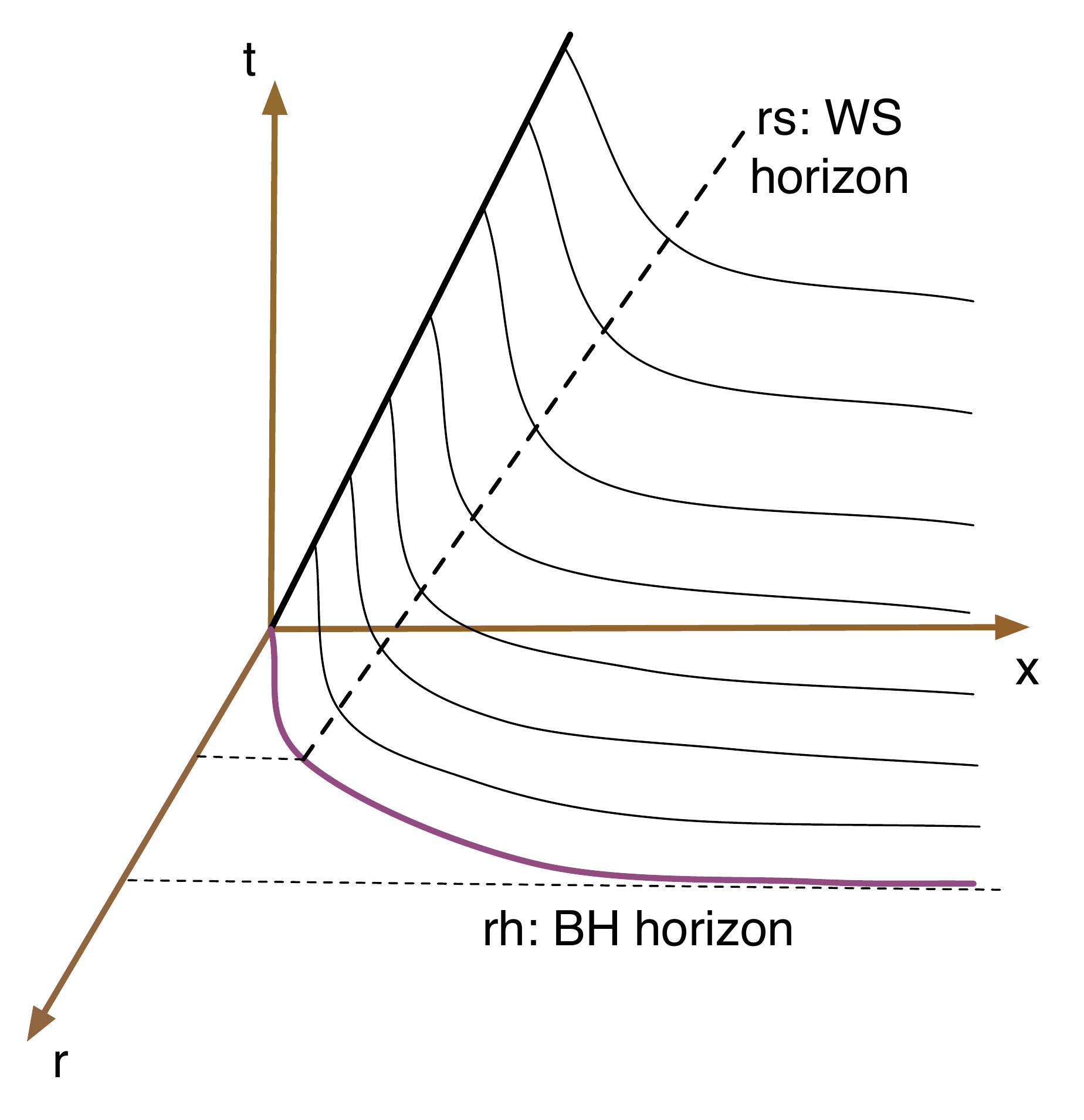}
 \end{center}
 \caption{Typical world-sheet geometry of the trailing string. Here $r_h$ denotes the horizon of the background geometry and $r_s<r_h$  denotes the horizon of the world-sheet metric.}
 \lab{figws}
\end{figure}
Let us note, in passing, that the temperature associated with the black world-sheet is given in terms of the background temperaure $T$ as, 
\be\lab{wsT} 
T_s = T (1-v^2)^{\frac14}\, .
\ee
\subsection{Drag force}
The string falls in the background horizon as in figure \ref{fighard} and loses its momentum. Calculating the world-sheet energy using the standard string theory formula 
\be\lab{wsmom}
\Pi_0 = - \frac{1}{2\pi \ell_s^2} \sqrt{-h} h^{\a\tau}\6_\b X_0\, ,
\ee  
one can calculate  \cite{GursoyBulk} the drag force in (\ref{hardforce}) as
\be\lab{dragf}
F_1 = dp^1/dt  = 1/v\,  d\Pi_0/dt = - \eta_D^{11} p^1 = -\frac{1}{2\pi\ell_s^2} \,\,v \,\, e^{2A(r_s)}\l(r_s)^{\frac43},  
\ee 
where $r_s$ is defined in (\ref{wshor}). 
One can further obtain relativistic and non-relativistic limits of this results, arriving at the following analytic expressions: 
\bea
F &=& -{\ell^2\over \ell_s^2} \sqrt{{45 ~T s(T)}\over 4N_c^2}{v\over \sqrt{1-v^2}\left(-{b_0\over 4}\log\left[1-v^2\right]\right)^{4\over 3}}+\cdots, \qquad v\to 1 \, ,\\
F&=& -{\ell^2\over \ls^2}\left({45\pi ~s(T)\over N_c^2}\right)^{2\over 3}{\l(r_h)^{4\over 3}\over 2\pi}v+\cdots, \qquad v\to 0\, ,
\eea
where $\ell$ is the AdS radius, $\ell_s$ is the string length and $b_0$ is the coefficient in (\ref{VUV}). 
Let us also mention the original result \cite{Gubsertrail, Kozcaz} for the conformal plasma for comparison:
\be\lab{dragconf}
F_{conf} = \frac{\pi}{2} \sqrt{\l} T^2 \frac{v}{\sqrt{1-v^2}} \, .
\ee
Here $\lambda$ is the 't Hooft coupling of the conformal plasma, that is a parameter of the theory. We compare 
the result of ihQCD (\ref{dragf}) to the conformal result (\ref{dragconf}) for a standard choice \cite{GubserCompare}  $\l=5.5$ in figures \ref{figdrag1}.    
\begin{figure}
 \begin{center}
 \includegraphics[scale=1]{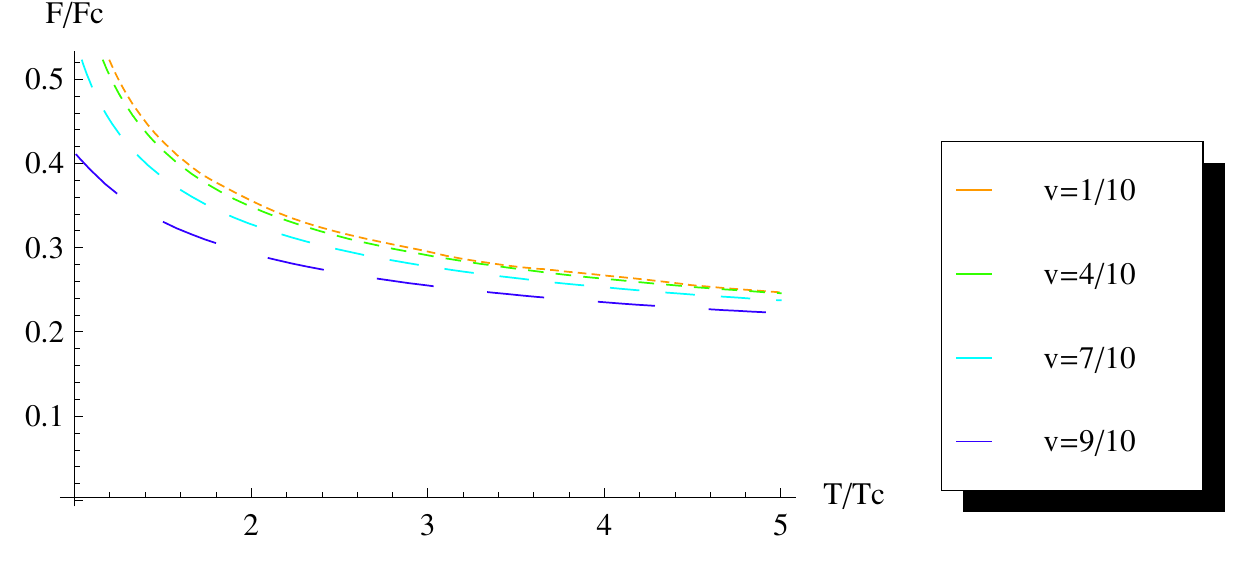}
\includegraphics[scale=1]{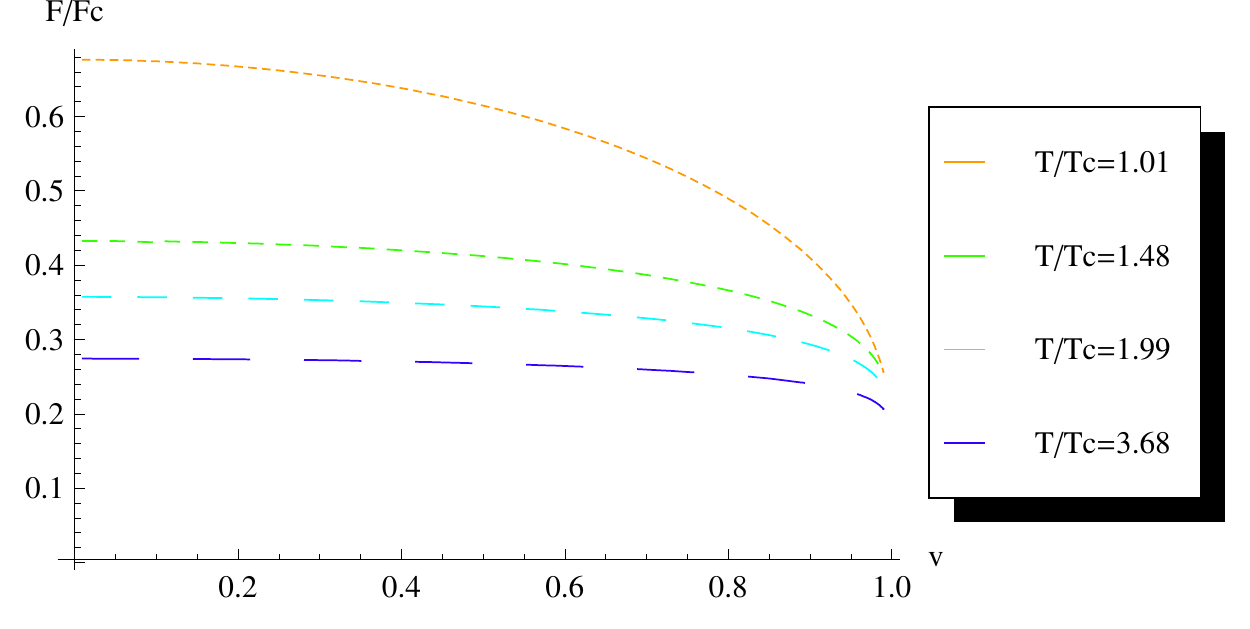}
 \end{center}
 \caption{Comparison of the ihQCD result for the drag force for the various values of $v$ and $T$ to the conformal result  (\ref{dragconf}) for 
 $\l=5.5$. One clearly sees the effects of asymptotic freedom captured by the ihQCD result.} 
 \lab{figdrag1} 
\end{figure}
From these figures we clearly observe the effect of asymptotic freedom captured by the ihQCD plasma, as in the ihQCD model which takes into account the asymptotic freedom in QCD, the drag force decreases with increasing $v$ and $T$.  
\subsection{Diffusion constant}
Finally let us give an overview of the holographic calculation of the diffusion constant $\kappa^{ij}$ in    (\ref{hardforce}). This coefficient measures the rate the momentum carried away by the fluctuations of the plasma, that is modelled by Langevin diffusion in (\ref{hardforce}). The Langevin force couples to fluctuations in the quark location $\delta X^\mu$ through the source term in the action of the probe quark 
\be\lab{qaction} 
S_q = S_0 + \int d\tau \delta X_\mu(\tau) \xi^\mu(\tau)\, , 
\ee
where $S_0$ is the free quark action and $\tau$ is the proper time on the world-line of the quark. How is this picture represented in the holographic dual theory? Fluctuations of the quark location should be the same as the fluctuations  of the trailing  string in figure \ref{fighard} on the boundary. As the boundary of the geometry is identified with boundary of the world-sheet of the trailing string (identifying the proper time $\tau$ of the quark with $\sigma^0$ coordinate on the world-sheet) we have a holographic picture on the world-sheet itself! In other words the fluctuations $\delta X^\mu(\tau)$ should be identified with the leading source term 
in  the boundary expansion of the string fluctuations $\delta X^\mu(r)$ and its response $\xi$ should be associated with the subleading term in $\delta X^\mu(r)$. The diffusion coefficient, that is given by the Wightman two-point function in (\ref{hardforce}), should then be obtained from a standard holographic calculation of the two-point functions. In thermal field theory this Wightman function is related to the retarded Green's function as \cite{GursoyLangevin}:
\be\lab{wightman}
\langle \xi^i(\omega) \xi^j(\omega) \rangle = -\coth\le(\frac{\omega}{2T}\ri) \textrm{Im} G_R^{ij}(\omega)\, .
\ee
We decompose the fluctuations and the corresponding Langevin force as transverse and longitudinal $\xi^i = (\xi_\perp, \xi_\parallel)$ with respect to quark momentum $\vec{p}$ (which we take in the $X^1$ direction above) and the diffusion constants in (\ref{hardforce}) is obtained by the Kubo formula
\be\lab{Kubodif} 
\kappa_\perp  = \lim_{\omega\to 0} \langle \xi_\perp(\omega) \xi_\perp(\omega) \rangle, \qquad
\kappa_\parallel  = \lim_{\omega\to 0} \langle \xi_\parallel(\omega) \xi_\parallel(\omega) \rangle \, .
\ee
 We then calculate the retarded Green's function in holography by solving the string fluctuation equations 
for $\delta X^\mu(r)$ on the world-sheet geometry, that is itself a black-brane with a horizon at $r=r_s$, imposing non-normalizable boundary conditions at the boundary and infalling boundary consitions at the horizon $r=r_s$, substitute in (\ref{wightman}) and read off the diffusion constant from the Kubo formula (\ref{Kubodif}). The result of this calculation \cite{GursoyLangevin} is 
\be\lab{kappas}
\kappa_{\perp}  = \frac{2}{\pi \ell_s^2} b^2(r_s) T_s, \quad 
\qquad \kappa_{\parallel}  = \frac{32\pi}{ \ell_s^2} \frac{b^2(r_s)}{f'(r_s)^2} T^3_s\, .
\ee
Note that it is the world-sheet temperature $T_s$ in (\ref{wsT}) that enters these expressions. We can express the result for the transverse momentum loss in terms of physical parameters in the relativistic limit, 
 \be\lab{kappa} 
 \kappa_{\perp} \approx \frac{(45\pi^2)^{\frac34}}{\sqrt{2}\pi^2}\frac{\ell^2}{\ell_s^2}\frac{(sT)^{\frac34}}{(1-v^2)^{\frac14}}  \le(-\frac{b_0}{4}\log(1-v^2) \ri)^{-\frac43}\, .
 \ee
 The corresponding results for the original calculation \cite{Gubser:2006nz,CasalderreySolana:2007qw} for the conformal plasma instead read, 
 \be
\kappa_{\perp{\cal N}=4} = {\pi} \sqrt{\lambda_{{\cal N}=4}} \gamma^{1/2} T^3, \qquad 
\kappa_{\parallel{\cal N}=4} = {\pi} \sqrt{\lambda_{{\cal N}=4}} \gamma^{5/2} T^3 \, ,
\ee
where $\gamma$ is the Lorentz contraction factor: 
\be\lab{gamma}
\gamma = \frac{1}{\sqrt{1-v^2}}\, .
\ee 
We note that these results satisfy 
\be\lab{ratdifconf} 
\frac{\kappa_{\parallel{\cal N}=4}}{\kappa_{\perp{\cal N}=4} \gamma^2 } = 1\, .
\ee  
One can understand the factor on the RHS as the boost factor associated with Lorentz contraction in the velocity direction. Hence, apart from this kinematic factor the conformal plasma does not distinguish between the longitudinal and the transverse momentum loss.  
It is interesting to note that in our non-conformal plasma we instead obtain a stronger result for this ratio  \cite{GursoyLangevin}:
  \be\lab{ratdifihqcd} 
\frac{\kappa_{\parallel}}{\kappa_{\perp}\gamma^2 } > 1\, ,
\ee  
that is then a universal prediction for strongly interacting  non-conformal plasmas  from holography. 

Finally we quote numerical results \cite{GursoyLangevin} obtained for the jet-quenching parameter in (\ref{qhat}) for a typical hard-probe, i.e. a {\em charm} quark traveling at $p = 10$GeV at $T=250$MeV:  
\be\lab{qhatres} 
\hat{q}_{\perp} = 5.2 (\textrm{direct}),\, 12.0 (\textrm{energy}),\, 13.1 (\textrm{entropy})\quad GeV^2/fm \, , 
\ee
Here the quotes next to the values corresponds to the various schemes used in comparison of the holographic results to QGP \cite{GubserCompare, GursoyLangevin}: ``direct'' scheme instructs to identify the temperature of the holographic plasma with that of the QGP, whereas ``energy'' and ``entropy'' scheme instructs to identify these quantities on the two sides. 

\section{ihQCD at finite B}
\label{finiteB}  

Interaction of electromagnetic fields with the quark-gluon plasma provide an entirely different set of phenomena and related observables. In this section we focus on the influence of external magnetic fields on the QGP. This is a situation realized in off-central heavy-ion experiments, see figure \ref{fig2}. When there is  non-vanishing impact parameter (off-central collisions) the charged ion beams, especially the spectator ions (see figure \ref{hard}) produce large magnetic fields at the center. The magnitude of this magnetic field depends on the experiment and the impact parameter, and a back of the envelope calculation using Biot-Savart law results in
\be\lab{magB} 
B \sim \gamma Z e \frac{e}{R^3} \sim 10^{18} - 10^{19}\, \textrm{G}\, ,  
\ee 
at the time of collision, $\tau=0$. Here $\gamma$ is the Lorentz factor of the collision, $\gamma \approx 100 \,  (1000)$ for RHIC (LHC), $R$ is the radius of ions, $R\sim 7$ fm, $b$is the impact parameter that is of the same order as $R$, $Z$ is the atomic number and $e$ is the electric charge. The values quoted above are the largest magnetic fields we know in the universe. As a comparison, the magnetic fields in neutron stars and magnetars are predicted to be around $10^{13}$ and $10^{15}$ respectively. Indeed the values quoted in (\ref{magB}) are huge, however, a unit more relevant  for the physics of QGP is $MeV$. Influence of these magnetic fields on the QGP physics should be measured by the ratio $eB/ m_\pi^2$ where $m_\pi \approx 135$ MeV is the rest mass of a pion. One finds 5 - 10 for this ratio or RHIC - LHC  at time of collision, hence strong magnetic effects are expected. 

Yet, the relevant time scale for QGP is not the collision time $\tau=0$ but the time when the QGP forms. This time is expected to be around $\tau = 0.3 - 1$ fm/c depending on the experiment, thus the question is how large is B at this time? To answer this question, one has to solve Maxwell's equations sourced by the spectator ion beams, in the presence of the QGP plasma. This calculation has been done in \cite{Kharzeev:2007jp, Skokov:2009qp,Tuchin:2010vs,Voronyuk:2011jd,Deng:2012pc,Tuchin:2013ie,McLerran:2013hla,Gursoy:2014aka}. The result in the last reference is plotted in figure \ref{magfield} as a function of proper time $\tau$ for different choices for the electric conductivity of the plasma. We observe that $B$ decays fast, because the source (mostly the spectator ions) creating this field moves away from the center of collision, yet it is sufficiently large at the time of formation of QGP (around $\tau = 0.5$ fm/c).     
\begin{figure}
 \begin{center}
\includegraphics[scale=1]{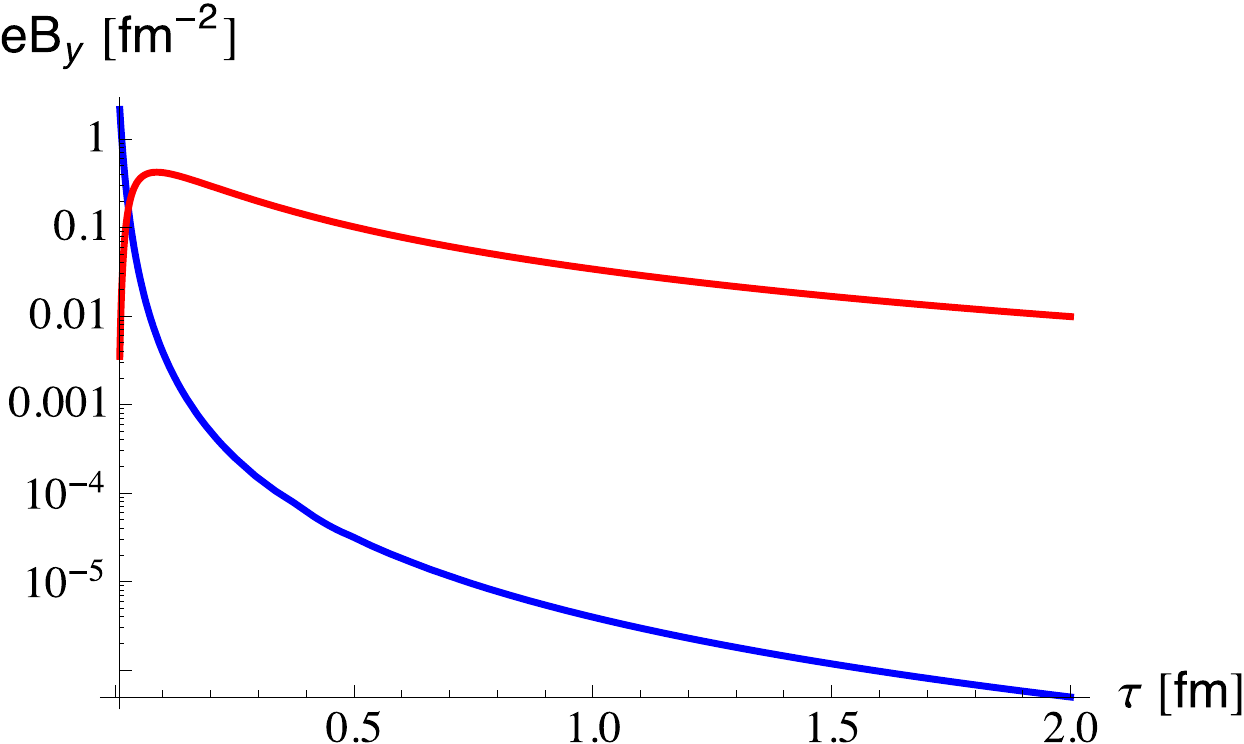}
 \end{center}
 \caption{Magnetic field at the center as a function of proper time resulting from spectator and participant ions in an off-central heavy ion collisions at LHC with impact parameter $b=7$ fm. Blue (red) curve is for electric conductivity $\sigma=0$ ($\sigma=0.023\, fm^{-1}$ ) respectively. Result is taken from \cite{Gursoy:2014aka}. }
 \lab{magfield}
\end{figure}

QCD under external magnetic fields hosts a range of interesting phenomena. In this review we shall discuss:  
\bet
\item Possibility of new phases on the $T-B$ plane
\item (Inverse) magnetic catalysis
\item Anomalous transport
\eet 
Comprehensive reviews for these and other phenomena, exist in the literature, see for example \cite{Kharzeev:2012ph, Kharzeev:2013jha, Miransky:2015ava}. In the next section we explain how to incorporate an external magnetic field in the picture of ihQCD, then we consider the phenomena above one by one in sections \ref{phaseBT}, \ref{IMC} and \ref{anotran} below. 
\subsection{Background at a finite magnetic field and temperature}
For simplicity, in this section we assume a plasma of infinite extent and a constant magnetic field in one direction that we take as the $x_3$ direction.  Magnetic field couples directly only to quarks in the QGP. Hence, for the same reasons as explained in section \ref{flavor}, its effects would be negligible in the large $N_c$ limit unless we also take $N_f\to\infty$ keeping the ratio fixed as in (\ref{vlim}). The relevant action is the same as in that section, that is (\ref{fullaction}) with $S_g$ given by (\ref{action}) and (\ref{pot}) and $S_f$ can be simplified down to (\ref{actfs}) that we reproduce here: 
\be
\label{act2}
S_f =-x\, M^3 N_c^2 \int d^5x V_f(\l,\tau) \sqrt{- \mathrm{det}\left(g_{\m\n} + w(\l)\, F^V_{\m\n} + \kappa(\l)\, \partial_{\m} \tau \,\partial_{\n} \tau\right) } \, .
\ee
The potentials $w$, $\kappa$ and $V_f$ are given in appendix \ref{appB}. The only difference from that section is that the $U(1)_V$ bulk gauge field is taken as  
\be
\lab{VmuB}
A^V_\mu =\left(0,-\frac{x_2 B}{2},\frac{x_1 B}{2},0,0\right)\, .
\ee
This choice indeed produces a constant magnetic field in the $x_3$ direction on the boundary at $r=0$. Because of this, the $SO(3)$ rotational symmetry of the plasma is broken down to an $SO(2)$ around $x_3$ direction. Thus the correct ansatz for the black-brane should be  
 \bea
 \lab{metB}
 ds^2&=&e^{2A(r)}\left(-f(r) dt^2+dx_1^2+dx_2^2+e^{2W(r)}dx_3^2+f(r)^{-1}dr^2\right), \\
 \phi&=&\f(r), \qquad \tau = \tau(r) \, \nn 
 \eea
 where we took into account the breaking of rotational symmetry by introducing a new metric function $W(r)$. There is a horizon at $r=r_h$ where $f$ vanishes and one has to require the same boundary asymptotics at $r\to 0$ as in the previous sections. In particular the new function $W \to 0$ as $r\to 0$. When we compare the physics that result from this action for $B\neq 0$ to the  physics at $B=0$, we have to make sure that the solution (\ref{metB}) for $B\neq 0$ and the solution (\ref{sol2}) for $B=0$ has exactly the same integration constants $T$, $\Lambda_{QCD}$, and quark mass $m_q$ that we set to zero in this review. 
 
 The ansatz (\ref{VmuB}) solves the Maxwell's equations automatically. One is left with solving the coupled non-linear system of Einstein's equations for the functions $A$, $f$ and $W$, the dilaton equation of motion for $\phi$ and the tachyon equation of motion for $\tau$. This is a formidable task, yet it is manageable by a numerical code. This has been achieved in this full system in \cite{Drwenski:2015sha, Demircik:2016nhr, GursoyIMC} and we will present the results of the last reference in the next section. 
 
 \subsection{Phase diagram of ihQCD under external magnetic field}
 \lab{phaseBT} 
 
 In the very recent paper \cite{GursoyIMC}, the phase diagram of ihQCD is studied on the phase space parametrized by temperature T and magnetic field B. The coupling of B to the background is controlled by the ratio of flavors to glue $x$ and the function $w$ that enters in (\ref{act2}). We allow for a one-parameter parametrization of the function $w$, as shown in appendix \ref{appB} parametrized by a positive real number $c$.  In this section we set the baryon chemical potential zero: $\mu=0$. 
 
 The phase diagram one obtains at finite T and B is qualitatively similar to the one at finite T and $\mu$. In particular, typically there exists three phases: confined - chiral symmetry broken, deconfined - chiral symmetry broken and deconfined - chiral symmetry restored. The first two are separated by a first order deconfinement transition line that we denote by $T_d(B)$ in this section. The last two are separated by a second order transition line that we denote by  $T_\chi(B)$. We plot these functions in figure \ref{figsep} for $x=1$ and various different choices for $c$.   
\begin{figure}
\includegraphics[scale=.8]{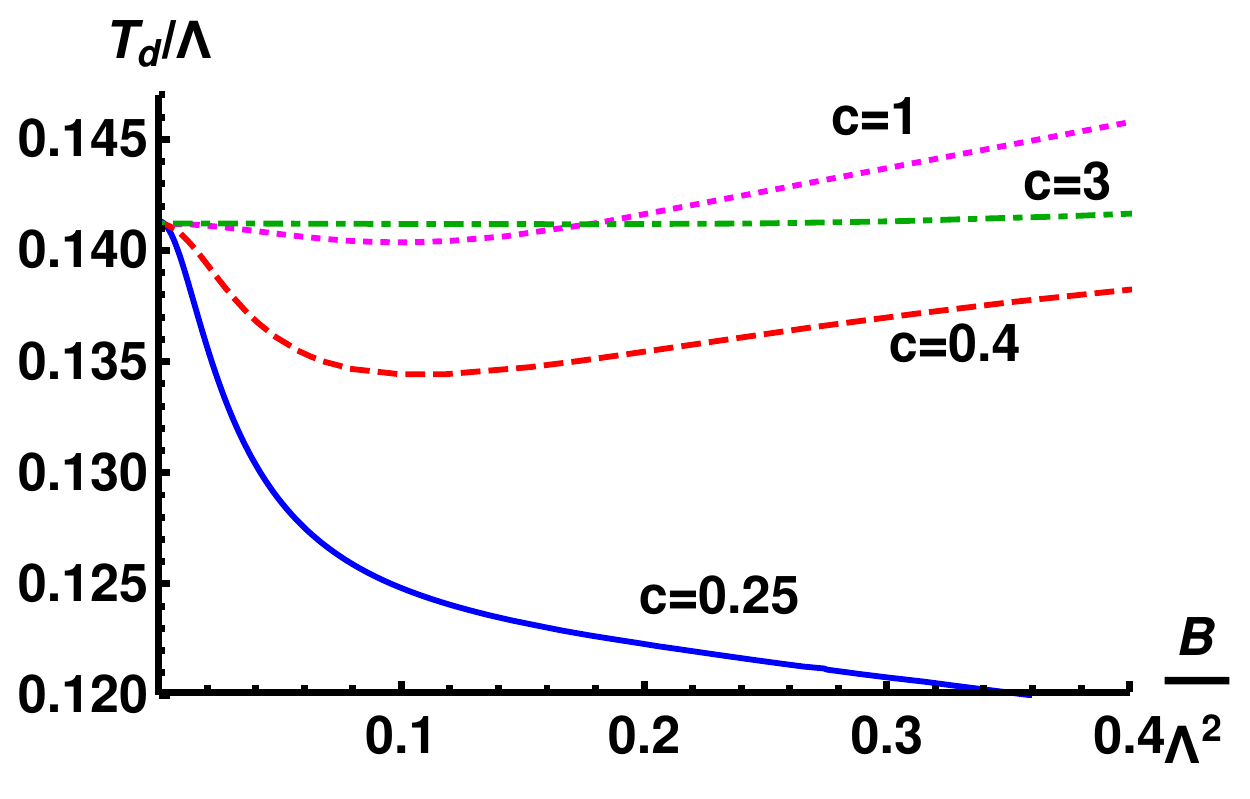}
\includegraphics[scale=.8]{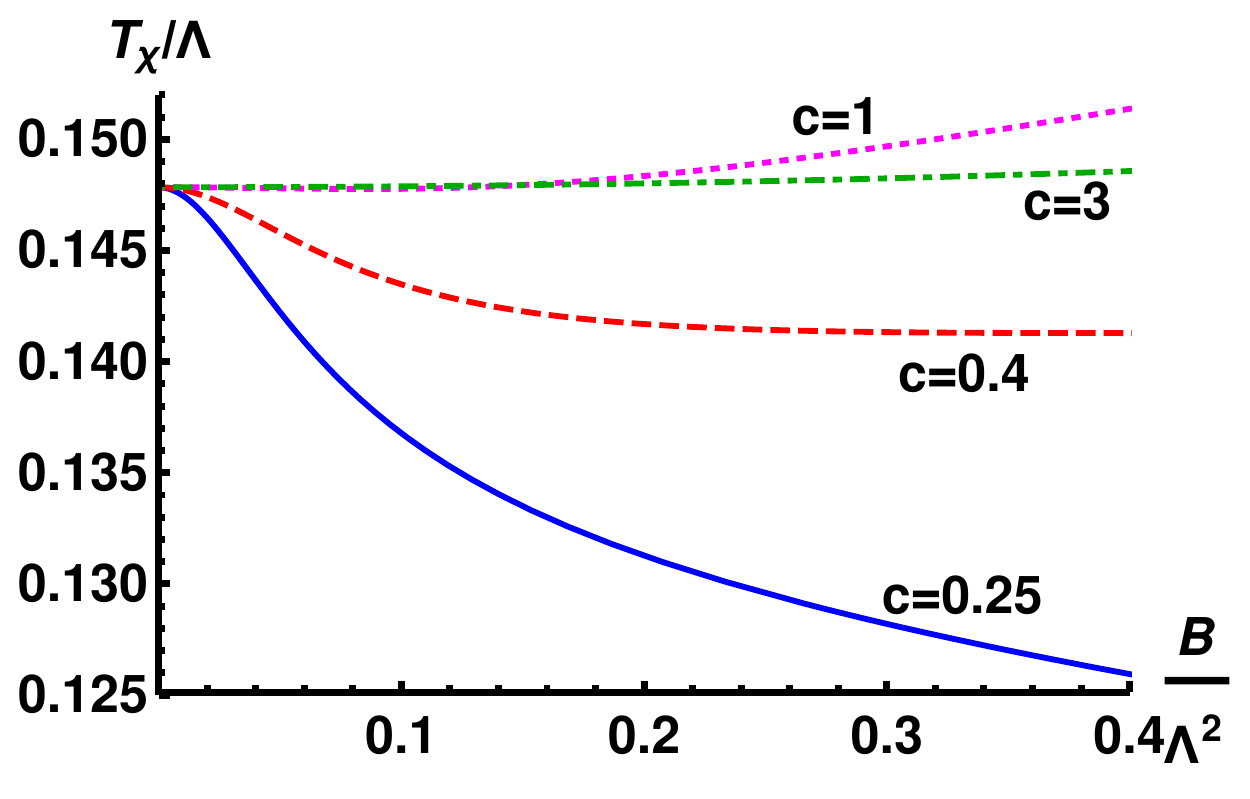}
 \caption{The deconfinement transition line (left) and the chiral symmetry restoration transition line (right) as a function of B for different choices of the parameter $c$ that parametrizes the response of the backgroud to the magnetic field in ihQCD. Figures reproduced from \cite{GursoyIMC}. }
 \lab{figsep}
\end{figure}
As we discuss in the next section a choice $c=0.4$ turns out to agree best with the recent lattice QCD results in \cite{Bali:2011qj,Bali:2011uf,Bali:2012zg,D'Elia:2012tr} regarding the phenomenon of ``inverse magnetic catalysis''. Therefore we fix $c=0.4$ below. We observe that both $T_d$ and $T_\chi$ exhibits a non-trivial profile in B. For smaller values of $c$ such as 0.4 they both start off decreasing for small values of B, reach a minimum at some intermediate value of B and increase thereof for larger B. The former behavior is typically associated with the phenomenon of inverse magnetic catalysis, see section \ref{IMC}. 

The phase diagram undergoes non-trivial changes when the number of flavors $x$ is varied. In figure \ref{pdB2} we show the diagram for the choice $c=0.4$ for the various values of $x$.  
\begin{figure}
\includegraphics[scale=.65]{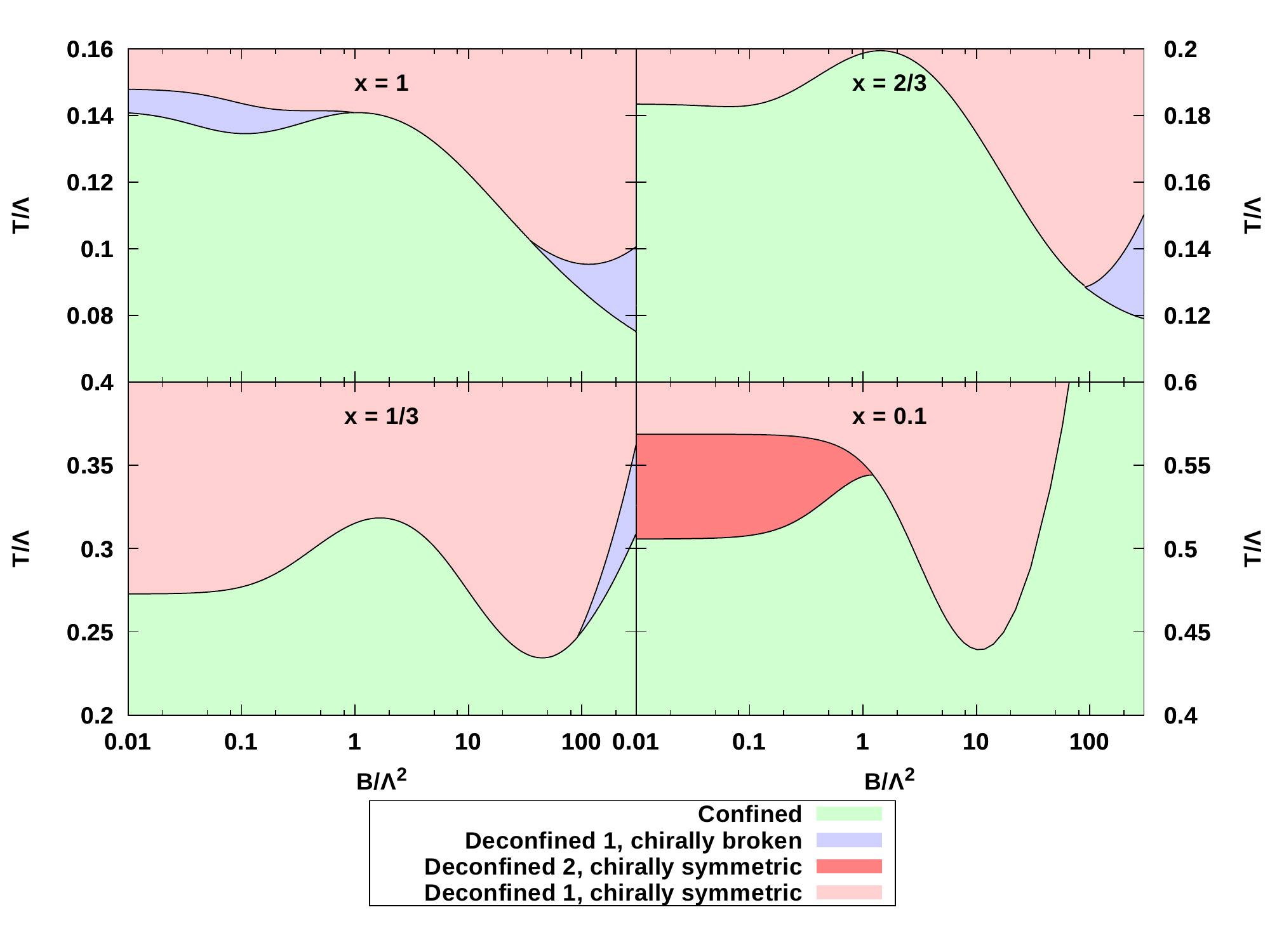}
 \caption{Phase diagram of ihQCD under an external magnetic field for the various values of the ratio $x = N_f/N_c$. Plots reproduced from paper \cite{GursoyIMC}.}
 \lab{pdB2}
\end{figure}
We observe that for very small values of $x$, such as $x=0.1$ the deconfined-chiral symmetry broken phase does not exist. Instead, there are two different deconfined - chiral symmetry restored phases shown by red and pink in these figures. These two phases correspond to different black-brane solutions on the gravity side that are separated by a first order phase transition. The physical meaning of the red phase and whether it is relevant to QGP physics is unclear to the author at the time of writing this review. When $x$ is increased, this second phase disappears, instead a deconfined - chiral symmetry broken phase (blue in figure \ref{pdB2}) arise at larger values of $B$. This happens around $x=1/3$ onward. Finally for even larger values of $x$ such as $x=1$, the same phase also appears for small values of $B$, in agreement with the $\mu\to 0$ edge of the phase diagram  in figure \ref{pdmu}.

\subsection{Inverse magnetic catalysis} 
\lab{IMC} 

The quark condensate in QCD behaves non-trivially under an external magnetic field. It is long known from perturbative QCD studies \cite{Gusynin:1994re, Gusynin:1994xp, Gusynin:1994va} that the condensate is strengthened when a magnetic field is turned on. This phenomenon is called the ``magnetic catalysis'' and one can qualitatively understand the reason behind this phenomenon as follows. Turning on a magnetic field results in Landau quantization of the fermions. In particular the momentum in the directions transverse to B is discretized and the separation between these discrete states increase with B. Landau quantization therefore restricts motion in the transverse directions. In particular for large values of B the dominant ground state has vanishing transverse momentum. This, in turn projects the physics of flavor in QCD to 1+1 dimensions for large B. On the other hand, it is well-known that the IR physics responsible for formation of condensates in general is stronger in 1+1 dimensions, resulting in an increase in the magnitude of the quark condensate with B. This suggestive argument can of course be shown to be the case by explicit calculations in perturbative QCD. 

The question then is what happens at strong coupling, such as the limit of QCD relevant for QGP physics. Recent lattice studies of QCD with 2+1 flavors \cite{Bali:2011qj,Bali:2011uf,Bali:2012zg,D'Elia:2012tr} show a more complicated behavior. It is found that the condensate again increases for small values of B in the confined phase up to a certain value of B, but it starts decreasing for larger values. This critical value of B depends on the temperature. Moreover, for temperatures above a certain value, slightly below the deconfinement crossover temperature, around  $150$~MeV, the condensate starts decreasing even for smaller B down to $B=0$. Therefore one finds that the strong coupling effects in QCD triggers the opposite effect, called the ``inverse magnetic catalysis''. The precise physical mechanism for this behavior is not completely clear at the time of writing this review. There are indications however from further lattice studies \cite{Bruckmann:2013oba, Bruckmann:2013ufa} that this complicated profile for the condensate results from a competition between two separate sources. Considering the path integral $\la \bar q q \ra$ one can identify these two sources as follows. First, there is a direct coupling to B of the fermion propagators inside the operator $\bar q q$ in the path integral. This source is called the ``valence quarks'' in \cite{Bruckmann:2013oba} and it always tends to strengthen the condensate, essentially for the same reason explained above for magnetic catalysis. There is a second source of coupling to B however, that comes from the quark determinant arising from the gluon path integral. This second source, called the ``sea quarks'' is weak at weak coupling compared to the first one above, hence it can be neglected, and one finds magnetic catalysis. However, it becomes stronger at  intermediate  or large values of the coupling constant, and it was argued in \cite{Bruckmann:2013oba, Bruckmann:2013ufa} that it dominates over the first source for relatively large values of B and T, leading to the inverse effect. These are only suggestive arguments however and it would be great to get a handle on the question in holographic QCD. 

The question has been addressed in the various papers in holography \cite{Zayakin:2008cy, Filev:2011mt, Erdmenger:2011bw, Preis:2012fh, Mamo:2015dea, Dudal:2015wfn, Rougemont:2015oea, Evans:2016jzo}, or with smeared backreacted flavor branes in the Veneziano limit~\cite{Jokela:2013qya} but most of these works are either for adjoint flavors or consider small values of fundamental quarks. Very recently, the question is addressed in \cite{GursoyIMC} for ihQCD in the Veneziano limit (\ref{vlim}). It is found that holography confirms, at least supports the valence vs. sea quark discussion in \cite{Bruckmann:2013oba, Bruckmann:2013ufa}. 

In figure \ref{figcond1} we show the phase transition curves for the deconfinement and chiral symmetry restoration transitions for a choice of $x=1$ and $c=0.4$ for the parametrization of function $w$ in (\ref{act2}).
In this and the following plots the dimensionful quantities are normalized with the integration constant $\Lambda$ that is proportional to the in intrinsic energy scale of QCD $\Lambda_{QCD}$. We observe that indeed both of these transition temperatures decrease with increasing B. In the deconfined - chiral symmetry broken phase, $T_d<T<T_\chi$, this means that it becomes easier to melt the condensate when B is increased. We also show contours of constant condensate in the same plot. This provides a direct confirmation that the condensate decreases with B at least for small enough B. One observes that the curves of constant condensate extend between the curves $T_d(B)$ and $T_\chi(B)$ continuously decreasing with increasing T and finally vanishing at $T_\chi$ leading to the second order chiral symmetry restoration transition discussed in the previous section.   
\begin{figure}
\includegraphics[scale=1]{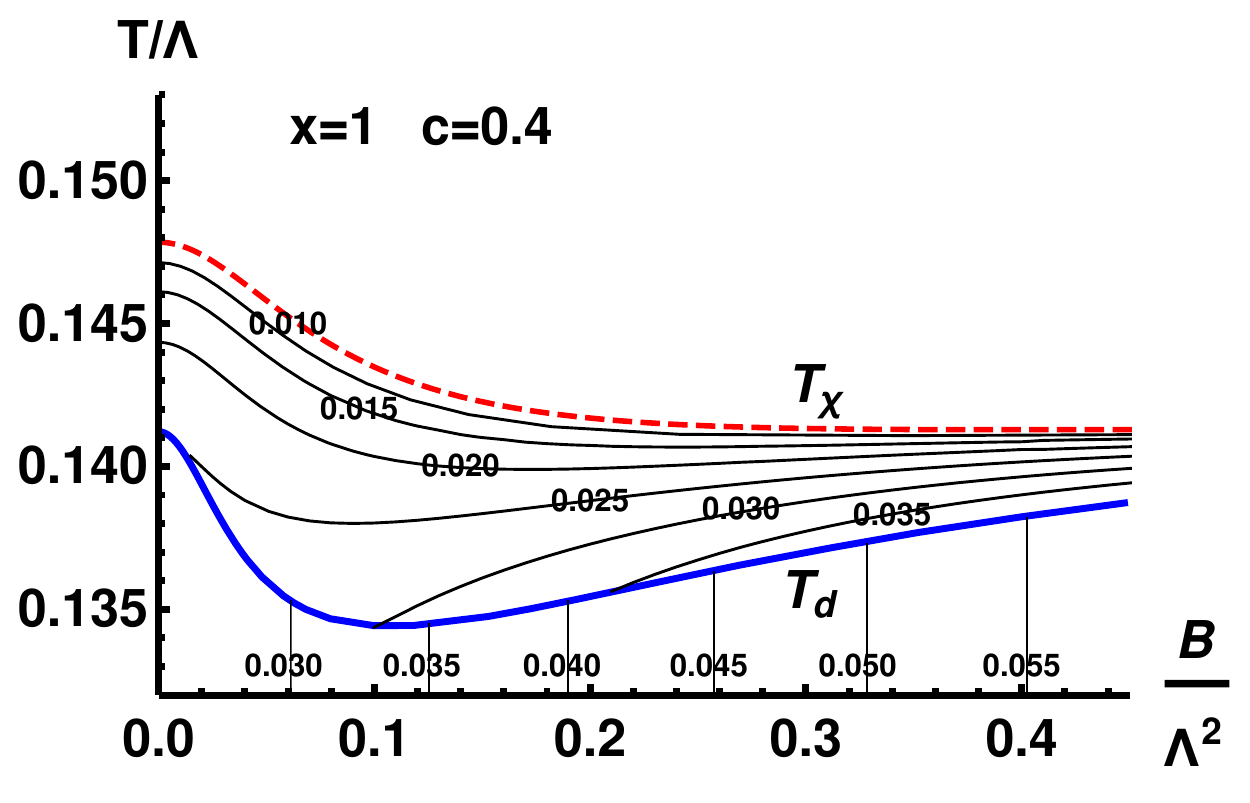}
 \caption{Phase diagram and curves of constant $\la \bar q q\ra$ in ihQCD for a choice of $x=1$ and $c=0.4$.  Plot reproduced from paper \cite{GursoyIMC}.}
 \lab{figcond1}
\end{figure}
The reason for vertical contours of constant condensate in the confined phase is an artifact of holographic QCD: the temperature dependence in the confined phase, that corresponds to the thermal gas solution (\ref{sol1}) cannot be seen in the large-N limit. This is because the temperature dependence in this solution is trivial (there is no blackening factor in (\ref{sol1}) and in order to capture this dependence one has to consider fluctuations of the background fields around the thermal gas background, leading to a correction of the free energy at order $1/N$. Therefore one has to regard the analysis in the confined phase as at $T=0$. 

Finally in figure \ref{figcond2} we plot the condensate, rather the renormalization invariant and dimensionless  combination  $\Delta \Sigma(T,B) =\Sigma(T,B)-\Sigma(T,0)$ where 
\be
\Sigma(T,B)= \frac{\langle \bar q q \rangle (T,B)}{\langle \bar q q \rangle (0,0)}= {1 \over \langle \bar q q \rangle (0,0) } \left(  \langle \bar q q \rangle (T,B)-\langle \bar q q \rangle (0,0) \right)+1 \, .
\ee
We observe that, in complete qualitative agreement with the lattice results described above, the condensate increases with B up to a certain value of the temperature around $T/\Lambda\approx 0.138$, and it starts decreasing for larger T up to the chiral symmetry restoration transition. Above this transition the condensate drops to zero of course, as demonstrated by the blue curve in figure \ref{figcond2}. 
\begin{figure}
\includegraphics[scale=1]{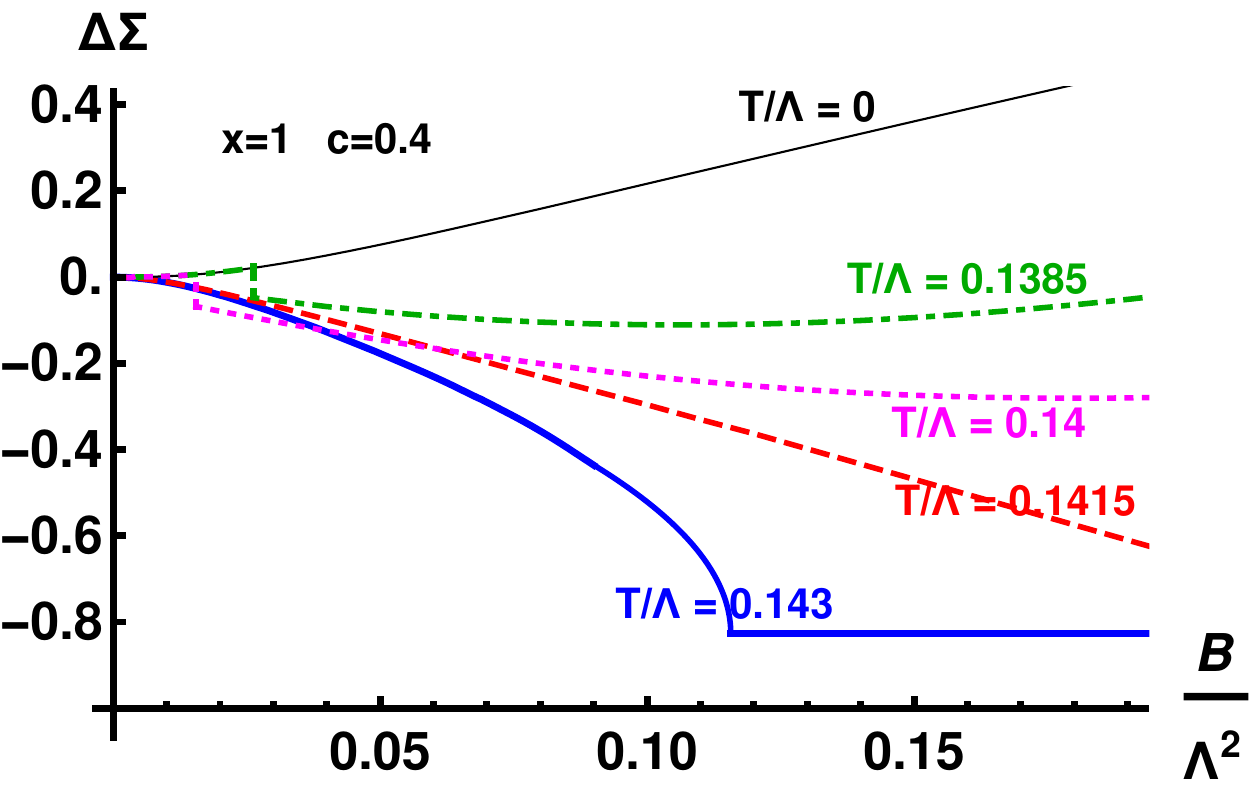}
 \caption{The normalized quark condensate as a function of B in the deconfined - chiral symmetry broken phase in ihQCD for $x=1$ and $c=0.4$ shows clear demonstration of inverse magnetic catalysis.  Plot reproduced from paper \cite{GursoyIMC}.}
 \lab{figcond2}
\end{figure}
The suggestion of  \cite{Bruckmann:2013oba, Bruckmann:2013ufa} for the physical mechanism behind the inverse magnetic catalysis relating it to the ``sea quarks'' as described above can also be tested in the context of ihQCD. The two sources of coupling of the condensate to B, the direct coupling called the valence quarks, and the indirect, glue induced coupling called the sea quarks can be identified in holography with two analogous sources as sollows: The condensate is determined by solving the tachyon equation motion. This equation depends on B again in two different ways. First, there is  the explicit dependence, that we identify with the valence quarks and there is the indirect dependence arising from dependence of the background functions that enter the on B, that we identify with the sea quarks. Various tests of this suggestion is made in \cite{GursoyIMC} by isolating either of the two dependences by playing with the values of B and $x$ and strong indications found supporting this suggestion. 

\subsection{Anomalous transport} 
\lab{anotran}

Another class of very interesting phenomena that occur in QCD under external magnetic fields is the anomalous transport. This, in general refers to new, dissipation free means of transport in QCD and other physical systems such as the Dirac and Weyl semimetals\footnote{See for example \cite{Li:2014bha} for a recent observation of anomalous transport in Dirac semimetals.}, induced by the well-known quantum anomalies of the axial current \cite{A,BJ} in the presence of a parity even vector source such as the external magnetic field or vorticity. Various comprehensive reviews of the subject exist, see \cite{CME2, Kharzeev:2012ph, Kharzeev:2013jha, Miransky:2015ava, Zakharov:2012vv} from a field theory point of view, and \cite{Landsteiner:2011cp} for a holographic point of view. The subject is treated in detail in this school by Karl Landsteiner whose lecture notes are available in \cite{Landsteiner:2016led}. In this section, we shall only touch upon a small corner of  the subject in relevance to the QGP physics: the {\em chiral magnetic effect} \cite{Vilenkin:1978is, Kharzeev:2007jp, CME2}. 
\begin{figure}
\includegraphics[scale=1]{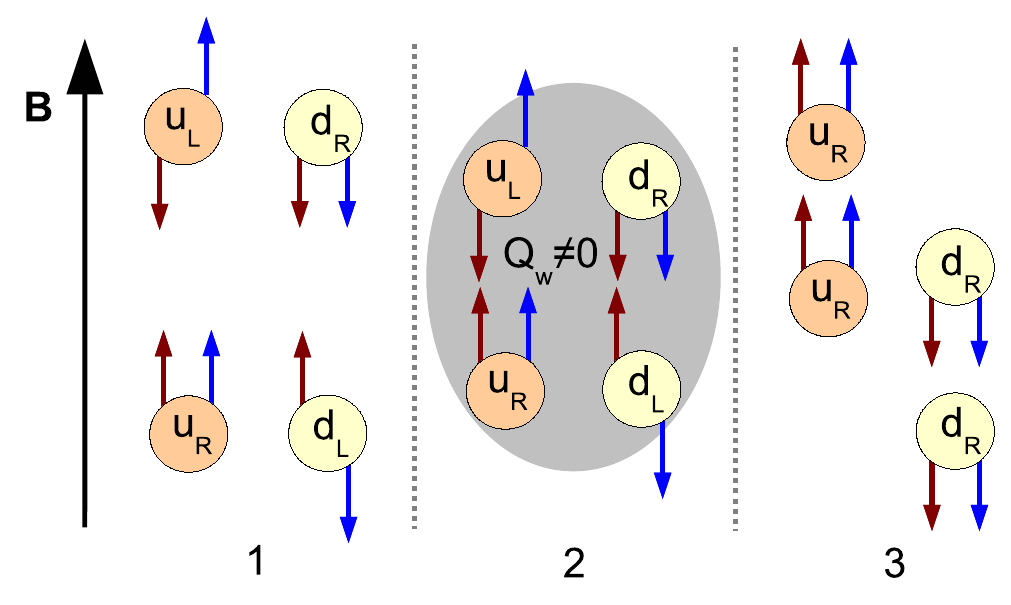}
 \caption{The mechanism that leads to the chiral magnetic effect. The horizontal axis denotes time. In phase  $t=1$ the vacuum contains no topologically non-trivial gluon fields. In phase $t=2$ a gluon configuration with non-trivial topology is generated leading to non-conservation of the axial charge. In phase $t=3$ this non-trivial gluon configuration decays, producing an imbalance in the axial change due to (\ref{anom}).  Plot reproduced from paper \cite{CME2}.}
 \lab{figCME}
\end{figure}

It is well-known that the classical conservation of axial charge in massless QCD is violated at the quantum level due to AVV triangle diagrams that lead to an electromagnetic anomaly and due to AGG triangle diagrams that lead to a QCD anomaly \cite{A,BJ}.:
\be\lab{anom}
J_A^\mu = \sum_{i=1}^{N_f} \bar{\psi}_i \gamma^\mu \gamma^5 \psi_i, \qquad   \partial_\mu J_A^\mu  = \epsilon_{\m\n\a\b} \le( c_1~F_V^{\m\n} F_V^{\a\b} + c_2~\tr (F^{\m\n} F^{\a\b})\ri)\, .
\ee
Here $J_A$ is the axial current that is classically conserved in the absence of quark masses.
Her $F_V$ and $F$ denotes the field strengths of the external electromagnetic fields, and the dynamical gluon fields respectively. $c_1$ and $c_2$ are the electromagnetic and QCD anomaly coefficients respectively. The last term in the RHS of the second equation in (\ref{anom}) is caused gluon field configurations with a non-trivial topology described by the topologic invariant, the gluon winding number: 
\be\lab{qw} 
Q_w = \frac{1}{24\pi^2}\epsilon_{\m\n\a\b} \int d^4x \tr (F^{\m\n} F^{\a\b})\, .
\ee
\begin{figure}
\includegraphics[scale=.53]{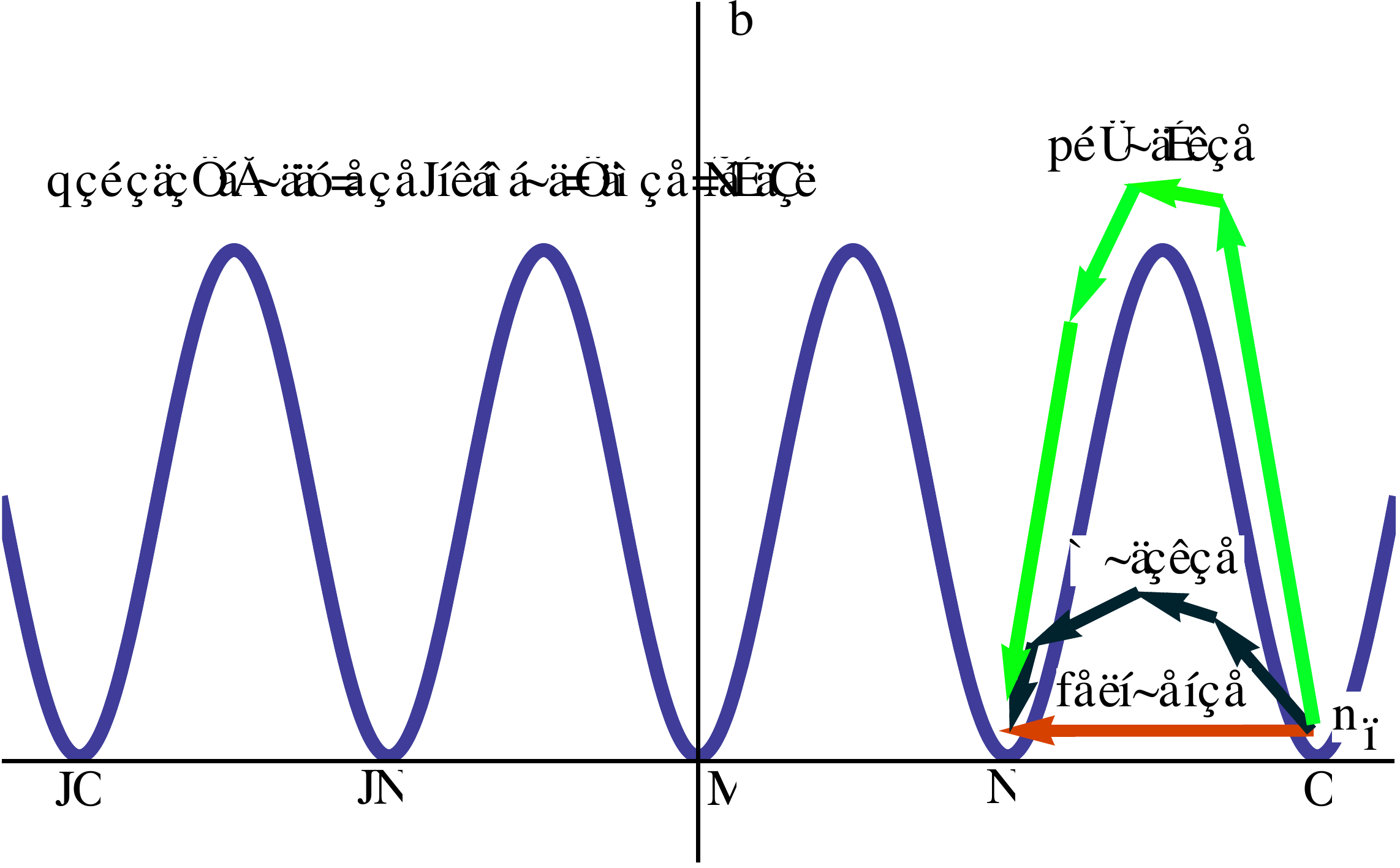}
 \caption{Non-perturbative processes that lead to change in gluon winding number. Sphalerons are the unstable gluon field configurations sitting on top of the potential.}
 \lab{figSph}
\end{figure}
 In QGP physics we are interested in temperatures much larger than the physical quark masses $m_q/T \ll 1$, hence the axial current is indeed effectively conserved in the QGP, at the classical level. However, the quantum anomaly in (\ref{anom}) is expected to result in interesting phenomena associated with anomalous transport. The chiral magnetic effect, CME for short, is one such major phenomenon. This effect is the generation of en electric current in the direction of an external magnetic field due to the anomalies in (\ref{anom}). 
 
 The mechanism in QCD that leads to CME is schematically described in figure \ref{figCME}. The spins of quarks are aligned with B due to the Zeeman effect. Since the quark masses can be neglected at high temperatures, in the absence of an external $\vec{E}\cdot\vec{B}$ term (the first term in (\ref{anom}) and in the absence of any gluon fields with non-trivial topology (the second term in (\ref{anom}) the axial charge is effectively conserved both at the classical and quantum levels. This means that helicity of these particles are also conserved and they will move parallel or anti-parallel to $B$ depending on their helicity $h = \vec{S} \cdot \vec{p}/|\vec{p}|$. Because there are equal number of left and right handed particles then there is no net generation of electric current in phase $t=1$. Now suppose that a gluon configuration with a non-trivial topology is generated in phase $t=2$ which decays in the phase $t=3$.  This would then convert some of the left (right) movers into right (left) movers due to the second term in the anomaly equation (\ref{anom})  leading to an imbalance of the axial charge. Then in phase $t=3$ we are back to the same configuration as in $t=1$ except that there is an axial imbalance. There is still effective conservation of the axial charge in phase $t=3$ both at the classical and the quantum level since the non-trivial gluon configuration decayed, but now there is a net electric current in the direction of $\vec{B}$. 

This electric current generated in the presence of an external magnetic field B can be shown to be 
\be\lab{CMEcur} 
\vec{J_V} = \sigma_B \vec{B}  = c_1 \mu_5 \vec{B}\, ,
\ee
both in field theory \cite{Kharzeev:2007jp} and in hydrodynamics \cite{SonSurowka}. Here an effective chemical potential $\mu_5$ for the axial charge tis introduced to take into account the non-conservation of this charge in (\ref{anom}). For example this $\mu_5$ will be non-zero if gluon configurations with non-trivial topology is  generated, such as phase $t=2$ of figure \ref{figCME}. 

The critical question then is: what mechanism are there in QCD that would lead to generation and decay of such non-trivial gluon configurations? Among possible sources, instantons, calorons and the sphalerons \cite{Manton:1983nd}, it was shown in \cite{Sphaleron1,Sphaleron2,Sphaleron3} that at high temperatures of order $T > \Lambda_{QCD}$ it is the latter, {\em sphaleron decays} constitute the prime source of generation and decay of such non-trivial gluon configurations. In figure \ref{figSph} the mechanism is described schematically. In this figure we plot the vacuum energy of QCD as a function of the gluon-winding number $Q_w$ (\ref{qw}). A sphaleron is an unstable field configuration, that corresponds to the maxima in this figure, that can be produced at high temperatures as the energetics allow this. As they are unstable, they decay by thermal fluctuations generating a net change in $Q_w$. In QCD this process is measured by a transport coefficient the so-called sphaleron decay rate (or Chern-Simons decay rate) \cite{Sphaleron1,Sphaleron2,Sphaleron3} as follows. Define the topological charge,
\be
\label{q}
q(x^{\mu})\equiv \frac{1}{16\pi^2}\textrm{tr} \left[F \wedge F\right] = \frac{1}{64\pi^2} \, \epsilon^{\mu\nu\rho\sigma} \textrm{tr} F_{\mu\nu} F_{\rho\sigma},
\ee
where $x^{\mu}=(t,\vec{x})$. In a state invariant under translations in space and time, the rate of change of $\ncs$ per unit volume $V$ per unit time $t$ is called the Sphaleron decay rate, denoted $\gcs$,
\be
\label{gcsdef}
\gcs \equiv \frac{\langle \left(\Delta \ncs\right)^2\rangle}{V t} = \int d^4x \, \left \langle q(x^{\mu}) q(0) \right \rangle_{\textrm{W}},
\ee
where the subscript W denotes the Wightman function. 

In the discussion that leads to (\ref{CMEcur}) the axial chemical potential $\mu_5$ was generated by such processes that lead to a non-trivial change in $Q_w$, hence $\mu_5$  is larger for larger $\ncs$. Therefore we need to determine the value of (\ref{gcsdef}) in order to assess the likelihood of observing CME in the heavy ion collisions. A QCD calculation at weak coupling leads to the result \cite{Arnold:1996dy} 
\be\lab{pgcs}
\gcs = 192.8 \alpha_s^5 T^4\, ,
\ee 
where $\alpha_s = g_s^2/(4\pi)$ is the interaction strength. 

How significant is $\ncs$ at strong coupling? This question is answered in  the context of holography first in \cite{SonStarinets}. In this paper the bulk field dual to the CP-odd operator (\ref{q}) is identified with the {\em bulk axion} $a(r,x)$ that is a CP-odd pseudo-scalar in the corresponding 5D ${\cal N=8}$ supergravity. It is a massless bulk field. The Wightman function in (\ref{gcsdef}) can be related to the corresponding retarded Green's function as in equation (\ref{wightman}), and the latter can be computed using the holographic prescription by solving this massless bulk field equation of motion with infalling boundary consitions at the horizon and non-normalizable boundary conditions at the boundary. In fact the latter UV value is nothing else but the $\theta$ parameter in the QCD Lagrangian $\theta \eps_{\m\n\a\b} \int d^4x \tr (G^{\m\n} G^{\a\b})$ 
\be\lab{UVa}
a(r,x) \to \kappa\, \theta, \qquad r\to 0\, ,
\ee
because, as mentioned above the bulk axion $a$ couples to the operator $\q \propto \eps_{\m\n\a\b}  \tr (G^{\m\n} G^{\a\b})$ on the boundary. Here $\kappa$ is another free parameter of the model. Using these boundary conditions, one obtains the following answer \cite{SonStarinets},   
\be\lab{confgcs}
\gcs\bigg|_{conf} = \frac{\lambda^2}{256\pi^3} T^4\, ,
\ee 
where $\l$ is the 't Hooft coupling in the ${\cal N}=4$ super Yang-Mills in the large-N limit. This transport coefficient is also expected to depend on $B$ when B is non-vanishing, that is the case relevant for CME. This holographic calculation for ${\cal N}=4$ super Yang-Mills at strong coupling in the presence of a non-trivial magnetic field was calculated in \cite{Basar:2012gh} using the dual black-brane background  constructed in \cite{D'Hoker:2009mm}. 

We are however interested in the analogous results for the strongly interacting, non-conformal plasma, described by the ihQCD model. This calculation was carried out in \cite{Gursoy:2012bt} for vanishing B and in \cite{Drwenski:2015sha} for finite B. We only summarize the crucial ingredients and the results of these papers below, referring the reader to these papers for details. 

The bulk-axion  field can be introduced in the ihQCD model by adding to (\ref{fullaction}) a kinetic term of the form \cite{ihqcd2,ihqcd4}:  
\be\lab{sac}
S_a =  M_p^3 \int d^5 x \sqrt{-g} Z(\f) g^{\m\n} \6_\m a(x,r)  \6_\n a(x,r) \, .
\ee
Note that this term in the action is suppressed as $1/N_c^2$ compared to the two terms in (\ref{fullaction}) consistently with the fact that the physics associated with the dual operator (\ref{q}) in QCD is $1/N_c^2$ suppressed in the large-N limit. Practically this means that we do not take into account the backreaction of (\ref{sac}) on the background that results from (\ref{fullaction}) this the field $a(r,x)$ can be treated as a perturbation on top of the ihQCD background obtained in the previous sections.  

We included a non-trivial, dilaton dependent kinetic potential $Z(\f)$ in (\ref{sac}). Its presence in general is expected in compactifications of $IIB$ supergravity down to 5D. It is argued to be also present in the effective action of non-critical string theory in \cite{ihqcd1} the ihQCD model is based on. Solving the for axion field equation  resulting from (\ref{sac}) on the black-brane background (\ref{sol2}) one obtains the analytic result:
 \be\lab{ihqcdgcs}
\gcs = \frac{\kappa^2}{N_c^2} \frac{sT}{2\pi} Z(\f_h)\, ,
\ee 
where $s$ is the entropy density and $\f_h$ is the value of the dilaton at the horizon, and the constant $\kappa$ is defined in (\ref{UVa}).  The UV asymptotics of the function $Z(\f)$ in the $\f\to -\infty$ limit is fixed by the value of the topological susceptibility $\chi_t = \6^2\eps(\theta)/\6\theta^2$ where $\eps(\theta)$ is the $\theta$-dependent vacuum energy that is identified with the on-shell action $S_a$ (\ref{sac}) in the bulk. As shown in \cite{ihqcd2} this requires 
\be\lab{ZUV}
Z \to Z_0 \approx 33.25/\kappa^2, \qquad \f\to -\infty\, ,
\ee
where $Z_0$ is fixed using the lattice data for this topological susceptibility \cite{ihqcd2,Gursoy:2012bt}. On the other hand the IR asymptotics of the function $Z(\f)$ can be fixed by ``glueball universality'' that originates from linear confinement \cite{ihqcd2}, i.e. requiring that the axionic glueball states (excitations of the operator $q$ in (\ref{q})) carry a mass $m_n^2 \propto n$ in the limit of large excitation number $n\ gg 1$. One finds that this asymptotic behavior follows if one requires \cite{ihqcd2} 
\be\lab{ZIR}
Z \to c_4 e^{4\f} \qquad \f\to +\infty\, ,
\ee
where $c_4$ is a constant. The profile of the function $Z(\f)$ for intermediate values of $\f$ is not completely fixed, but one finds good match with lattice data if one parametrizes this function as 
\be\lab{Zfull}
Z(\f) = Z_0\le(1+ c_1 e^\f +  c_4 e^{4\f}\ri)\, , 
\ee
depending on two parameters $c_1$ and $c_4$. These constants can then be fixed by matching the lattice data \cite{Morningstar:1999rf}. One still finds a large allowed range for these parameters \cite{Gursoy:2012bt}: 
\be\lab{rangesc1c4} 
0< c_1 < 5,\qquad 0.06< c_4 < 50\, .
\ee 
Thus we necessarily have large systematic errors for the physics associated to the CP-odd term (\ref{sac}) in ihQCD. 
\begin{figure}
\includegraphics[scale=1.35]{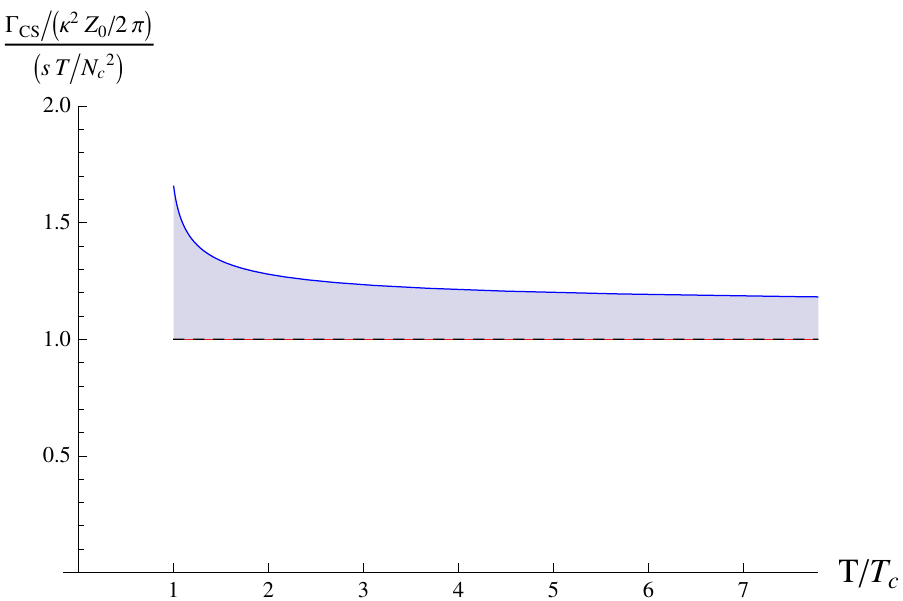}
 \caption{The sphaleron decay rate, properly normalized, as a function of temperature in ihQCD. The blue shaded region are the allowed values for this decay rate, the uncertainty arising from the uncertainty in fixing constand $c_1$ and $c_4$ in (\ref{rangesc1c4}). Plot taken from paper \cite{Gursoy:2012bt}.}
 \lab{figgcs}
\end{figure}
We show the result for the sphaleron decay rate $\gcs$ as a function of temperature in figure \ref{figgcs}. The allowed values for the decay rate is shown by the blue shaded region. The large systematic uncertainty follows from equation (\ref{rangesc1c4}) in parametrization of the function $Z(\f)$, (\ref{Zfull}). The decay rate shown in the plot is normalized by its value in the limit $\f_h\to \infty $ (large T)\footnote{The constant $\kappa$ that appears in this normalization is defined in (\ref{UVa}).}. We observe two salient features in figure \ref{figgcs}. First, that it is bounded from below as: 
\be\lab{boundgcs} 
\gcs(T) > \frac{\kappa^2}{N_c^2} \frac{s(T)T}{2\pi} Z_0,\qquad T>T_c\, , 
\ee
for all values of $T$ larger than the deconfinement transition temperature. Second, we observe\footnote{This can be shown analytically \cite{Gursoy:2012bt}.}) that it is a monotonically decreasing function of $T$. This means that the rate of sphaleron decays, hence rate of production of the axial chemical potential $\mu_5$ is largest above but close to the deconfinement temperature $T_c$, that is the regime most relevant to QGP physics. We emphasize that these are {\em universal features} that follow from ihQCD, regardless of the detailed choices made for the potentials $V$ and $Z$ that enter the ihQCD action.  
\begin{figure}
\includegraphics[scale=1]{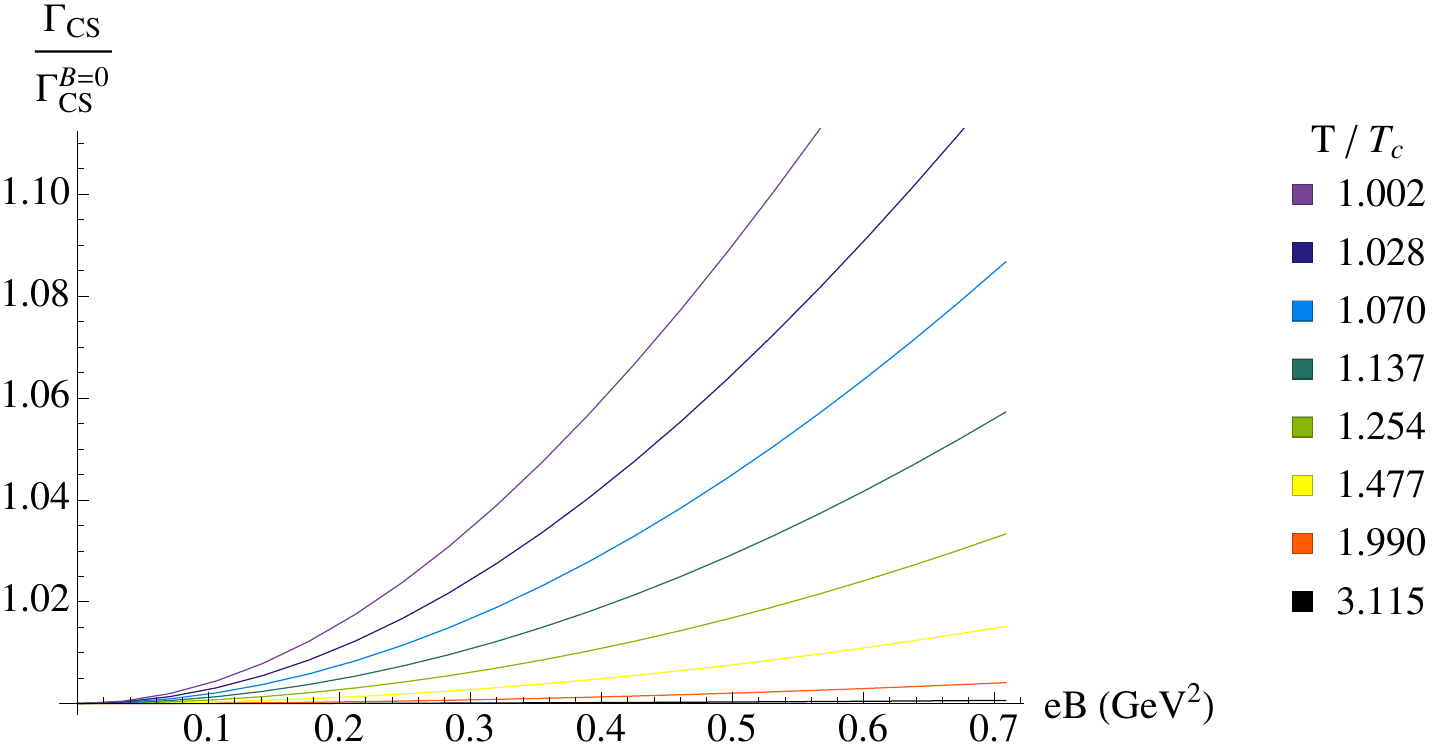}
 \caption{The sphaleron decay rate, normalized by its value at $B=0$ as a function of B for the various choices of T. Plot taken from paper \cite{Drwenski:2015sha}.}
 \lab{gcsB}
\end{figure}

It is tempting to compare the actual value for the decay rate we obtain from ihQCD for the non-conformal plasma, to the original conformal result (\ref{confgcs}). In the conformal case, for a typical choice one makes for $\l_t = 6\pi$ \cite{Gursoy:2012bt} one finds 
\be\lab{valconf} 
\frac{\gcs}{T^4}\bigg|_{conf}\approx 0.045\, .
\ee
On the other hand, if one calculates (\ref{ihqcdgcs}) at $T_c$ using values of parameters quoted above one finds 
\be\lab{valihqcd} 
2.8 > \frac{\gcs(T_c)}{T_c^4}\bigg|_{ihQCD} > 1.64\, .
\ee
This is much larger than the conformal value! We conclude that the rate of sphaleron decays, hence the production of CME in the non-conformal plasma modelled by ihQCD is much larger than the conformal plasma modelled by the AdS-Schwarzchild black-brane. 

Magnetic field dependence of the sphaleron decay rate was studied in \cite{Drwenski:2015sha}. For this calculation, one has to use the background with the flavor term that follows from (\ref{fullaction}). The analytic 
result (\ref{ihqcdgcs}) is still valid but one finds that now it is  a (different) function of T and B when the functions entering in this expression are expressed in terms of T and B. This is because $\f_h$ is now a function of both B and T, as the value of the dilaton at the horizon depends on the integration constants T and B chosen when solving the background field equations. Furthermore the entropy density $s$ that was a function of T before now becomes a function of also B, since the area of the horizon also depends on these integration constants. We show the result for the sphaleron decay rate in figure \ref{gcsB}.  

We again observe that the value of the decay rate increases with $T$ for any value chosen for $B$. On top of that we also observe that it also increases with increasing B for any choice of B. This leads to the holographic prediction that presence of a magnetic field further strengthens the rate of Sphaleron decays, hence fortifies the axial chemical potential $\mu_5$. One therefore expects a stronger production rate for the chiral magnetic effect for larger values of the magnetic field. 

\newpage

\section{Conclusion and a look ahead} 

In these lectures we aimed at a self-contained introduction to applications of the gauge/gravity duality in QCD, with emphasis on the quark-gluon plasma produced in the heavy ion collision experiments. We have explained the construction of the improved holographic QCD model, explained how to fix its parameters by comparison to lattice QCD data, the structure of the vacuum state and the thermal states, calculation of thermodynamic observables and comparison to the lattice data, the hydrodynamics and the transport coefficients such as the bulk and the shear viscosities, holographic treatment of energy loss of  hard probes in QGP, and finally the QGP under external magnetic fields. We argued that the model should be trusted up to a certain UV scale above which weak coupling effects are expected to invalidate the holographic correspondence. The model seem to successfully capture all the salient features of QCD in the IR and match very well with the thermodynamic observables and the hadron spectra calculated on the lattice.

Our predictions in the IR regime comes in two different flavors: the qualitative predictions and the quantitaive predictions.  The most important qualitative predictions are: a holographic connection between linear confinement and a discrete and gapped hadron spectrum, presence of a deconfinement temperature at finite T for any holographic gauge theory that exhibits linear confinement at zero T, a universal increase in transport coefficients such as the bulk viscosity and the sphaleron decay rate as temperature approaches the deconfinement temperature from above, and universal bounds on diffusion constants that describe the energy-momentum loss of probe quarks in the plasma. On top of this the quantitative predictions of the specific holographic model with parameters fixed by comparison to lattice data are also interesting. 

We have deliberately left out the various important topics:  

\bet
\item Fixing the improved holographic model in the Veneziano limit with large number of flavors is very important if we want it to agree with all available lattice data, not only qualitatively but also quantitatively. One needs to take into account the various aspects of flavor physics in this quest: the meson spectra, thermodynamic functions at finite baryon chemical potential, physics under external magnetic fields, anomalous transport etc. This task is hard but rewarding: once the model in this regime is completely fixed, then interesting predictions for the various other observables can be made. 

\item One open field of research is the full phase diagram of the improved holographic theory in the Veneziano limit on the full phase space parametrized by $(T,\mu,B)$. It may be very interesting to see if there are new, previously unknown phases in this phase space. It is also very important to obtain, at least qualitatively, the shape of the phase separation surfaces between, confined/deconfined, chirally symmetric/chirally broken etc. phases in this space. 

\item Fixing the CP-odd sector of the holographic model. We have seen in the last section that the kinetic term $Z(\f)$ of the bulk axion is weakly constrained in our theory, due to lack of lattice data. One potentially fruitful idea is to nail this function down by comparing {\em Euclidean} correlation functions that involve the operator $\tr F \wedge F$ with future lattice data. This requires establishing a long-term collaboration with lattice experts. 

\item Related to the previous point one may consider testing the holographic theory by comparing the Euclidean correlators of energy-momentum tensor, the topological charge operator and the $\tr F^2$ operator. In particular holographic calculation of the spectral densities associated with these operators may be very useful in analytically continuing the lattice data for the Euclidean correlators with the final aim of calculating the real-time correlators on the lattice. 

\item We have described how to calculate the glueball and the meson spectra in the holographic model leaving out  baryon spectrum in the improved holographic theory. The baryon sector of the theory is harder to treat in the holographic dual model and it constitutes a sub-field that needs to be developed. 

\item Another extremely important field that needs more attention is the process of {\em thermalization} in strongly coupled {\em non-conformal} gauge theories, such as the QCD. Indeed, one of the most important open problems in the heavy-ion physics is the precise mechanism(s) behind the rapid thermalization of the system of quarks and gluons produced in the heavy ion collisions into a nearly thermal state of the quark-gluon plasma. The study of holographic thermalization has started with the pioneering work of Chesler and Yaffe \cite{Chesler:2008hg}. This work and most of the subsequent developments focused on thermalization and out-of equilibrium physics in conformal rather than non-conformal plasmas. The latter only started attracting attention recently with the works \cite{Ishii:2015gia, Buchel:2015saa, Fuini:2015hba, Janik:2015waa,  Janik:2016btb, Gursoy:2015nza, Du:2015zcb, Attems:2016ugt, Gursoy:2016ggq}. We have not reviewed these developments in this review due to lack of space. 

\item Finally there are many open problems concerning the strongly interacting QGP under external magnetic fields. One of the most important is the question of renormalization in anomalous transport coefficients. In particular radiative or non-perturbative corrections to chiral magnetic and chiral vortical conductivities in the presence of dynamical gluon or photon field to the axial anomaly is an open problem that  can be explored using holographic methods \cite{Gursoy:2014ela, Jimenez-Alba:2014iia}. Another open problem related to magnetic fields is to use holographic methods to understand the physical reasons behind the inverse magnetic catalysis. A paper coming closest to this is \cite{GursoyIMC} but there is still much to be done to understand this phenomenon both on the field theory and the holography sides. Finally, in a very recent study \cite{Demircik:2016nhr}, a breakdown of hydrodynamical approximation in the presence of a magnetic field is observed. Whether this is an artifact of the holographic model or a similar effect can be observed in realistic systems, if so, whether this breakdown may have significant  consequences for the physics of QGP under magnetic fields are to be understood in future work. 

\eet

As a final word, I would like say that the quest for developing a realistic holographic model for QCD and QGP physics  has been and continues to be a great scientific journey that in the end provided us with a useful analytic tool in elucidating the problems that haunted the high energy community for a long time. I should apologize for my unintended omission of the various important references on the subject 
other than the ones contained in the References.

\section*{Acknowledgements}

The improved holographic QCD theory we reviewed in these lecture notes is developed in collaboration with many authors, especially Elias Kiritsis and Francesco Nitti. In addition to this I would like to acknowledge collaboration and useful discussions with Mohammad Ali-Akbari, Francesco Bigazzi, Alex Buchel, Aldo Cotrone, Tuna Demircik, Ioannis Iatrakis, Aron Jansen, Matti Jarvinen, Karl Landsteiner, Georgios Michalogiorgakis, Govert Nijs, Carlos Nunez, Andy O'Bannon, Marco Panero, Ioannis Papadimitriou, Giuseppe Policastro, Andreas Schafer, Wilke van der Schee and especially Liuba Mazzanti. Subsequent developments in the subject, in the context of V-QCD (the Veneziano limit of improved holographic QCD) resulted from the various works by Timo Alho, Daniel Arean, Francesco Bigazzi, Roberto Casero, Aldo Cotrone, Ioannis Iatrakis, Matti Jarvinen,  Keijo Kajantie, Elias Kiritsis, Carlos Nunez, Angel Paredes, Cobi Sonnenschein and Kimmo Tuominen. This work was supported, in part by  the Netherlands Organisation for Scientific Research (NWO) under VIDI grant 680-47-518, and the Delta-Institute for Theoretical Physics (D-ITP) that is funded by the Dutch Ministry of Education, Culture and Science (OCW). I am grateful to the organizers of the 56th Cracow School of Theoretical Physics in Spring 2016 in Zakopane, Poland for their hospitality. 

\newpage


\appendix
\section{Scalar variables}
\label{appA}%

First rewrite the Einstein's equations for the black-brane background in the domain-wall coordinate system 
\be\lab{sol2dw} 
 ds^2 =e^{2A(u)}(\frac{du^2}{f(u)}  + \delta_{ij}dx^i dx^j) - f(u) dt^2  \, ,
\ee
that is related to (the Lorentzian version of ) (\ref{sol2}) by a coordinate transformation $du = \exp(A{r}) dr$. The Einstein's equations (\ref{EE2}) in this coordinate system read: 
\be\lab{EE2dw} 
A''  = -\frac49 (\f')^2, \qquad 3 A'' + 12 (A')^2  + 3 A' \frac{f'}{f}= \frac{e^{2A}}{f}V(\Phi),\qquad f''+ 4A'f'=0\, .
\ee
Now define 
\be\lab{Ddef} 
D(\f) \equiv A'\, ,
\ee
and use the chain rule for derivatives to solve the first equation in (\ref{EE2dw}) for $D$: 
\be\lab{solD} 
D(\f) = -\frac{1}{\ell} e^{-\frac43\int^\f_0 d\f X(\f)}\, ,
\ee
where we used the definition (\ref{XYdef}). Then again using the chain rule in the third equation in (\ref{EE2dw})  and taking into account the definitions on obtains the equation of motion for the Y scalar variable (\ref{Y1eq}). Now, use the chain rule to rewrite  the second equation in (\ref{EE2dw}) in terms of $X$, $Y$, $D$ and their derivatives, take the logarithmic derivative of this equation with respect to $\f$, use the solution (\ref{solD}) and the equation (\ref{Y1eq}) derived above and simplify to obtain the equation of motion for the X scalar variable, 
equation (\ref{X1eq}). All in all we derived 
\bea\lab{X1eqapp}
\frac{dX}{d\f} & = & -\frac43~(1-X^2 + Y)\le(1+\frac{3}{8}\frac{1}{X}\frac{d\log V}{d\f}\ri)\, ,\\
\lab{Y1eqapp}
\frac{dY}{d\f} & = & -\frac43~(1-X^2 + Y) \frac{Y}{X}\, .
\eea
These are two first order equations. The total degree of the system of Einstein's equations is 5. The rest of the equations follow from (\ref{solD}) and the definitions (\ref{XYdef}) as 
\bea
\lab{Aeqapp} 
A' &=& -\frac{1}{\ell} e^{-\frac43\int^\f_0 d\f X(\f)}\, \\
\lab{phieqapp} 
\f' &=& -\frac{3X}{\ell} e^{-\frac43\int^\f_0 d\f X(\f)}\, \\
\lab{feqapp} 
g' &=& -\frac{4Y}{\ell} e^{-\frac43\int^\f_0 d\f X(\f)}\, 
\eea
where we defined $g = \log f$. These equations complete the system. The corresponding equations for the thermal gas solution (\ref{sol1}) can be obtained from these by setting $Y=0$.

\section{The Potentials}
\label{appB}%

In this appendix we list the potentials of the V-QCD model. We define $\lambda = \exp \f$. The potentials read: 

\begin{eqnarray}
\label{Vf0SB}
V_g(\lambda)&=&{12\over \mathcal{L}_0^2}\biggl[1+{88\lambda\over27}+{4619\lambda^2
\over 729}{\sqrt{1+\ln(1+\lambda)}\over(1+\lambda)^{2/3}}\biggr]\, , \\
 V_{f0}& =& {12\over \mathcal{L}_{UV}^2}\biggl[{\mathcal{L}_{UV}^2\over\mathcal{L}_0^2}
-1+{8\over27}\biggl(11{\mathcal{L}_{UV}^2\over\mathcal{L}_0^2}-11+2x \biggr)\lambda\nn\\
 &&+{1\over729}\biggl(4619{\mathcal{L}_{UV}^2\over \mathcal{L}_0^2}-4619+1714x - 92x^2\biggr)\lambda^2\biggr] \, , \nn \\
 \kappa(\l) &=& {[1+\ln(1+\l)]^{-1/2}\over[1+\frac{3}{4}(\frac{115-16x }{27}-{1\over 2})\l]^{4/3}} \,, \quad\quad a(\l)=\frac{3}{2 \, \mathcal{L}_{UV}^2} \, ,
\label{kappaa}
 \end{eqnarray}
where $\mathcal{L}_{UV}$ is  the AdS radius, so that the boundary expansion of the metric is 
$ A \sim \ln \left( { \mathcal{L}_{UV} / r} \right)+\cdots \, .$
The radius depends on $x$ as 
\be
\mathcal{L}_{UV}^3 = \mathcal{L}_0^3 \left( 1+{7 x \over 4} \right) \, .
\label{adsrad}
\ee
The function $w$ is parametrized by a single parameter $c$ 
\be
w(\l)=\kappa(c\l) =  \frac{( 1+\log(1+ c \, \l))^{-{1\over 2}}}{\left(1+ {3 \over 4} \left({115-16 x \over 27}-{1\over 2} \right) c \,\l  \right)^{4/3}}       \, ,
\label{wl}
\ee
where $x$ is the ratio of the number of flavors to color. 

\end{document}